\documentclass{aa}
\usepackage{graphicx}

\def\deltamin{\mbox{\,$\delta_{\rm min}$}}
\def\hminus{\mbox{\,$h^{-1}$}}
\def\betap{\mbox{\,$\beta_{\rm app}$}}
\def\masyr{\mbox{~mas~yr$^{-1}$}}

\def\lsim{\!\!\!\phantom{\le}\smash{\buildrel{}\over
{\lower2.5dd\hbox{$\buildrel{\lower2dd\hbox{$\displaystyle<$}}\over
                                  \sim$}}}\,\,}
\def\gsim{\!\!\!\phantom{\ge}\smash{\buildrel{}\over
{\lower2.5dd\hbox{$\buildrel{\lower2dd\hbox{$\displaystyle>$}}\over
                               \sim$}}}\,\,}

\def\et{et al.}                        

\newcommand\mconc[1]{\multicolumn{1}{c}{#1}}
\newcommand\mctwc[1]{\multicolumn{2}{@{}c@{}}{#1}}
\newcommand\mcthc[1]{\multicolumn{3}{c}{#1}}

\newcommand\mcfic[1]{\multicolumn{5}{c}{#1}}

\begin{document}

   \title{Absolute kinematics of radio source components in the complete S5
polar cap sample}

   \subtitle{II. \fontshape{it}{First} and second epoch maps at 15\,GHz}

   \author{M.A.\ P\'erez-Torres\inst{1,2}
\and J.M.\ Marcaide\inst{3}
\and J.C.\ Guirado\inst{3}
\and E.\ Ros\inst{4}
          }

   \offprints{M.A.\ P\'erez-Torres, torres@iaa.es}

   \institute{Istituto di Radioastronomia, Via P. Gobetti 101, 40129 Bologna, Italy
\and 	 
   Instituto de Astrof\'{\i}sica de Andaluc\'{\i}a (CSIC), Apdo. Correos
	 3004, E-18080 Granada, Spain
\and
Departament d'Astronomia i Astrof\'{\i}sica, Universitat de Val\`encia,
E-46100 Burjassot, Val\`encia, Spain
\and
Max-Planck-Institut f\"ur Radioastronomie,
Auf dem H\"ugel 69, D--53121 Bonn, Germany
}

\date{Submitted: 10 March 2004 / Accepted: 27 July 2004}

\abstract{
We observed the thirteen extragalactic radio sources  
of the complete S5 polar cap sample at 15.4\,GHz with the Very Long 
Baseline Array, on 27 July 1999 (1999.57) and 15 June 2000 (2000.46).
We present the maps from those two epochs, along
with maps obtained from observations of the 2\,cm VLBA survey
for some of the sources of the sample, making a total of 40 maps.
We discuss the apparent morphological changes  displayed by
the radio sources  between the observing epochs.
Our VLBA observations correspond to the first two epochs at 15.4\,GHz 
of a program to study the absolute kinematics of the radio source 
components of the members of the sample, by means of phase delay
astrometry at 8.4\,GHz, 15.4\,GHz, and 43\,GHz.

~~~~Our 15.4\,GHz VLBA imaging allowed us to disentangle the 
inner milliarcsecond structure of some of the sources, thus resolving
components that appeared blended at 8.4\,GHz.
For most of the sources, we identified the brightest 
feature in each radio source with the core. 
These identifications are supported by the spectral index
estimates for those brightest features, which are in general flat, 
or even inverted.
Most of the sources display core-dominance in the overall
emission, as given by the core-to-extended ratio in the 15.4\,GHz frame, $Q$.
We find that three of the sources have their most inverted spectrum component
shifted with respect to the origin in the map, which approximately 
coincides with the peak-of-brightness at both 15.4\,GHz and 8.4\,GHz. 

}

\authorrunning{P\'erez-Torres et al.}
\titlerunning{Absolute kinematics of radio source components. II.}

\maketitle


\section{Introduction\label{sec:introduction}}

This paper is the second of a series aimed at conducting absolute
kinematics studies of a complete sample of extragalactic radio sources, 
using astrometry techniques.
The sample consists of the thirteen sources from the S5 polar cap
sample, selected by Eckart \et\ (\cite{eck86,eck87}) from
the  S5 survey (K\"uhr \et\ \cite{kuh81}),
that satisfy the following criteria:
(i) $\delta\ge70^{\circ}$, 
(ii) $|b_{\rm II}|\ge10^{\circ}$,
(iii) $S_{\rm 5\,GHz}\ge1$\,Jy at the epoch of the survey, and 
(iv) $\alpha_{\rm 2.7,\,5\,GHz}\ge-0.5$ ($S\sim\nu^{+\alpha}$).  
Hereafter, we shall refer to those  thirteen radio sources as 
``the complete S5 polar cap sample".

We described the goals of our astrometric program in  
Ros \et\ (\cite{ros01}, hereafter Paper I),  
and presented 8.4\,GHz VLBA maps obtained at two
different epochs, 1997.93 and 1999.41. 
We refer the reader to that paper for further details.
We recall here that this ongoing astrometric program 
is intended to test the absolute kinematics of the innermost regions
of the sources of the complete S5 polar cap sample, 
with precisions better than $\sim$(20 to 100)\,$\mu$as, 
depending on the observing frequency.
We expect to obtain such accurate results only after 
the mapping and astrometric reduction of our observations are completed.
The determination of the detailed kinematics 
of the compact components of the complete S5 polar cap sample
will probably be a decisive element in our understanding 
of the activity around the cores of these compact
radio sources, and in a definitive test of the standard jet model.

We present here our 15.4\,GHz maps obtained with the Very Long Baseline Array
on 1999.57 and 2000.46.  
We also present maps obtained from observations of
some of the sources of the complete sample that are targets of the
2\,cm VLBA survey (Kellermann \et\ \cite{kellermann98},
Zensus \et\ \cite{zen02}). 
We discuss in detail the apparent changes in 
the brightness distribution of 
some of the radio sources during the time covered by the 
15.4\,GHz observations, as well as a comparison with the results
presented at 8.4\,GHz (Paper I).
Throughout the paper, we use a Hubble constant 
$H_0$=$65\,h$\,km\,s$^{-1}$\,Mpc$^{-1}$ and a deceleration parameter
$q_0$=$0.5$.

\begin{table*}[bhtp]
\begin{flushleft}
\caption{Map parameters for {sources of} the complete 
S5 polar cap sample at 15.4\,GHz.}
\label{table:results-mapping}
\[
\centering
\resizebox{\textwidth}{!}{%
  \begin{tabular}{r@{~~}c@{$\times$}rrrc@{~}c@{~~}c@{$\times$}crrr@{~}c}
& \mcfic{----------- Epoch 1999.57 -----------} &
& \mcfic{----------- Epoch 2000.46 -----------} 
& 
\\
& \mcthc{Beam FWHM$^{\rm a}$} & & & 
& \mcthc{Beam FWHM$^{\rm a}$} & & 
\\
\mconc{\bf Source}  
& \mctwc{Size} & {P.A.} & {$S_{\rm peak}$} & {$S_{\rm tot}$$^{\rm (b)}$} 
& {\scriptsize rms level$^{\rm c}$}
& \mctwc{Size} & {P.A.} & {$S_{\rm peak}$} & {$S_{\rm tot}$$^{\rm (b)}$}
& {\scriptsize rms level$^{\rm c}$}
\\
& \mctwc{[mas]} & & {\scriptsize [mJy/beam]} & {\scriptsize [mJy]} 
& {\scriptsize [mJy/beam]}
& \mctwc{[mas]} & & {\scriptsize [mJy/beam]} & {\scriptsize [mJy]}
& {\scriptsize [mJy/beam]}
\\ \hline
{\bf \object{QSO\,0016+731}} & 0.619 & 0.396 & 6\fdg4	 &  487 &  647 & 0.5 & 0.643 & 0.392 &   6\fdg9  & 694 & 817 & 0.6 \\
{\bf \object{QSO\,0153+744}} & 0.860 & 0.352 & --6\fdg9  &  163 &  316 & 1.0 & 0.853 & 0.345 & --4\fdg9  & 169 & 312 & 0.7 \\
{\bf \object{QSO\,0212+735}} & 0.899 & 0.371 & --3\fdg3	 & 1025	& 2529 & 0.5 & 0.869 & 0.366 & --3\fdg4  &1032 & 2507 & 0.6 \\
{\bf \object{BL\,0454+844}}  & 0.425 & 0.362 & 57\fdg5	 &   59	&  120 & 0.4 & 0.393 & 0.370 &  70\fdg9   & 71 & 131 & 0.4 \\
{\bf \object{QSO\,0615+820}} & 0.451 & 0.401 & 22\fdg6	 &  129	&  434 & 0.7 & 0.475 & 0.392 &  16\fdg4  & 148 & 410 & 0.7 \\
{\bf \object{BL\,0716+714}}  & 0.853 & 0.344 & --4\fdg9	 & 1024	& 1107 & 0.8 & 0.845 & 0.326 & --2\fdg7  & 968 & 1051 & 0.7 \\
{\bf \object{QSO\,0836+710}} & 0.885 & 0.345 & --15\fdg9 & 1259	& 1766 & 0.8 & 0.898 & 0.336 & --12\fdg5 & 1147& 1728 & 0.6 \\
{\bf \object{QSO\,1039+811}} & 0.621 & 0.390 & --18\fdg3 &  737	&  926 & 0.6 & 0.546 & 0.365 & --11\fdg7 & 584 & 826 & 0.9 \\
{\bf \object{QSO\,1150+812}} & 0.626 & 0.310 &--39\fdg0	 &  316 &  946 & 0.8 & 0.659 & 0.345 & --37\fdg2 & 358 & 907 & 0.6 \\
{\bf \object{BL\,1749+701}}  & 1.044 & 0.359 &--7\fdg3	 &  229	&  382 & 0.7 & 1.268 & 0.405 & --16\fdg8 & 261 &  405 & 0.9 \\
{\bf \object{QSO\,1803+784}}  & 0.571 & 0.374 &--3\fdg0	 & 1965 & 2878 & 0.8 & 0.533 & 0.426 & 15\fdg3   & 1524& 2313 & 0.9 \\
{\bf \object{QSO\,1928+738}} & 1.090 & 0.376 & 15\fdg9	 & 1492 & 3006 & 1.0 & 1.371 & 0.394 & 8\fdg1	 & 1214& 2889 & 0.8 \\
{\bf \object{BL\,2007+777}}  & 0.487 & 0.429 & --30\fdg1 &  384	&  825 & 0.4 & 0.488 & 0.422 &--49\fdg1  & 680 & 1258 & 0.5 \\ \hline
\end{tabular}
}
\]
\begin{list}{}{
\setlength{\leftmargin}{0pt}
\setlength{\rightmargin}{0pt}
}
\item[$^{\rm a}$] The restoring beam is an elliptical Gaussian
with full-width-half-maximum (FWHM) axes $a\times b$.
{For each source, the} position angle (P.A.) 
stands for the direction of the
major {axis} measured north through east.
\item[$^{\rm b}$] Total flux density recovered in the {hybrid}
mapping process.
\item[$^{\rm c}$] Contours in the maps of the figures shown  in 
Sects.~\ref{subsec:0016} to \ref{subsec:2007}
are the tabulated rms values times $(-3,3,3\sqrt{3},9,\cdots)$.
\end{list}
\end{flushleft}
\end{table*}


\section{Observations\label{sec:observations}}
We observed the complete S5 polar cap sample at 15.4\,GHz on 27 July 1999 (epoch
1999.57) and on 15 June 2000 (epoch 2000.46) with the VLBA, each time for 24 hours. 
For the first epoch, we used the standard VLBA recording mode 128--8--1
(data rate of 128 Mbps, recording in 8 IFs of 16 channels each), 
in left circular polarization, thus yielding a total bandwidth of 64\,MHz. 
For scheduling reasons, in our second epoch we used the 
standard recording VLBA mode 64--4--1
(data rate of 64 Mbps, recording in 4 IFs of 8 channels each), 
also in left circular polarization, 
yielding a total bandwidth of 32\,MHz.
All data were correlated at the 
VLBA Array Operations Center of the National Radio 
Astronomy Observatory (NRAO) in Socorro, New Mexico, 
using a basic integration time of 4\,s. 
All radio sources were detected and provided fringes for all baselines.  
We cycled around in {groups} of three or four radio sources, 
with duty cycles of about 5\,min.
Every scan was 55\,s long, and antenna slew times were $\lsim$20\,s.
We replaced one or two members of the group
{of observed sources} by new
ones every two hours approximately, until all thirteen sources
of the sample had been observed. 
We tracked the clock behaviour---of relevance for the astrometric 
analysis---by including some scans of 
\object{BL\,0454+844} in all groups observed during the
first (second) half of the experiment 
on 27 July 1999 (15 June 2000), and of \object{BL\,2007+777} in 
all {groups} observed during the second (first) half.
 
After fringe-fitting the correlator output, we obtained correlation amplitudes 
for each source. We calibrated the visibility amplitudes in
a standard manner, using the antenna gain and
system temperature information provided by each station.  
We then used the Caltech VLBI Package
(Pearson \cite{pea91}) and the difference
mapping software {\sc difmap} (Shepherd \et\ \cite{she94})
to obtain the radio images shown in this paper.


\section{Imaging results\label{sec:results}}

We imaged all thirteen radio sources using standard hybrid mapping techniques.  
We found minor amplitude calibration problems for the Saint Croix
antenna at epoch 1999.57; 
the noise level was high for most of the observing run.
Nevertheless, we used the data for Saint Croix, since it 
provides long baselines, and hence high resolution.
We show the maps obtained in Figs.\ \ref{fig:map0016} through \ref{fig:map2007}. 
In all figures, east is left, and north is up.
{We list the main parameters of those maps in Tables~\ref{table:results-mapping}
 and \ref{table:2cm-survey}.}

We also present images of several sources of the S5 polar cap sample
obtained from observations of the VLBA 2\,cm Survey that were relatively 
close in time to our observations.
We used the same data reduction and standard hybrid 
mapping techniques that were used with our own data.
To distinguish the 2\,cm VLBA survey maps from ours, 
we have labelled the latter ``2\,cm survey''.

We now proceed to discuss the apparent morphological changes 
displayed by each of the radio sources of the complete polar cap sample, as
seen from 15.4\,GHz VLBA observations made in
1999.57 and 2000.46, complemented with maps of some of the sources, 
which  we obtained from 
observations of the 2\,cm VLBA Survey 
(Kellermann \et\ \cite{kellermann98}; Zensus \et\ \cite{zen02}; see
also http://www.nrao.edu/2cmsurvey).
To quantify the morphological changes 
displayed by the radio sources, 
we used the tasks {\sc modfit} and {\sc erfit} of
the Caltech package (Pearson \cite{pea91}) to
fit the visibility data to a model of elliptical components with
Gaussian brightness profiles.  
Those models are listed in Tables~\ref{table:modelfit} 
and \ref{table:2cm-survey}.

We compare our observations with those taken
at  8.4\,GHz with the VLBA in 1997.93,
and in particular in 1999.41 (very close to our first epoch,
1999.57), and presented in Paper I.
These yield the spectral index, $\alpha$ 
($S\sim\nu^\alpha$) for those 15.4\,GHz
source components that have clear counterparts at 8.4\,GHz.
We refer the reader to Paper I for a comprehensive discussion of 
historical details for each radio source.

\begin{figure*}[htbp]
\label{fig:comp_positions}
\hspace{40pt}
 \includegraphics[angle=0,width=0.85\textwidth]{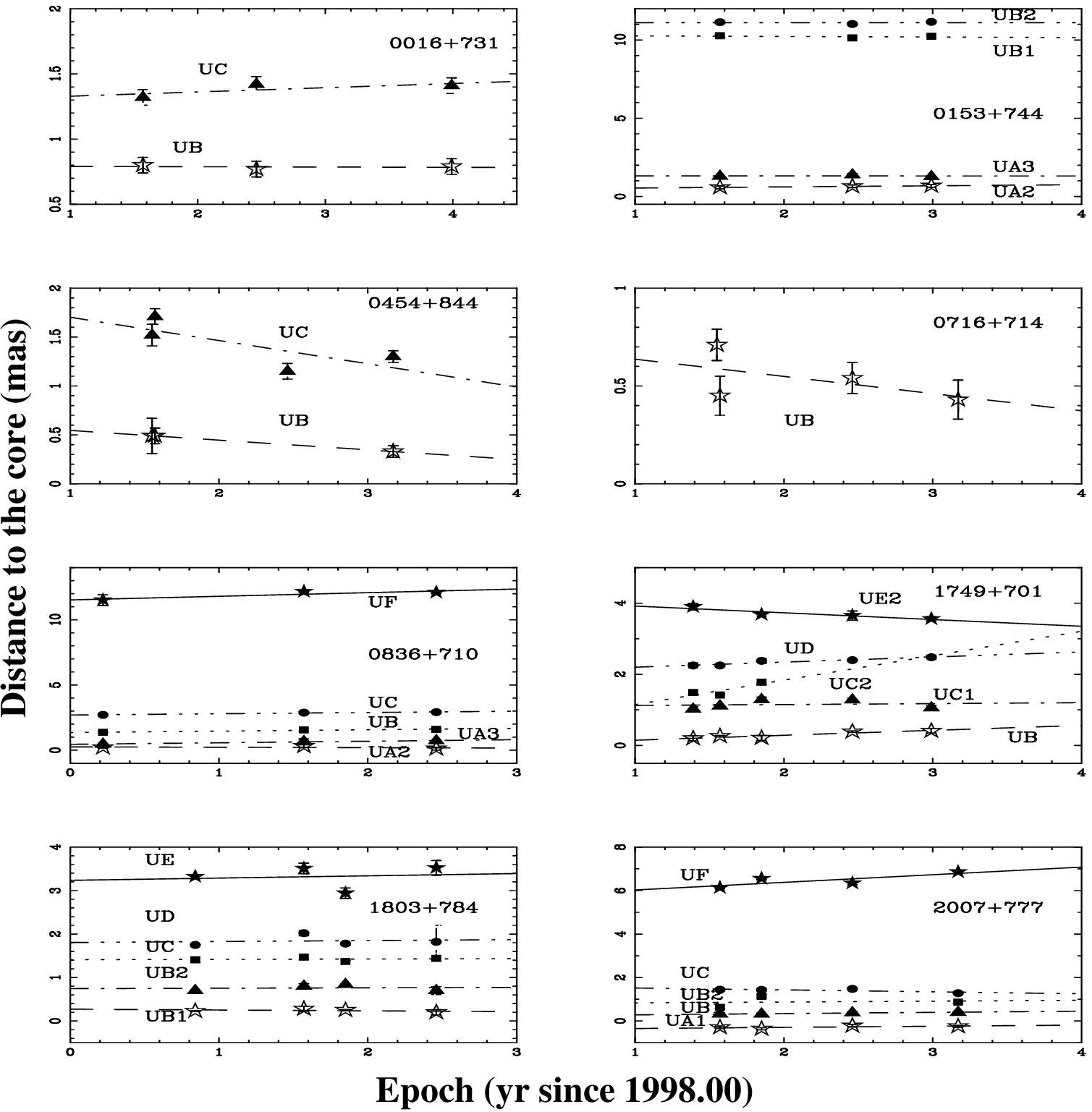}
\caption{
Angular distance to the core vs. time elapsed since 1998.0.
The plots show the change in angular separation with time of source 
features for which we have measured an angular velocity from observations
at three or more epochs. The lines are linear fits to the data, 
the slope representing the proper motion, $\mu$, tabulated in 
Table~\ref{tab:proper_motion}.
}
\end{figure*}

We also show in Fig.~\ref{fig:comp_positions} the distance to the core for
sources of the S5 polar cap sample that were observed at least three times, 
and in Table~\ref{tab:proper_motion} we present proper motions 
obtained from model fitting source component positions at different epochs.
We list proper motions only for those components
that were identified in at least three epochs, and whose proper motion
exceeded 1$\sigma$.


\begin{table*}[htbp]
\begin{flushleft}
\caption{Elliptical Gaussian component model parameters for the 
radio sources
of the complete S5 polar cap sample.  
}
\renewcommand{\baselinestretch}{0.75}
\label{table:modelfit}
\[
\centering
\resizebox{0.88\textwidth}{!}{%
\begin{tabular}{@{}l@{}c@{~}r@{$\pm$}l@{~}r@{$\pm$}l@{~}r@{$\pm$}l@{~~}r@{$\pm$}l@{~~}r@{$\pm$}l@{}r@{$\pm$}l@{~}r@{$\pm$}l@{}r@{$\pm$}l@{~}r@{$\pm$}l@{~}r@{$\pm$}l@{~~}r@{$\pm$}l@{~~}r@{$\pm$}l@{~~}r@{$\pm$}l@{}r@{$\pm$}l@{}}
    &  & 
\multicolumn{14}{c}{
--------------------- 
Epoch 1999.57
--------------------- 
} &
\multicolumn{14}{c}{
---------------------
Epoch 2000.46
--------------------- 
} \\
Source  &
Comp.\   
    & \mctwc{$S$} & \mctwc{$r$} & \mctwc{$\theta$} & \mctwc{$a$} & \mctwc{$b/a$} & \mctwc{$\phi$} 
    & \mctwc{$S$} & \mctwc{$r$} & \mctwc{$\theta$} & \mctwc{$a$} & \mctwc{$b/a$} & \mctwc{$\phi$} \\
    &
    & \mctwc{[mJy]} & \mctwc{[mas]} & \mctwc{} & \mctwc{[mas]} & \mctwc{} & \mctwc{} 
    & \mctwc{[mJy]} & \mctwc{[mas]} & \mctwc{} & \mctwc{[mas]} & \mctwc{} & \mctwc{} \\ 
\hline
%
%
%
0016+731 &
 UA  &  510& 2 & \mctwc{0}  & \mctwc{--} & 0.11&0.05 &  0.75&0.06 & $-61$& 7
     &  718& 2 & \mctwc{0}  & \mctwc{--} & 0.09&0.00 &  0.65&0.09 & $-94$& 4 \\  
&UB  &   96&13 &  0.80&0.06 & $137$&1  & 0.57&0.04 &  0.89&0.05 & $107$&16
     &  59&15  &  0.77&0.06 & $139$&1  & 0.61&0.04 &  0.96&0.17 & $ 17$&258 \\
&UC  &   43&13 &  1.32&0.06 & $124$&1  & 0.84&0.07 &  0.82&0.09 & $155$&31 
     &   41&14 &  1.42&0.06 & $129$&2  & 0.75&0.15 &  0.81&0.18 & $105$&19 \\ 
\hline
%
%
0153+744 &
 UA1 &  176&13  & \mctwc{0}  & \mctwc{--} & 0.12&0.07 & 0.15&1.41 & $63$&21
     &  188& 2  & \mctwc{0}  & \mctwc{--} & 0.12&0.01 & 0.54&5.93 & $81$&12 \\  
&UA2 &   50&17  & 0.58&0.09  & $70$&2   & 0.51&0.22 & 0.33&0.21 & $-80$&19
     &   38& 4  & 0.65&0.09  & $65$&1   & 0.48&0.08 & 0.32&1.32 & $89$&10 \\
&UA3 &   25& 5  & 1.29&0.09  & $83$&1   & 0.63&0.08 & 0.33&0.11 & $139$&9
     &   22& 2  & 1.38&0.09  & $95$&2   & 1.00&0.13 & 0.33&0.07 & $-23$&5.4 \\ 
&UB1 &   43& 4  & 10.27&0.09 & $153$&1  & 1.14&0.12 & 0.87&0.12 & $-66$&22
     &   36& 5  & 10.13&0.09 & $154$&1  & 1.35&0.12 & 0.65&0.12 & $34$&10 \\
&UB2 &   20& 4  & 11.14&0.09 & $151$&1  & 0.73&0.12 & 0.59&0.10 & $-32$&9
     &   30& 5  & 11.02&.09  & $151$&1  & 0.90&0.12 & 0.66&0.15 & $-73$&18 \\
\hline
%
%
0212+735 &
UA  & 1123&71 & 0.55&0.09  & $-49$&2 & 0.25&0.02 & 0.43&0.03 & $-52$&3
    & 1233&31 & 0.59&0.09  & $-44$&2 & 0.30&0.01 & 0.31&0.06 & $-60$&3 \\
&UB1 &  674&97 & \mctwc{0}  & \mctwc{--}& 0.52&0.10 &  0.53&0.14 & $-28$&6 
    &  411&118& \mctwc{0}  & \mctwc{--}& 0.59&0.10 &  0.48&0.13 & $-55$&6\\
&UB2 &  595&160& 0.26&0.09  & $+90$&7 & 0.29&0.03 & 0.82&0.14 & $-172$&8
    &  708&111& 0.25&0.09  & $+70$&11& 0.32&0.03 & 0.89&0.04 & $-141$&38 \\
&UC &   88&17 & 0.90&0.16  & $97$&2  & 1.75&0.37 & 0.44&0.11 & $103$&5
    &   74&13 & 0.90&0.12  & $93$&2  & 1.37&0.23 & 0.43&0.11 & $89$&5 \\
&UD &   43&4  & 2.51&0.09  & $101$&1 & 0.79&0.06 & 0.46&0.11 & $118$&6
    &   47&3  & 2.49&0.09  & $ 99$&1 & 0.95&0.08 & 0.66&0.06 & $132$&7\\
&UG &   26&2  & 13.81&0.09 & $ 93$&1 & 1.87&0.18 & 0.41&0.12 & $111$&6
    &   32&3  & 13.71&0.09 & $ 92$&1 & 2.29&0.23 & 0.43&0.09 & $115$&6\\
\hline
%
0454+844 &
UA   &  93&5 & \mctwc{0}&  \mctwc{--}& 0.37&0.01 &  0.49&0.04 & $-1$&2 
     & 121&1 & \mctwc{0}&  \mctwc{--}& 0.35&0.01 &  0.27&0.03 & $-2$&1    \\
&UB$^{\rm a}$  &  20&5 & 0.49&0.08 & $137$&3  & 0.65&0.12 &  0.73&0.16 & $-31$&19 
             &   \mctwc{} & \mctwc{}& \mctwc{}& \mctwc{}& \mctwc{}& \mctwc{}  \\
&UC$^{\rm a}$  &  10&1 & 1.71&0.08 & $172$&1  & 1.01&0.09 &  0.58&0.08 & $-48$&8 
               &  19&1 & 1.15&0.08 & $158$&1  & 1.77&0.09 &  0.31&0.04 & $-31$&4   \\
\hline
%
%
0615+820 &
 UA1 &  221&8 & \mctwc{0}& \mctwc{--} & 0.34&0.01 & 0.89&0.03 & $-18$&19 
     &  242&4 & \mctwc{0}& \mctwc{--} & 0.41&0.01 & 0.46&0.05 & $138$&1\\
&UA2 &  131&6 & 0.58&0.05 & $-73$&2   & 1.35&0.03 & 0.50&0.04 & $37$&2
     &   71&3 & 0.84&0.05 & $-99$&1   & 1.35&0.04 & 0.31&0.03 & $17$&2\\
&UA3 &   80&4 & 0.71&0.05 & $-147$&1  & 0.25&0.04 & 0.14&0.32 & $-119$&9 
q     &  107&3 & 0.74&0.05 & $-152$&1  & 0.25&0.03 & 0.10&0.93 & $40$&4 \\
\hline
%
%
0716+714 &
 UA  &  1058&5   & \mctwc{0}& \mctwc{--}& 0.17&0.01 & 0.21&0.05 & $22$&2
     &   981&17  & \mctwc{0}& \mctwc{--}& 0.12&0.01 & 0.10&0.88 & $36$&8 \\
&UB  &   53&5    & 0.71&0.08 & $17$&1     & 1.49&0.09 & 0.16&0.02 & $7$&2
     &   62&17   & 0.45&0.10 & $30$&4     & 0.77&0.10 & 0.11&0.11 & $4$&4 \\
\hline
%
0836+710 &
UA1  & 1307&121 &  \mctwc{0}& \mctwc{--} & 0.20&0.02 & 0.21&0.21 & $+25$&9 
     & 1238&124 &  \mctwc{0}& \mctwc{--} & 0.31&0.03 & 0.31&0.07 & $+34$&3 \\
&UA2 &  213&93  &  0.31&0.11& $-138$&6   & 0.21&0.21 & 0.98&1.47 & $-160$&180
     &  226&120 &  0.12&0.10& $-138$&39  & 0.32&0.10 & 0.76&1.14 & $-137$&62 \\
&UA3 &   61&66  &  0.66&0.20& $-129$&4   & 0.30&0.22 & 0.91&0.96 & $+32$&48
     &   84&7   &  0.73&0.09& $-129$&1   & 0.33&0.04 & 0.51&0.32 & $48$&19\\
&UB  &   29&4   &  1.55&0.09& $-139$&1   & 0.48&0.13 & 0.79&0.40 & $+59$&71
     &   30&4   &  1.60&0.09& $-139$&1   & 0.65&0.15 & 0.57&0.22 & $+48$&19  \\
&UC  &  100&2   &  2.88&0.09& $-144$&1   & 0.69&0.03 & 0.68&0.04 & $+30$&6
     &  101&2   &  2.92&0.09& $-144$&1   & 0.67&0.04 & 0.67&0.03 & $+13$&4  \\
&UF  &   53&3   & 12.16&0.09& $-148$&1   & 2.76&0.17 & 0.46&0.05 & $-8$&4
     &   53&2   & 12.09&0.09& $-147$&1   & 2.58&0.14 & 0.38&0.03 & $-1$&2  \\ \hline
%
%
1039+811 &
UA    & 883&2  & \mctwc{0}&  \mctwc{--}& 0.27&0.01  & 0.34&0.01 & $100$&1 
      & 785&2  & \mctwc{0}&  \mctwc{--}& 0.32&0.01  & 0.43&0.01 & $101$&1 \\
&UB$^{\rm a}$  &  29&6  & 1.12&0.07 & $-68$&2   & 1.13&0.24  & 0.26&0.11 & $-80$&6 
               &  48&2  & 1.68&0.06 & $-65$&1   & 2.22&0.15  & 0.11&0.04 & $109$&1 \\
&UC$^{\rm a}$  &  23&4  & 2.38&0.11 & $-67$&1   & 1.36&0.18  & 0.09&0.12 & $-92$&4 
               &   \mctwc{} & \mctwc{}& \mctwc{}& \mctwc{}& \mctwc{}& \mctwc{}  \\ \hline
%
%
1150+812 &
UA  &  405&35 & \mctwc{0} & \mctwc{--} & 0.25&0.03 & 0.32&0.17 & $63$&5
    &  355&72 & \mctwc{0} & \mctwc{--} & 0.14&0.07 & 0.20&0.19 & $60$&12 \\
&UB1&  157&38 & 0.33&0.06 & $233$&2   & 0.19&0.07 & 0.17&0.15 & $69$&17
    &  172&78 & 0.34&0.06 & $239$&3   & 0.29&0.12 & 0.84&0.41 & $31$&102 \\
&UB2&  113&14 & 0.65&0.06 & $221$&2   & 0.65&0.04 & 0.44&0.05 & $-15$&5
    &   78&29 & 0.74&0.09 & $218$&3   & 0.48&0.14 & 0.17&0.50 & $23$&12 \\
&UC &  102&5  & 1.97&0.06 & $170$&1   & 2.15&0.76 & 0.23&0.25 & $-26$&1
    &  130&10 & 1.89&0.09 & $182$&1   & 1.62&0.15 & 0.35&0.04 & $172$&1 \\
&UD &  153&3  & 2.21&0.06 & $183$&1   & 0.66&0.02 & 0.76&0.02 & $-6$&3
    &  158&14 & 2.31&0.06 & $178$&1   & 0.90&0.02 & 0.62&0.04 & $-71$&2 \\
&UE &   55&4  & 3.59&0.13 & $168$&1   & 2.10&0.13 & 0.74&0.06 & $-59$&10
    &   28&4  & 4.23&0.20 & $162$&1   & 3.11&0.46 & 0.26&0.05 & $-26$&2 \\ \hline
%
1749+701 &
UA & 198&6 & \mctwc{0} & \mctwc{--}& 0.26&0.21 &  0.22&0.17 & $-17$&15
    & 218&8 & \mctwc{0} & \mctwc{--}& 0.42&0.27 &  0.19&0.15 & $-36$&19  \\
&UB& 78&6  & 0.26&0.02 & $-48$&3   & 0.43&0.23 &  0.11&0.10 & $-70$&25
    & 105&8 & 0.39&0.02 & $-48$&2   & 0.31&0.17 &  0.09&0.07 & $-88$&23 \\
&UC1 &  12&1 & 1.11&0.04 & $-61$&3   & 0.28&0.20 &  0.30&0.16 & $-169$&35
    &  16&1 & 1.28&0.05 & $-63$&2   & 0.45&0.27 &  0.56&0.23 & $143$&28 \\
&UC2 &  15&2 & 1.42&0.04 & $-77$&2   & 0.42&0.23 &  0.63&0.19 & $48$&15
    &   \mctwc{} & \mctwc{}& \mctwc{}& \mctwc{}& \mctwc{}& \mctwc{}  \\
&UD &  31&2 & 2.25&0.02 & $-73$&2   & 0.74&0.32 &  0.47&0.21 & $51$&8
    &  47&2 & 2.40&0.02 & $-70$&1   & 0.66&0.30 &  0.43&0.17 & $71$&14 \\
&UE1 &  55&3 & 3.19&0.12 & $-57$&1   & 3.09&0.16 &  0.42&0.15 & $-44$&7
    &   \mctwc{} & \mctwc{}& \mctwc{}& \mctwc{}& \mctwc{}& \mctwc{}  \\
&UE2 &   \mctwc{} & \mctwc{}& \mctwc{}& \mctwc{}& \mctwc{}& \mctwc{}
    &  28&2 & 3.65&0.13 & $-54$&2   & 2.40&0.22 &  0.21&0.14 & $87$&11\\
\hline
%
%
1803+784 &
UA   & 1920&7  & \mctwc{0}& \mctwc{--}& 0.13&0.01 & 0.81&0.06 & $-47$&15 
     & 1233&70 & \mctwc{0}& \mctwc{--}& 0.09&0.01 & 0.54&0.41 & $6$&15  \\
&UB1 &  494&6  & 0.28&0.06 & $-76$&2  & 0.32&0.05 & 0.67&0.08 & $82$&7
     &  538&61 & 0.20&0.06 & $-72$&5  & 0.36&0.04 & 0.56&0.07 & $86$&8  \\
&UB2 &  194&31 & 0.80&0.06 & $-77$&1  & 0.45&0.06 & 0.55&0.08 & $126$&4
     &  225&51 & 0.70&0.09 & $-79$&2  & 0.84&0.14 & 0.28&0.05 & $110$&3  \\
&UC  &  187&7  & 1.47&0.06 & $-90$&1  & 0.57&0.02 & 0.63&0.03 & $66$&3 
     &  198&32 & 1.44&0.06 & $-90$&1  & 0.50&0.03 & 0.68&0.04 & $79$&5  \\
&UD  &  38&6   & 2.02&0.06 & $-93$&1  & 0.92&0.07 & 0.60&0.09 & $164$&6
     &  47&9   & 1.82&0.38 & $-91$&3  & 1.06&0.34 & 0.49&0.19 & $70$&10  \\
&UE  &   41&4  & 3.51&0.14 & $-92$&1  & 3.18&0.27 & 0.39&0.05 & $102$&4
     &   70&5  & 3.52&0.17 & $-92$&1  & 4.44&0.31 & 0.35&0.04 & $91$&2 \\ \hline
%
%
1928+738 &
UA   &  941&25 &  0.26&0.11  & $-20$&6 & 0.20&0.10 & 0.26&0.13 &  $-6$&1
     &  774&8  &  0.39&0.14  & $-16$&1 & 0.52&0.01 & 0.11&0.03 & $-13$&2 \\
&UB1 & 656&27 & \mctwc{0}& \mctwc{--} & 0.23&0.01 & 0.16&0.08 & $-19$&2 
     & 602&20 & \mctwc{0}& \mctwc{--} & 0.52&0.01 & 0.11&0.03 & $-14$&1 \\
&UB2 &  571&30 &  0.38&0.11 & $155$&1 & 0.34&0.02 & 0.61&0.03 & $-28$&1
     &  660&13 &  0.32&0.14 & $143$&1 & 0.38&0.09 & 0.27&0.02 & $-13$&1 \\
&UC  &  390&2  &  1.15&0.11 & $152$&1 & 0.34&0.02 & 0.61&0.03 &   $6$&2
     &  331&1  &  1.26&0.14 & $152$&1 & 0.31&0.08 & 0.72&0.14 & $-38$&3\\
&UD  &  184&3  &  2.50&0.11 & $162$&1 & 0.91&0.02 & 0.52&0.02 &  $19$&2
     &  184&6  &  2.55&0.14 & $158$&1 & 0.57&0.04 & 0.44&0.04 &$-165$&2\\  
&UE  &  157&2  &  2.96&0.11 & $173$&1 & 1.27&0.02 & 0.17&0.05 & $-14$&2
     &  181&2  &  3.07&0.14 & $173$&1 & 1.22&0.12 & 0.31&0.01 & $-13$&1\\
&UF  &  40&2   &  3.93&0.11 & $162$&1 & 0.72&0.05 & 0.13&0.32 & $-46$&9
     &  63&3   &  3.82&0.14 & $163$&1 & 1.38&0.10 & 0.39&0.05 & $-32$&6\\
&UG  &  19&1   &  5.57&0.11 & $164$&1 & 0.61&0.05 & 0.23&0.23 & $-79$&13
     &  11&2   &  5.26&0.14 & $168$&2 & 1.59&0.45 & 0.17&0.16 &  $21$&7\\
&UI  &  76&3  &  11.46&0.11 & $167$&1 & 2.60&0.10 & 0.65&0.05 & $-15$&5
     &  93&3  &  11.58&0.14 & $167$&1 & 2.46&0.16 & 0.84&0.08 &   $6$&13\\
\hline
%
2007+777 &
 UA1 & 283&12  & 0.29&0.05  & $96$&2  & 0.18&0.04 &  0.22&0.21 & $39$&25  
     & 516&15  & 0.22&0.05  & $85$&1  & 0.14&0.04 &  0.25&0.08 & $82$&20\\
&UA2 & 223&26  & \mctwc{0}& \mctwc{--}& 0.20&0.02 &  0.12&0.12 & $93$&8
     & 278&14  & \mctwc{0}& \mctwc{--}& 0.14&0.08 &  0.17&0.09 & $-45$&3\\
&UB1$^{\rm a}$ & 146&21  & 0.32&0.05  & $-84$&2 & 0.15&0.04 &  0.50&0.41 & $37$&20 
               & 323&2   & 0.38&0.05  & $-82$&1 & 0.21&0.03 &  0.61&0.02 & $100$&2\\
&UB2$^{\rm a}$ & 94&22   & 0.63&0.07  & $-86$&1 & 0.62&0.11 &  0.65&0.12 & $107$&10 
               &  \mctwc{} & \mctwc{}& \mctwc{} & \mctwc{}& \mctwc{}& \mctwc{}  \\
&UC  & 51&2    & 1.44&0.05  & $-88$&1 & 0.34&0.17 &  0.59&0.08 & $-66$&7
     & 112&1   & 1.48&0.05  & $-89$&1 & 0.67&0.01 &  0.53&0.01 & $87$&1\\
&UF  & 38&1    & 6.13&0.05  & $-90$&1 & 1.94&0.08 &  0.44&0.03 & $-53$&2
     & 25&1    & 6.34&0.07  & $-95$&1 & 3.19&0.16 &  0.28&0.03 & $107$&2\\
\hline
\end{tabular}
}
\]
\begin{scriptsize}
\begin{list}{}{
\setlength{\leftmargin}{0pt}
\setlength{\rightmargin}{0pt}
}
\renewcommand{\baselinestretch}{0.8}
\item[Note:]
For each source component, we list its
flux density, separation and position angle relative to 
the component fixed at the origin ($r=0, \theta =0$),
major axis, axis ratio, and position angle of the major axis.
{The uncertainties given here have been estimated from the exploration}
of the parameter space ($S, r, \theta, a, b/a$, and $\phi$), 
and from empirical considerations (e.g., the nominal errors of the component positions 
are in general smaller than 10$\mu$\,as, but to be conservative we
increased this figure up to  a tenth of a beam width).
\item[
$^{\rm a}$
]
This component is model fitted as double only in the
first epoch.  In the second epoch the model fit algorithm merged both components
into a single one.
\renewcommand{\baselinestretch}{1.0}
\end{list}
\end{scriptsize}
\end{flushleft}
\end{table*}


%
\begin{table}[htb!]
\begin{flushleft}
\caption{Map parameters 
and elliptical Gaussian component model parameters
for selected sources of the complete
S5 polar cap sample from the 2\,cm Survey
\label{table:2cm-survey}
}
\[
\centering
\resizebox{\columnwidth}{!}{%
\begin{tabular}{@{}c@{}c@{~}r@{$\pm$}l@{~~}r@{$\pm$}l@{~~}r@{$\pm$}l@{~~}r@{$\pm$}l@{~~}r@{$\pm$}l@{~~}r@{$\pm$}l@{}}
{\bf Source} 
& Comp.         & \mctwc{$S$}   & \mctwc{$r$}   & \mctwc{$\theta$} &
\mctwc{$a$} & \mctwc{$b/a$} & \mctwc{$\phi$}  \\
& & \mctwc{[mJy]}   & \mctwc{[mas]} & \mctwc{~} &
\mctwc{[mas]} & \mctwc{~} & \mctwc{~}  \\
\noalign{\smallskip} \hline \noalign{\smallskip} 
{\bf \object{0016+731}} 
& \multicolumn{13}{@{}r@{}}{\it \hrulefill Epoch 2000.99, (0.647$\times$0.452, 2\fdg4)$^\mathrm{a}$, 731$^\mathrm{b}$, 811$^\mathrm{c}$, 0.5$^\mathrm{d}$ }\\
&  UA  &  742&1  & \mctwc{0} & \mctwc{--} & 0.09&0.01 & 0.30&0.08 & -81&1 \\
&  UB  &  35&6  & 0.79&0.06 & 140&1 & 0.55&0.05 & 0.88&0.08 & 178&28 \\
&  UC  &  36&6  & 1.41&0.06 & 131&1 & 0.76&0.07 & 0.84&0.09 &  88&21 \\
\noalign{\smallskip} \hline \noalign{\smallskip} 
{\bf \object{0153+744}} 
& \multicolumn{13}{@{}r@{}}{\it \hrulefill 
Epoch 2000.99, (0.833$\times$0.446, 8\fdg9), 188, 288, 0.7} \\
& UA1 &  190&1 &  \mctwc{0} & \mctwc{--} & 0.13&0.01 & 0.16&15.0 & 74&8 \\
&  UA2 &  28&2  &  0.68&0.08 & 69&1 & 0.31&0.05 & 0.20&0.25 & -96&15\\
&  UA3 &  17&1  &  1.27&0.08 &  87&2 & 0.63&0.10 & 0.11&0.13 & 153&10\\
&  UB1 &  42&3  & 10.24&0.08 & 153&1 & 1.01&0.10 & 0.79&0.11 & 131&15\\
&  UB2 &  14&3  & 11.16&0.08 & 151&1 & 0.53&0.09 & 0.86&0.26 & 57&125\\
\noalign{\smallskip} \hline \noalign{\smallskip} 
{\bf \object{0454+844}} 
& \multicolumn{13}{@{}r@{}}{\it \hrulefill 
Epoch 1999.55, (0.687$\times$0.492, -14\fdg8), 126, 186, 0.6} \\ 
& UA  &  144&12 & \mctwc{0} & \mctwc{--} & 0.34&0.02 & 0.58&0.06 & 18&5\\
& UB  &  24&13  & 0.49&0.18 & 134&7 & 0.78&0.29 & 0.20&0.17 & -22&9\\
& UC  &  18&3   & 1.52&0.11 & 168&2 & 1.34&0.24 & 0.47&0.10 & 19&7\\
& \multicolumn{13}{@{}r@{}}{\it \hrulefill 
Epoch 2001.17, (0.641$\times$0.411, 15\fdg4),  187,  257, 0.7}\\
&  UA  &  201&6 & \mctwc{0} & \mctwc{--} & 0.23&0.01 & 0.30&0.10 & 16&3\\
&  UB  &  39&5  & 0.33&0.06 & 138&2 & 0.23&0.21 & 0.30&0.27 & 75&12 \\
&  UC  &  22&1  & 1.30&0.06 & 158&1 & 0.74&0.03 & 0.70&0.08 & 83&8\\
\noalign{\smallskip} \hline \noalign{\smallskip} 
{\bf \object{0716+714}} 
& \multicolumn{13}{@{}r@{}}{\it \hrulefill 
Epoch  1999.55, (0.849$\times$0.551, 8\fdg3),  1176, 1254, 0.6}\\
&  UA  & 1174&8 & \mctwc{0} & \mctwc{--} & 0.14&0.01 & 0.37&0.03 & 34&3\\
&  UB  &   63&9 & 0.54&0.08 &  16&1 & 0.89&0.11 & 0.13&0.05 & 11&2\\
&  UC  &   13&1 & 1.85&0.09 &  9&1  & 1.08&0.33 & 0.66&0.35 & 157&16\\
&  UD  &    7&1 & 3.69&0.12  & 20&2 & 1.86&0.37 & 0.38&0.17 & 89&13\\
& \multicolumn{13}{@{}r@{}}{\it \hrulefill 
Epoch 2001.17, (0.770$\times$0.451, 31\fdg9), 572,  643, 0.3}\\
&  UA  &  549&33 & \mctwc{0} & \mctwc{--} & 0.12&0.05 & 0.27&0.06 & 28&4\\
&  UB  &   71&34 & 0.43&0.10 &  22&2 & 0.35&0.16 & 0.47&0.26 & 16&9\\
&  UC  &   15&2  & 1.94&0.09 &  19&1  & 1.27&0.16 & 0.38&0.05 & 8&4\\
&  UD  &    8&2  & 3.95&0.35  & 15&2  & 3.28&0.65 & 0.22&0.07 & 27&5\\
\noalign{\smallskip} \hline \noalign{\smallskip} 
{\bf \object{0836+710}} 
& \multicolumn{13}{@{}r@{}}{\it \hrulefill 
Epoch  1998.22, (0.681$\times$0.442, -12\fdg5),  1378, 2298, 1.0} \\
&  UA1  & 1672&26   & \mctwc{0} & \mctwc{--} & 0.38&0.01 & 0.69&0.01 & 33&19\\
&  UA2  &  228&33   & 0.21&0.04 & -84&2 & 0.21&0.03 & 0.63&0.26 & -139&42\\
&  UA3  &   60&9    & 0.47&0.04 & -114&2 & 0.34&0.21 & 0.35&1.0 & 69&13 \\
&  UB   &   38&1    & 1.37&0.04 & -135&1 & 0.39&0.26 & 0.84&0.93 & 174&19\\
&  UC  &   159&1    & 2.71&0.04 & -141&1 & 0.69&0.01 & 0.81&0.01 & 19&2\\
&  UF  &    72&3    &11.61&0.41 & -147&1 & 3.17&0.14 & 0.41&0.03 & 2&5\\
\noalign{\smallskip} \hline \noalign{\smallskip} 
{\bf \object{1749+701}}  
& \multicolumn{13}{@{}r@{}}{\it \hrulefill 
Epoch  1999.39, (0.657$\times$0.419, -12\fdg6), 246,  415, 0.5} \\
&  UA &  196&3  &\mctwc{0} & \mctwc{--} & 0.15&0.01 & 0.22&0.13 & -17&4\\
&  UB &   95&3  & 0.20&0.07 & -48&1 & 0.32&0.01 & 0.32&0.10 & -63&2\\ 
&  UC1&   28&1  & 1.02&0.07 & -62&1 & 0.43&0.36 & 0.61&0.07 & -84&7\\ 
&  UC2&   22&1  & 1.49&0.07 & -75&1 & 0.37&0.02 & 0.86&0.14 & 65&24\\
&  UD &   30&1  & 2.25&0.07 & -73&1 & 0.68&0.03 & 0.80&0.05 & -25&6\\
&  UE1 &   26&2  & 2.80&0.08 & -59&1 & 1.43&0.07 & 0.90&0.08 & -1&10\\
&  UE2 &   22&3  & 3.90&0.09 & -53&1 & 1.79&0.12 & 0.77&0.07 & 84&11\\
& \multicolumn{13}{@{}r@{}}{\it \hrulefill 
Epoch 1999.85, (0.715$\times$0.446, -25\fdg9),  299,  446, 0.4} \\
&  UA &  229&4  & \mctwc{0} & \mctwc{--} & 0.10&0.02 & 0.24&0.13 & -39&6\\
&  UB &   103&4 & 0.21&0.07 & -49&1      & 0.41&0.01 & 0.20&0.04 & -54&2\\ 
&  UC1  &   25&1  & 1.29&0.07 & -67&1 & 0.67&0.02 & 0.20&0.14 & 78&3\\ 
&  UC2  &   15&1  & 1.78&0.07 & -77&1 & 0.83&0.06 & 0.38&0.06 & 120&4\\
&  UD  &   34&1  & 2.38&0.07 & -70&1 & 0.75&0.02 & 0.66&0.04 & 15&4\\
&  UE1  &   \mctwc{} &  \mctwc{} &  \mctwc{} \\
&  UE2  &   37&1  & 3.68&0.04 & -53&1 & 1.86&0.04 & 0.84&0.04 & 154&8\\
& \multicolumn{13}{@{}r@{}}{\it \hrulefill 
Epoch 2000.99, (0.640$\times$0.441, 20\fdg4), 271,  434, 0.5}\\
&  UA  &  245&1 & \mctwc{0} & \mctwc{--} & 0.05&0.20 & 0.25&0.26 & -8&17 \\
&  UB  &   111&4 & 0.41&0.06 & -51&1 & 0.43&0.01 & 0.31&0.03 & -52&1\\ 
&  UC1 &   12&1  & 1.06&0.06 & -69&1  & 0.61&0.05 & 0.93&0.51 & -62&84\\ 
&  UC2 &   \mctwc{} &  \mctwc{} &  \mctwc{} \\
&  UD  &   36&1  & 2.48&0.06 & -71&1 & 0.78&0.02 & 0.93&0.04 & 56&25\\
&  UE1 &   \mctwc{} &  \mctwc{} &  \mctwc{} \\
&  UE2 &   34&2  & 3.56&0.06 & -53&1 & 3.03&0.13 & 0.74&0.06 & 8&9 \\

\noalign{\smallskip} \hline \hline
\end{tabular}
}
\]
\end{flushleft}
\end{table}

\setcounter{table}{2}
\begin{table}
\begin{flushleft}
\caption{{\sl Continued}}
\[
\centering
\resizebox{\columnwidth}{!}{%
\begin{tabular}{@{}c@{}c@{~}r@{$\pm$}l@{~~}r@{$\pm$}l@{~~}r@{$\pm$}l@{~~}r@{$\pm$}l@{~~}r@{$\pm$}l@{~~}r@{$\pm$}l@{}}
{\bf Source} 
& Comp.         & \mctwc{$S$}   & \mctwc{$r$}   & \mctwc{$\theta$} &
\mctwc{$a$} & \mctwc{$b/a$} & \mctwc{$\phi$}  \\
& & \mctwc{[mJy]}   & \mctwc{[mas]} & \mctwc{~} &
\mctwc{[mas]} & \mctwc{~} & \mctwc{~}  \\

\noalign{\smallskip} \hline \noalign{\smallskip} 
{\bf \object{1803+784}}
& \multicolumn{13}{@{}r@{}}{\it \hrulefill 
Epoch 1998.84, (0.731$\times$0.425, -8\fdg8), 1498, 2291, 0.9} \\
&  UA  &  1310&46 & \mctwc{0} & \mctwc{--} & 0.12&0.01 & 0.86&0.03 & -74&36\\
&  UB1 &   452&43 &  0.23&0.01 & -88&1 & 0.28&0.01 & 0.75&0.29 & 76&39\\ 
&  UB2  &  157&4  &  0.70&0.01 & -80&1 & 0.81&0.28 & 0.46&0.47 & 102&55\\
&  UC  &   193&3  & 1.41&0.01  & -93&1 & 0.33&0.35 & 0.79&0.02 & 17&3\\
&  UD  &   109&3  & 1.75&0.01  & -99&1 & 0.59&0.28 & 0.70&0.14 & 165&3\\
&  UE  &    52&2  & 3.32&0.05  & -93&1 & 2.93&0.11 & 0.43&0.03 & 106&25\\
& \multicolumn{13}{@{}r@{}}{\it \hrulefill 
Epoch 1999.85, (0.660$\times$0.465, -27\fdg9), 1777, 2518, 1.1} \\
&  UA  &  1531&27 & \mctwc{0} & \mctwc{--} & 0.10&0.01 & 0.74&0.08 & -63&15\\
&  UB1 &   494&25 &  0.25&0.01 & -80&1 & 0.28&0.17 & 0.77&0.05 & 119&17\\ 
&  UB2 &   145&4  &  0.84&0.01 & -76&1 & 0.48&0.02 & 0.45&0.03 & 131&4\\
&  UC  &   201&5  &  1.37&0.01 & -88&1 & 0.52&0.01 & 0.38&0.10 & 64&13\\
&  UD  &    50&3  &  1.78&0.03 & -98&1 & 0.75&0.48 & 0.52&0.41 & 127&23\\
&  UE  &    44&1  &  2.94&0.04 & -96&1 & 1.67&0.95 & 0.70&0.52 & 107&6\\
\noalign{\smallskip} \hline \noalign{\smallskip} 
{\bf \object{2007+777}} 
& \multicolumn{13}{@{}r@{}}{\it \hrulefill 
Epoch 1999.85, (0.653$\times$0.447, 0\fdg6),  428,  929, 0.4} \\
&  UA1  & 413&1   & 0.36&0.01 & 93&1 & 0.17&0.01 & 0.28&0.03 & 91&25\\
&  UA2 &  163&1   & \mctwc{0} & \mctwc{--} & 0.20&0.01 & 0.26&0.07 & 90&23\\
&  UB1  & 230&1   & 0.32&0.01 & -86&1 & 0.13&0.01 & 0.75&0.24 & 65&18\\
&  UB2  &  75&2   & 1.14&0.01 & -91&1 & 0.52&0.23 & 0.82&0.01 & 152&3\\
&  UC  &   24&1   & 1.43&0.01 & -88&1 & 0.43&0.17 & 0.38&0.10 & -81&3\\
&  UF  &   28&1   & 6.55&0.02 & -94&1 & 1.30&0.05 & 0.73&0.06 & 83&7 \\
& \multicolumn{13}{@{}r@{}}{\it \hrulefill 
Epoch 2001.17, (0.617$\times$0.421, -0\fdg9), 393,  948, 0.4}\\
&  UA1  & 330&7 & 0.25&0.01 & 87&1 & 0.17&0.01 & 0.22&0.15 & 85&1\\
&  UA2 &  189&4   & \mctwc{0} & \mctwc{--} & 0.38&0.01 & 0.24&0.03 & 90&1\\
&  UB1  & 213&3   & 0.40&0.01 & -87&1 & 0.22&0.01 & 0.37&0.05 & 104&2\\
&  UB2  & 115&1   & 0.87&0.01 & -76&1 & 0.52&0.01 & 0.38&0.03 & 104&1\\
&  UC  &   77&1   & 1.28&0.01 & -96&1 & 0.46&0.01 & 0.54&0.03 & -79&2\\
&  UE  &   7&1    & 5.08&0.03 & -106&1 & 0.82&0.08 & 0.31&0.29 & 112&7\\ 
&  UF  &   15&1   & 6.86&0.06 & -95&1  & 2.08&0.15 & 0.29&0.06 & 71&3\\
\noalign{\smallskip} \hline \hline
\end{tabular}
}
\]
\begin{scriptsize}
\begin{list}{}{
\setlength{\leftmargin}{0pt}
\setlength{\rightmargin}{0pt}
}
\item[Note:]
For each source component, we list its
flux density, separation and position angle relative to 
the component fixed at the origin ($r=0, \theta =0$),
major axis, axis ratio, and position angle of the major axis.
{The uncertainties given here have been estimated from the exploration}
of the parameter space ($S, r, \theta, a, b/a$, and $\phi$), 
and from empirical considerations (e.g., the nominal errors of the component positions 
are in general smaller than 10$\mu$\,as, but to be conservative we
increased this figure up to  a tenth of a beam width).
\item[$^{\rm a}$] The restoring beam is an elliptical Gaussian
with FWHM axes $a\times b$ [mas].
For each source, the position angle (P.A.) 
stands for the direction of the
major {axis}, measured north through east.
\item[$^{\rm b}$] $S_{\rm peak}$: brightness peak [mJy/beam].
\item[$^{\rm c}$] $S_{\rm tot}$: total flux density recovered in the {hybrid}
mapping process [mJy].
\item[$^{\rm c}$] rms: root-mean-square noise in the image [mJy]. Contours in the maps of the figures shown  in Sects.~\ref{subsec:0016} to \ref{subsec:2007}
are the tabulated rms values times $(-3,3,3\sqrt{3}\cdots)$.
\end{list}
\end{scriptsize}
\end{flushleft}
\end{table}

\subsection{\object{QSO\,0016+731}\label{subsec:0016}}

\begin{figure}[htbp]
\begin{tabular}{@{}cc@{}}
\includegraphics[width=0.40\textwidth]{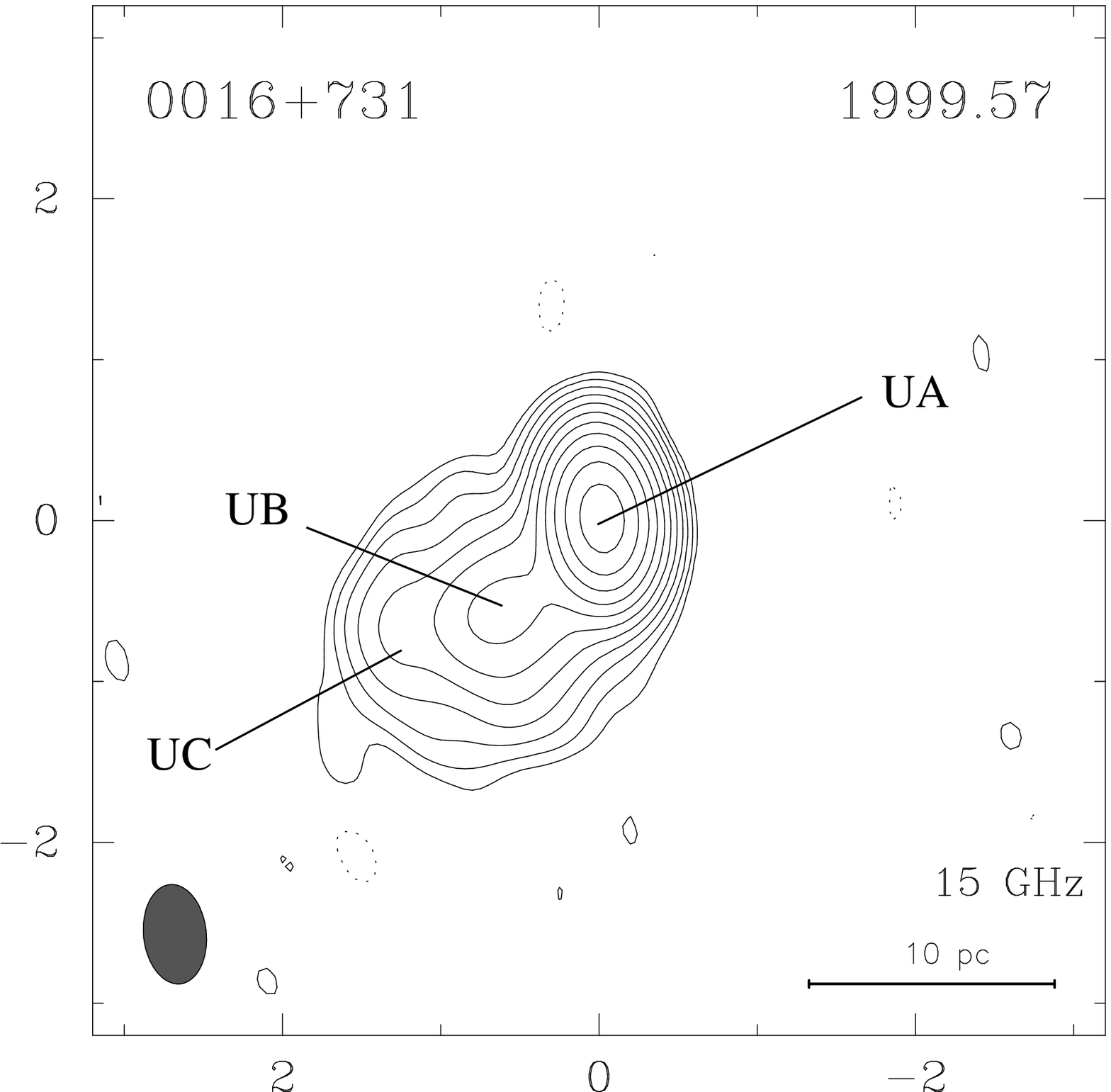} \\ 
\includegraphics[width=0.40\textwidth]{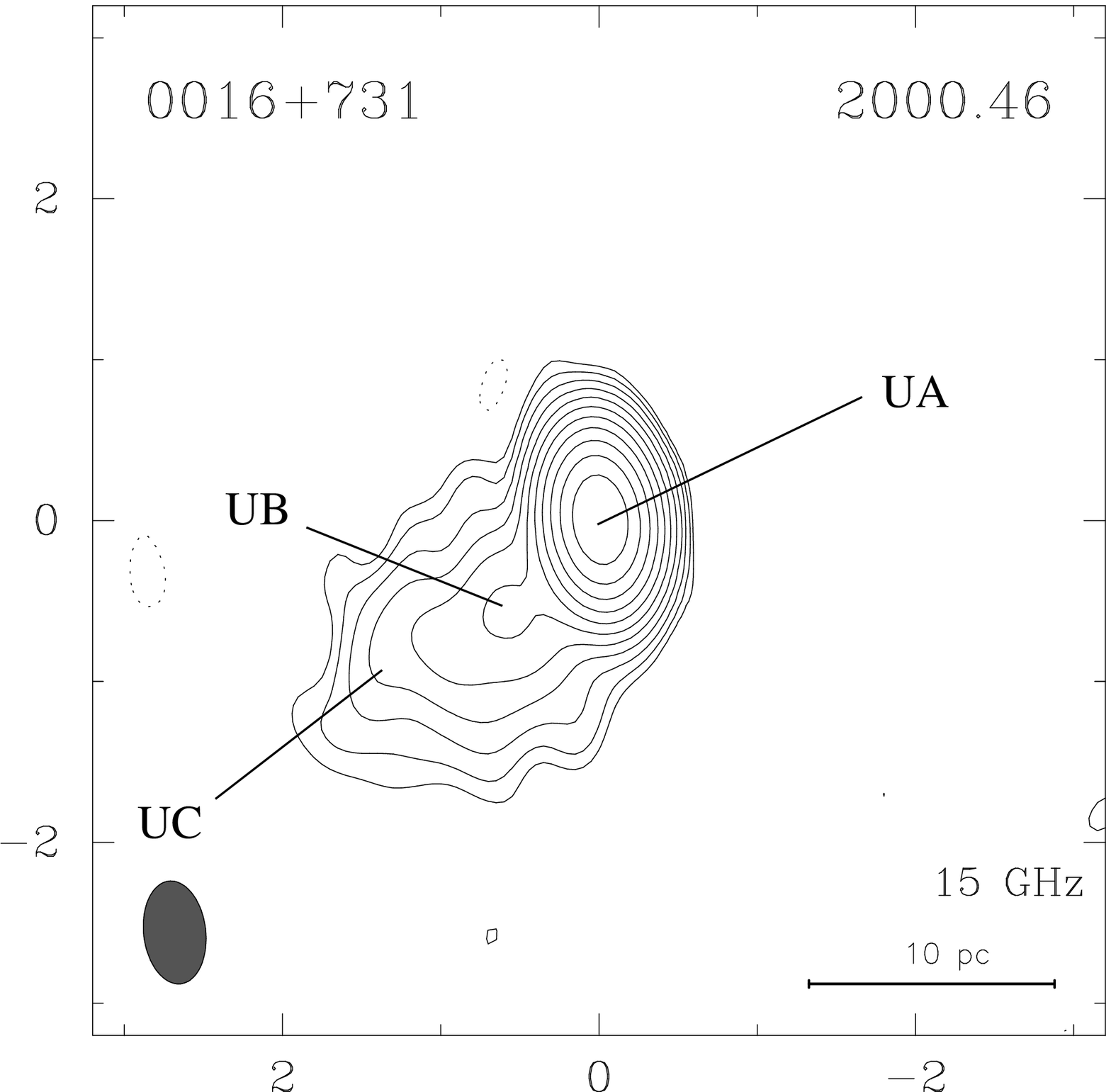} \\
\includegraphics[width=0.40\textwidth]{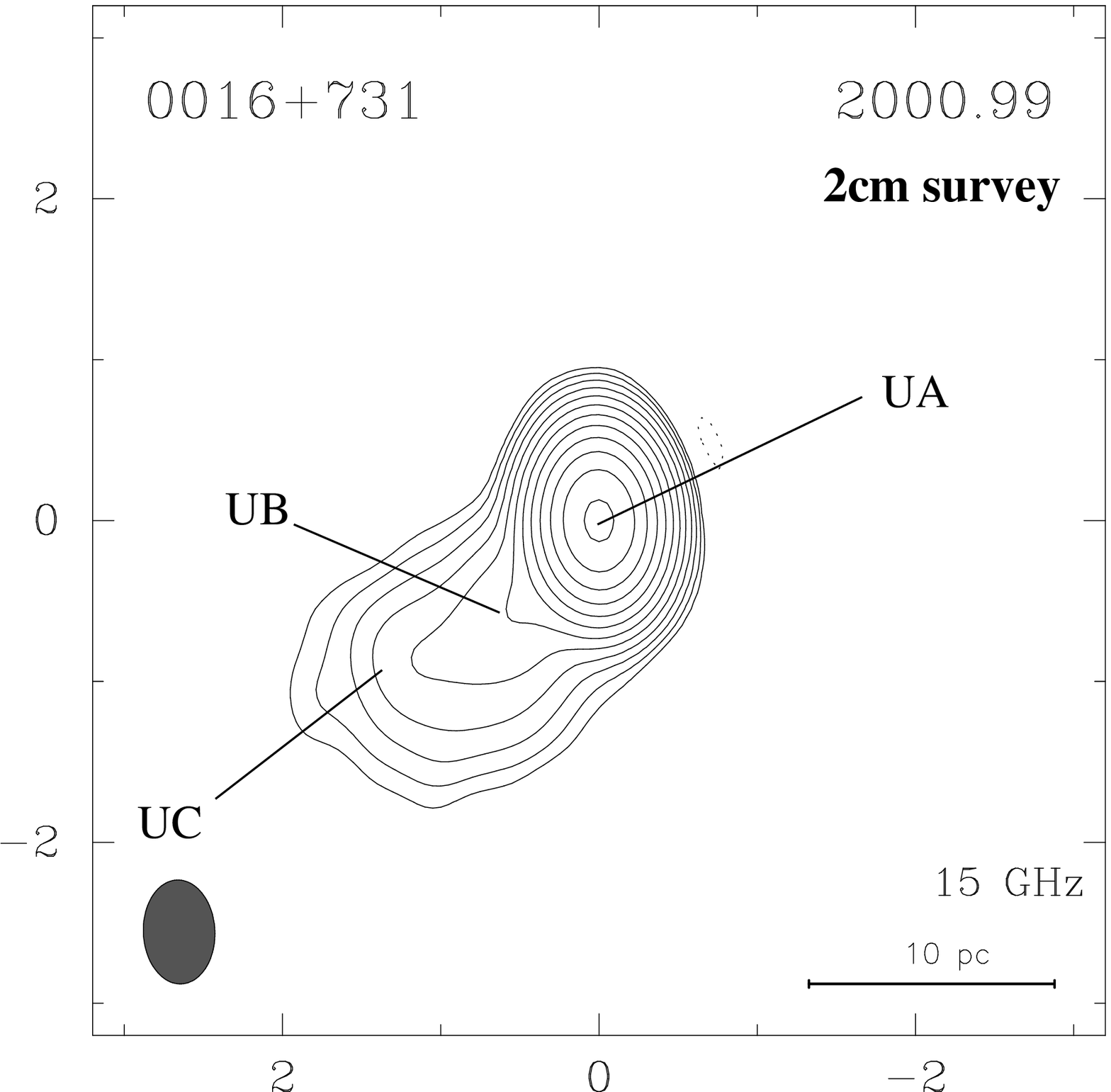} & \\
\end{tabular}
\caption[]{
VLBA images of \object{QSO\,0153+744}
on 27 July 1999 (1999.57), 15 June 2000 (2000.46),
and 28 December 2000 (2000.99).
See Tables 1--3 for contour levels, beam sizes (bottom left in the maps),
peak flux densities, and component parametrization.
Axes are relative $\alpha$ and $\delta$ in mas.
\label{fig:map0016}
}
\end{figure}

The source \object{QSO\,0016+731} 
(Fig.\ \ref{fig:map0016}, $z$=1.781)
shows at 15.4\,GHz a core-jet structure at
an angle of $\sim$130\degr.  
The modeling of the radio source provides a good fit
with three components: a compact, strong one
(UA), which likely corresponds to the core, and two extended, 
weak components (UB and UC), directed to the east-southeast, 
and which correspond to the jet. 
The total flux density increased significantly from 647\,mJy to 817\,mJy 
(26\% change) between our two observing epochs.
The corresponding monochromatic luminosity is 
$L_{\rm 15\,GHz}\approx$(1.3 to 1.6)$\times 10^{39}\,$W.
The increase in the total flux {density} of the source was
due to a strong increment in the flux of the 
westernmost component (UA, {the brightest
feature} in Fig.\ \ref{fig:map0016}).
Correspondingly, the  observed, uncorrected ratio of the core flux density 
to the extended flux density in the 15.4\,GHz frame, $Q$,
increased from 4 up to 11 between 1999.57 and 2000.99.
Our map of \object{QSO\,0016+731} from observations of the VLBA 2cm survey
on 2000.99 (bottom panel in Fig.\ \ref{fig:map0016}) shows that the 
total flux density of the source did not change significantly for over six
months ($\lsim$1\,\% decrement).

In Paper I, the lower resolution of the observations prevented 
us from clearly distinguishing which component of \object{QSO\,0016+731} 
(either XA, or XB in Fig.\,2) could be identified with the core,
a much easier task at 15.4\,GHz.
We suggest that component UA in our maps corresponds to component
XA in Paper I. 
In addition, it seems clear that component XB is indeed
a blend of two components, UB and UC. 
Component XC has no counterpart in our maps.
The second observing epoch at 8.4\,GHz (1999.41) was close in time
to our first observing epoch at 15.4\,GHz (1999.57).
Assuming that the 8.4\,GHz flux density of the source did not change
between 1999.41 and 1999.57, we obtain a global spectral index, $\alpha$, 
of 0.89. 
The spectral index of component UA/XA is much more inverted,
$\alpha$=1.99, while $\alpha =-0.68$ for the jet structure
(UB+UC/XB). 
\footnote{We have kept the name convention used in Paper I. 
Thus, ``UA'' represents the first component (``A'')
at the observing frequency band (``U'', corresponding to 15.4\,GHz).
We have also tried to assign to a given feature of a source the same name 
as used in Paper I. However, this has not always been possible. 
For example, we have no component labeled
UE, nor UF for \object{QSO\,0212+735} (see Sect.~\ref{subsec:0212}).
This means we found no counterpart at 15.4\,GHz for components XE and XF.}

\subsection{\object{QSO\,0153+744}\label{subsec:0153}}
\begin{figure}[htbp]
\begin{tabular}{@{}cc@{}}
\includegraphics[width=0.40\textwidth]{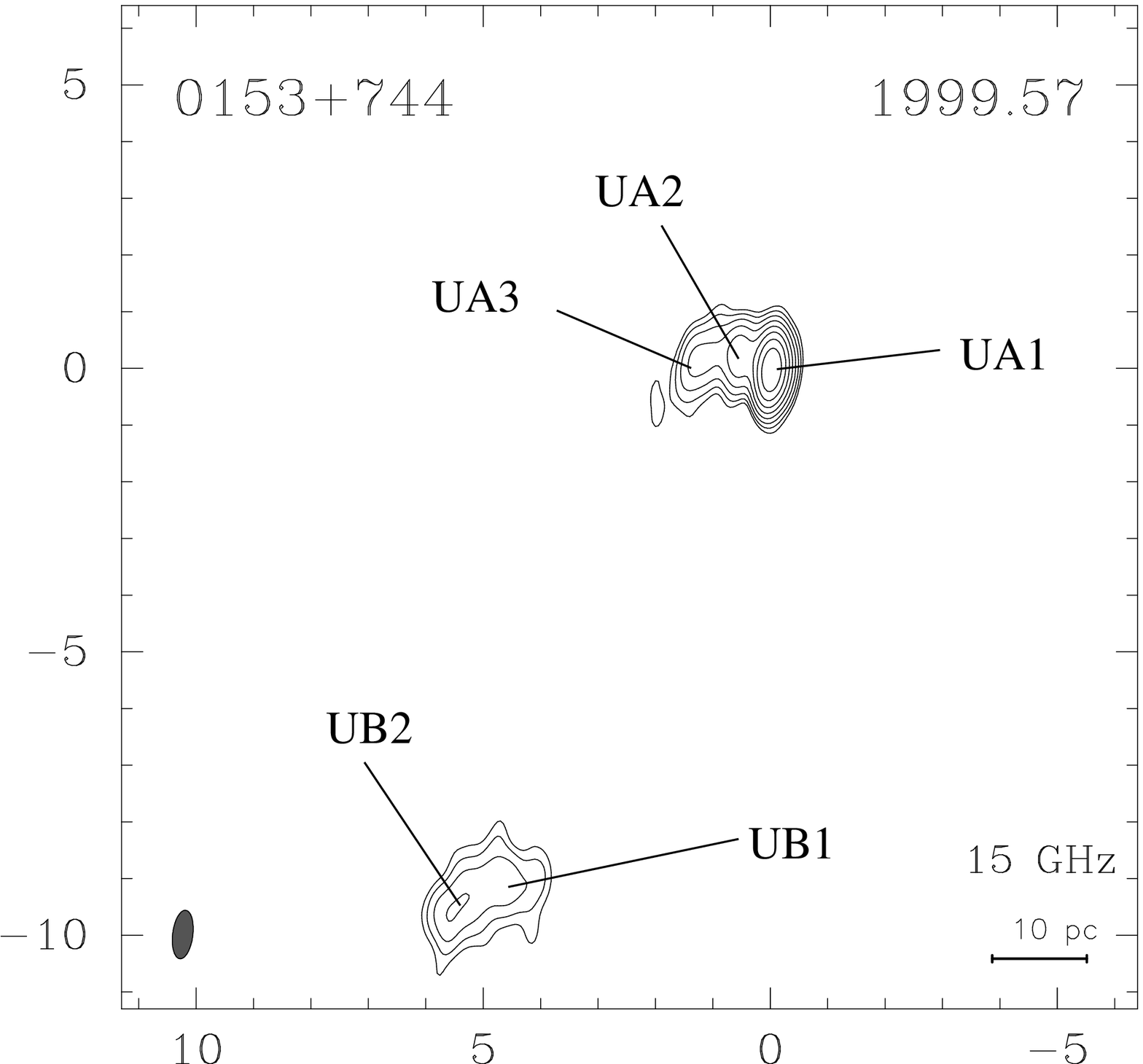} \\
\includegraphics[width=0.40\textwidth]{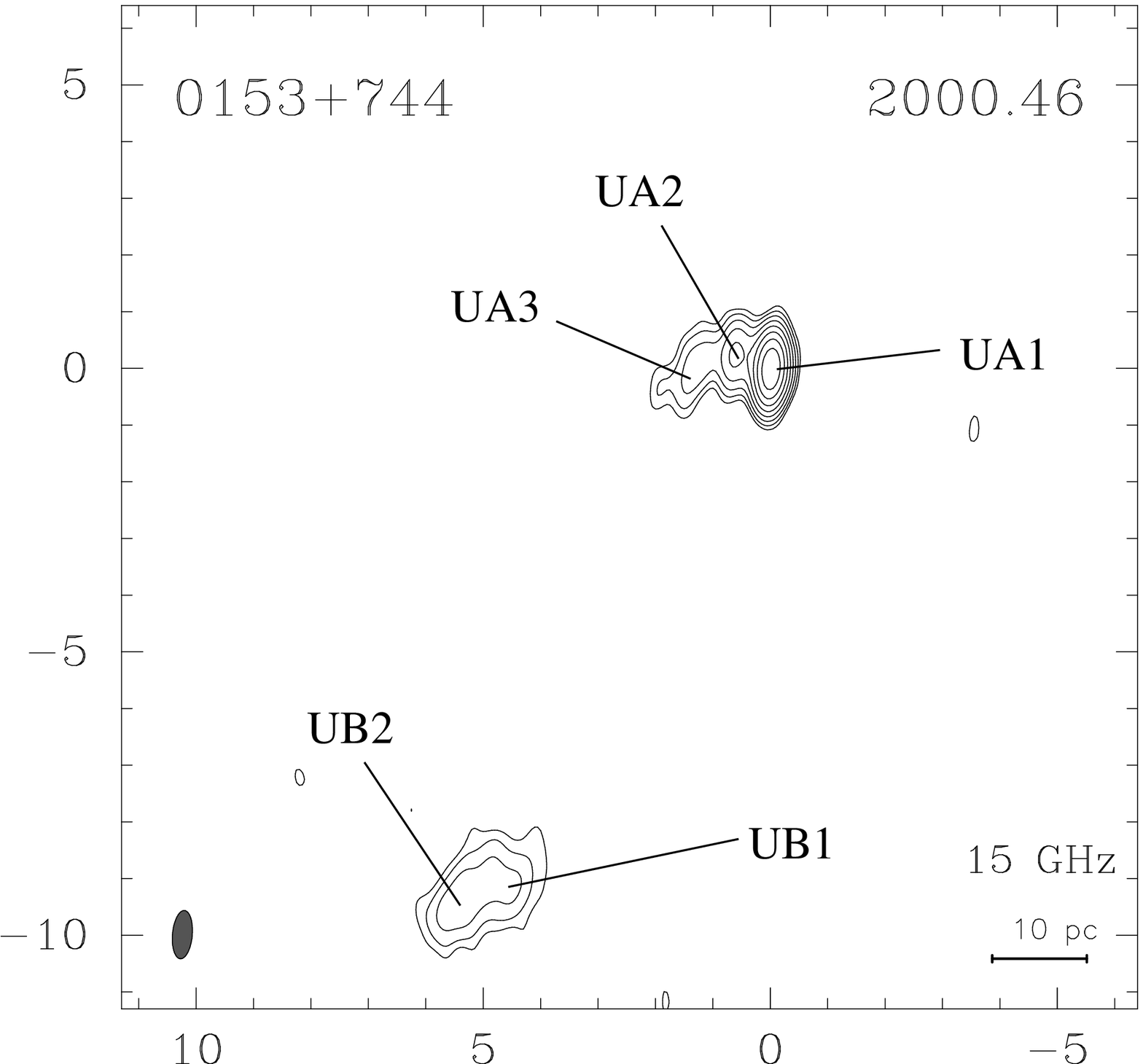} \\
\includegraphics[width=0.40\textwidth]{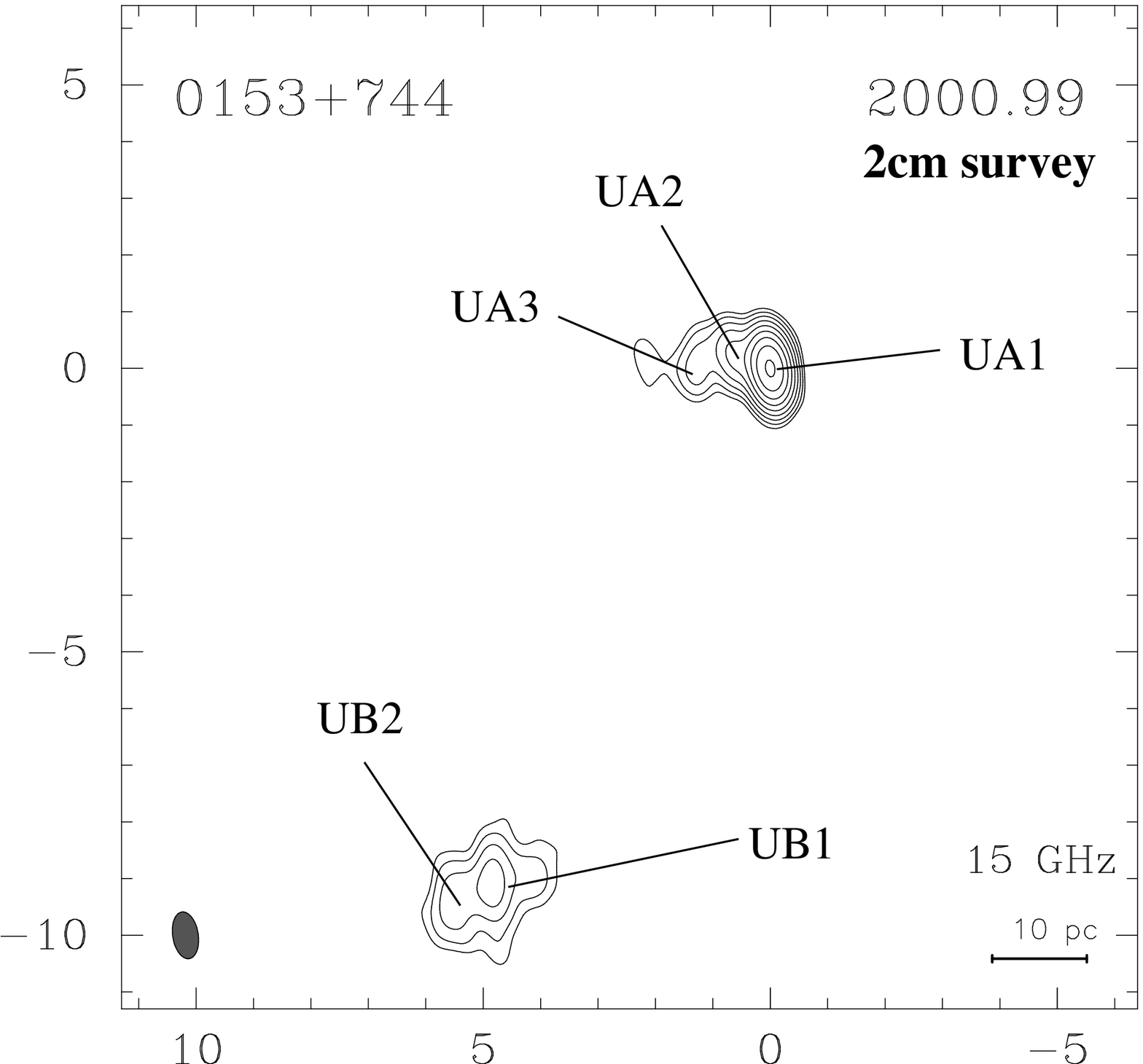} & \\
\end{tabular}
\caption[]{ 
VLBA images of \object{QSO\,0153+744}
on 27 July 1999 (1999.57), 15 June 2000 (2000.46),
and 28 December 2000 (2000.99).
See Tables 1--3 for contour levels, beam sizes (bottom left in the maps),
peak flux densities, and component parametrization.
Axes are relative $\alpha$ and $\delta$ in mas.
\label{fig:map0153}
}
\end{figure}

\object{QSO\,0153+744} 
(Fig.\ \ref{fig:map0153}, $z$=2.341) is a one-sided
radio source (e.g., Hummel \et\ \cite{hummel88}), 
whose 15.4\,GHz VLBA emission is 
dominated by two components, UA and UB,  
separated by $\sim$10\,mas ($\sim$61\hminus\,pc). 
The main component, UA(likely the core), is composed of three sub-components, UA1, 
and UA2 and UA3 (the innermost regions of the jet).
UA2 and UA3 trace a smooth change in the direction of the jet, 
changing from 
65\degr\,$\pm$\,3\degr 
at a distance 
$r = (0.65\pm0.05)$\,mas [$\sim$ (3.6 to 4.2)\hminus\,pc], 
to 88\degr $\pm$ 8\degr 
at a distance $r=(1.35\pm0.05)$\,mas.
No emission is detected above 3$\sigma$ (3\,mJy), up to 
a distance of $\sim$10 mas ($\sim$61\hminus\,pc), where component 
UB traces the outermost 
regions of the jet. At this distance, the jet is at an angle $\sim$150\degr.  

The total flux density stayed remarkably stable between 1999.57 and 
2000.46, decreasing from (316$\pm$3)\,mJy to (312$\pm$2)\,mJy. 
It decreased by 8\% between 2000.46 and 2000.99, down to (288$\pm$2)\,mJy,
implying $L_{\rm 15\,GHz}\,\approx$(1.0 to 1.1)$\times 10^{39}\,$W.
At all three epochs, component UA contributed $\approx$80\% 
to the total flux density, while
component UB contributed the remaining $\approx$20\%.
The brightest sub-component of UA, UA1, 
slightly increased its flux density in spite of the total flux decrease.
On the contrary, both UA2 and UA3 continued to fade between 
1999.57 and 2000.99 (e.g., UA2 decreased its flux density by 
$\approx$50\%).
The value of $Q$ (taken to be the ratio of emission 
from UA1 to the rest of the emission) 
increased from 1.26 in 1999.57 to 1.94 in 2000.99, 
indicating that the compact core marginally dominates the total
emission, and tended to increase this dominance.

%

Our model fit indicates a proper motion for UA2 of
$\mu\,=(71\pm6)\,\mu$as/yr [$\betap$=(4.7 to 0.4)\,\hminus].
UA3, located at a radial distance of (1.3 to 1.4)\,mas at an angle
90\degr, does not show evidence for any proper motion.
The extended, optically thin emission of the jet is characterized by 
two components, UB1 and UB2, that lie at an angular distance of
(10.15$\pm$0.15)\,mas
and (11.1$\pm$0.1)\,mas,
respectively, and at an angle 
of $\sim$150\degr.

Since \object{QSO\,0153+744} did not show evidence of 
intra-day variability (IDV) (A. Kraus, private communication), we 
assumed that the 8.4\,GHz VLBA flux density of the source did not change
between 1999.41 and 1999.57, and obtained a global spectral index 
$\alpha=-0.79$.
We also obtain $\alpha\approx-0.01$ for component UA/XA (a perfect
flat spectrum for the innermost regions of the core-jet structure). 
The sub-component XA1/UA1 has $\alpha = 0.52$, which supports
our suggestion that  this sub-component might be the core.
We noticed
that $\alpha\approx-2.05$ for XB/UB, which corresponds to
a synchrotron spectrum partially suppressed by an external medium.
In this case, this would indicate that at a distance of 
$\sim$60\,pc from the putative core the outer regions of the
jet of \object{QSO\,0153+744} interact strongly with their
surrounding medium.
Finally, 
we failed
to detect the jet components that in Paper I were found to 
form an arc between XA/UA and XB/UB, which results in 
upper limits of 
$\alpha_{XC/UC}\lsim-2.24$, 
$\alpha_{XD/UD}\lsim-3.09$, 
$\alpha_{XE/UE}\lsim-3.09$, 
at the 3\,$\sigma$ level.
The spectral behaviour displayed by component XB/UB, as well
as the non-detection of any significant motion, 
is typical of a stationary region of the jet, thus confirming
the finding by Hummel \et\ (\cite{hummel97}).

\begin{table}
 \caption[]{Proper motions$^\dagger$ in the S5 polar cap sample}
  \label{tab:proper_motion}
$$
 \begin{array}{ccrr}
   \hline\noalign{\smallskip}
  \multicolumn{1}{c}{\rm Source} & 
  \multicolumn{1}{c}{\rm Comp.} & 
  \multicolumn{1}{c}{\mu} & 
  \multicolumn{1}{c}{\beta_{\rm app}} \\ 
  \multicolumn{2}{c}{} & 
  \multicolumn{1}{c}{\rm [\mu\,as/yr] } & 
  \multicolumn{1}{c}{\rm [v_{app}/c ]} \\ 
	\cline{1-4}\noalign{\smallskip} 
0016+731 & {\rm UC}  &   32 \pm  31    &   1.9 \pm 1.8  \\ 
0153+744 & {\rm UA2} &   71 \pm   6    &   4.7 \pm 0.4  \\ 
0454+844 & {\rm UB}  & -100 \pm   3    &   -5.1 \pm 0.2 \\ 
         & {\rm UC}  & -217 \pm 140    &   -11.0 \pm 7.4  \\ 
%
%
0716+714 & {\rm UB }&  -105 \pm  100   & -1.6 \pm 1.5  \\
0836+710 & {\rm UA2 }&  -13 \pm   6    & -0.8 \pm 0.4  \\
	 & {\rm UA3 }&  118 \pm   7    &  7.6 \pm 0.5  \\
	 & {\rm UB }&  110 \pm   16    &  7.1 \pm 1.0  \\
	 & {\rm UC }&  101 \pm   16    &  6.5 \pm 1.0  \\
1749+701 & {\rm UB  }&  141 \pm   25   &  5.1 \pm 0.9  \\
	 & {\rm UC2 }&  732 \pm   603  &  26.5 \pm  21.8  \\
	 & {\rm UD  }&  161 \pm    19  &   5.8 \pm 0.7   \\
	 & {\rm UE2 }&  -150 \pm   62  & -5.4 \pm 2.2   \\
1803+784 & {\rm UB1  }&  17 \pm   16   &  0.6 \pm 0.5  \\
	 & {\rm UB2  }&  133 \pm  24   &  4.4 \pm 0.8  \\
2007+777 & {\rm UA1}  &   52 \pm   48  &  1.0 \pm 1.0  \\
	 & {\rm UB1}  &   56 \pm   11  &  1.1 \pm 0.2  \\
	 & {\rm UC }  &  -88 \pm   61  &  -1.8 \pm 1.2 \\
	 & {\rm UF }  &  350 \pm   203 &  7.0 \pm  4.0 \\
   \hline\noalign{\smallskip}
 \end{array}
$$
\begin{list}{}{}
\item[] {$^\dagger$Proper 
        motions are listed for components observed at three or more epochs, 
	whose least-squares fit showed a motion larger than 1$\sigma$, which is the
        quoted uncertainty. See details in the main text 
        and Fig.~\ref{fig:comp_positions}.
         }
\end{list}
\end{table}


\subsection{\object{QSO\,0212+735}\label{subsec:0212}}

\begin{figure}[htbp]
\vspace{228pt}
\includegraphics{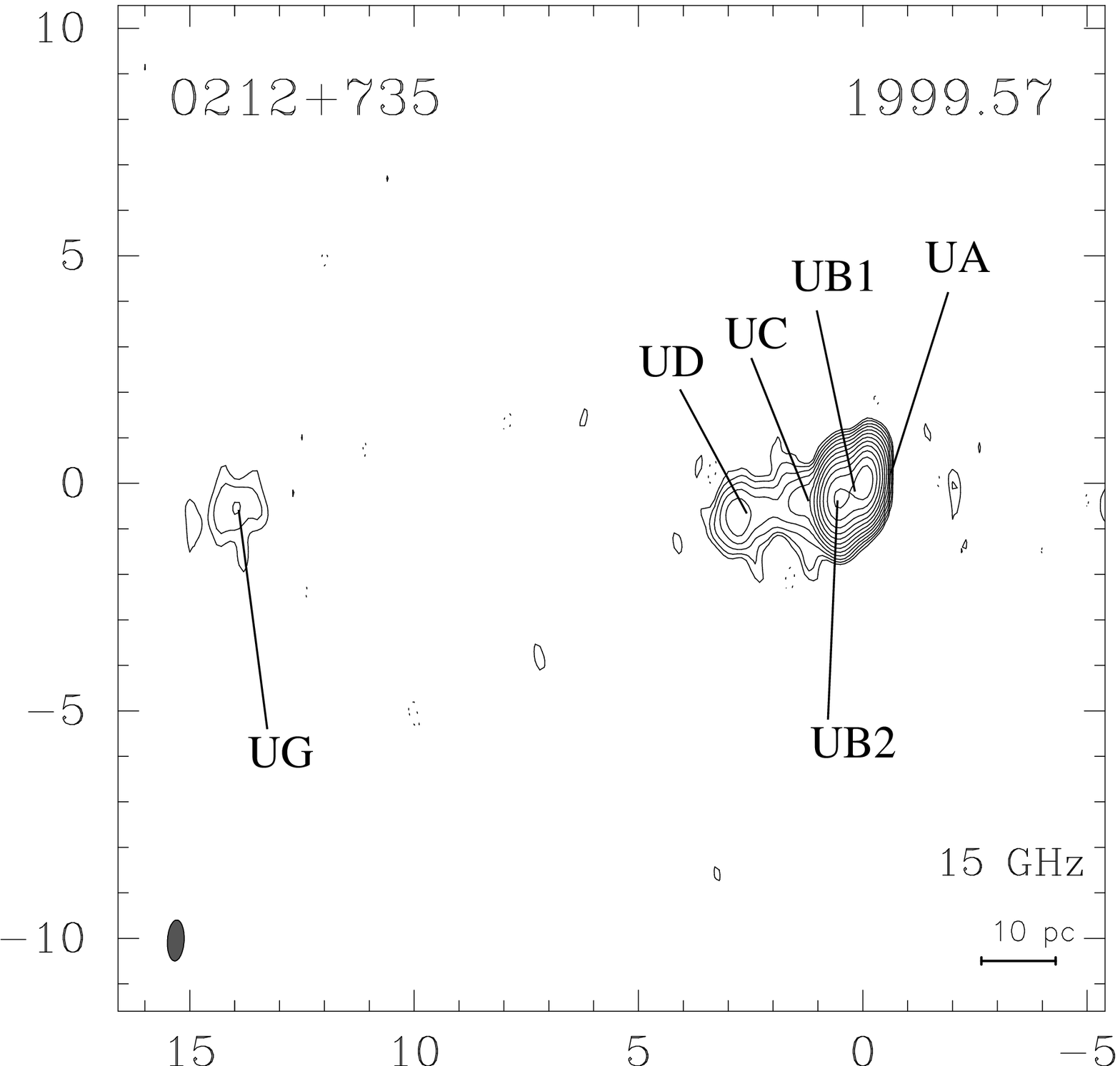}
\vspace{228pt}
\includegraphics{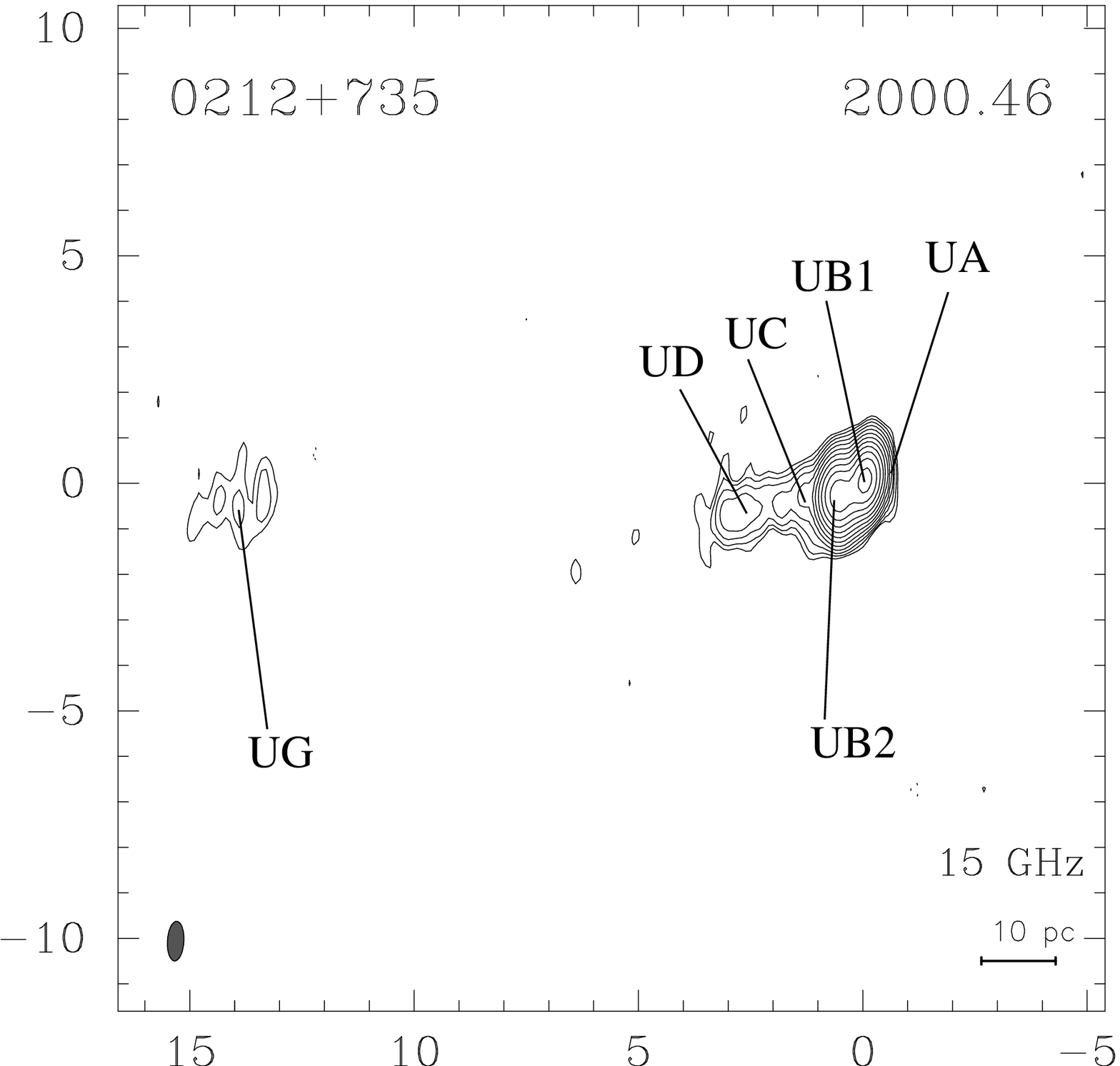}
\caption{
VLBA images of \object{QSO\,0212+735}, 
from observations on
27 July 1999 (1999.57) and  15 June 2000 (2000.46).  
Axes are relative $\alpha$ and $\delta$ in mas.  
See Table 1 for contour levels, 
synthesized beam size (bottom left in the figure), and
peak flux densities.
{See Table~\ref{table:modelfit} for component parametrization.}
\label{fig:map0212}
}
\end{figure}

\object{QSO\,0212+735} (Fig.\ \ref{fig:map0212}, $z$=2.367)
is the most distant source of the complete S5 polar cap sample.
Our 15.4\,GHz  VLBA maps show a jet-like structure 
extending
 eastwards up to 14\,mas ($\approx$85\hminus\,pc, 
the longest jet in the sample).
The total flux density of \object{QSO\,0212+735} 
remained virtually constant between our two observing epochs.
Our maps and model fit show two distinct regions in 
\object{QSO\,0212+735}: the inner core-jet region, 
formed by components UA, UB1 and UB2; 
and the extended, weaker region of the jet, composed 
of UC, UD, and UG.

The value of $Q$ (UA emission to extended emission), increased
from 0.8 to 1.0 between 1999.57 and 2000.46.
This is one of the lowest values of $Q$ found in the sample,
indicating that the extended emission 
contributes about the same amount as the emission from the core.
This is so because component UB (which we suggest is at the base of
the jet, and does not belong to the core region) is as strong as 
the core itself. 
(If for $Q$ we take the emission from UA+UB to the extended emission, 
we obtain 17.5 and 15.2, in 1999.57 and 2000.46,
respectively, and we would say the source is ``core''-dominated.)
The flux density of \object{QSO\,0212+735} puts it among the three
brightest sources of the sample and, given its redshift, it is the 
most powerful source with a monochromatic luminosity of 
$L_{\rm 15\,GHz}$=(9.2 to 9.3)$\times 10^{39}$\,W.

The flux density variations in each individual component of the innermost and
outermost regions are marginally consistent ($\sim$2\,$\sigma$), 
with no change between the first and second epochs.
The innermost region is the one that shows the most 
interesting
features:
component UA, the brightest component at both epochs and probably the core,  
lies at a distance of 0.6\,mas (3.6\hminus\,pc) from the origin,
at an angle of $\sim-47\degr$.
Component UB1 is at the origin, while 
component UB2 is at a distance $r\approx$0.26\,mas
at angle $\approx$90\,\degr.
The remaining components, UC, UD, and UG, lie at distances of
$\approx$0.9\,mas, 2.5\,mas, and 13.8\,mas
($\approx$83\hminus\,pc) respectively, and within an angle
of 90\degr to 100\,\degr at both epochs.
Our model fit for \object{QSO\,0212+735} gives no evidence
for any significant proper motion either in the 
innermost or the outermost region. 
(We note that \object{QSO\,0212+735} is one of only three sources in the sample
for which the brightest component is not at the origin of the coordinates; 
see Table~\ref{table:results-mapping}.)

%

Assuming that the 8.4\,GHz flux density of the source did not change
between 1999.41 and 1999.57, we obtain a global value of $\alpha=-0.07$.
A detailed comparison of our model fit at 15.4\, GHz for the first epoch
with the model fit made at 8.4 GHz for the second epoch (see Paper
I), gives the following results.
The position of component XA nicely corresponds to that of UA.
The inferred spectral index is then $\alpha$=0.78, a rather
inverted spectrum, as we would expect if UA/XA were the core.
Component XB is very likely a blend of two components (UB1 and
UB2), which yields $\alpha_{\rm UB/XB}=-0.45$.
The positions at 
8.4\,GHz and 15.4\,GHz
for the remaining components are the
same within the uncertainties.
For those components, we find a progressive steepening of the spectral
index with distance from the jet. Namely, 
$\alpha_{\rm UC/XC}=-0.49$, 
$\alpha_{\rm UD/XD}=-1.00$,
and 
$\alpha_{\rm UG/XG}=-1.76$.
The position of component UG/XG has remained remarkably stable, both
in the X- and U-band between 1997.93 and 2000.46, at a nominal
position of 13.8\,mas at angle 93\degr. 
Most likely, component UG/XG is a stationary region of the source.
We do not see emission from components UE+UF 
(clearly detected at 8.4\,GHz 
as XE+XF; see Paper I) above 3\,$\sigma$, which
implies a very steep spectrum for those components.


\subsection{\object{BL\,0454+844}\label{subsec:0454}}

\begin{figure*}[htbp]
\begin{tabular}{@{}cc@{}}
\includegraphics[width=0.45\textwidth]{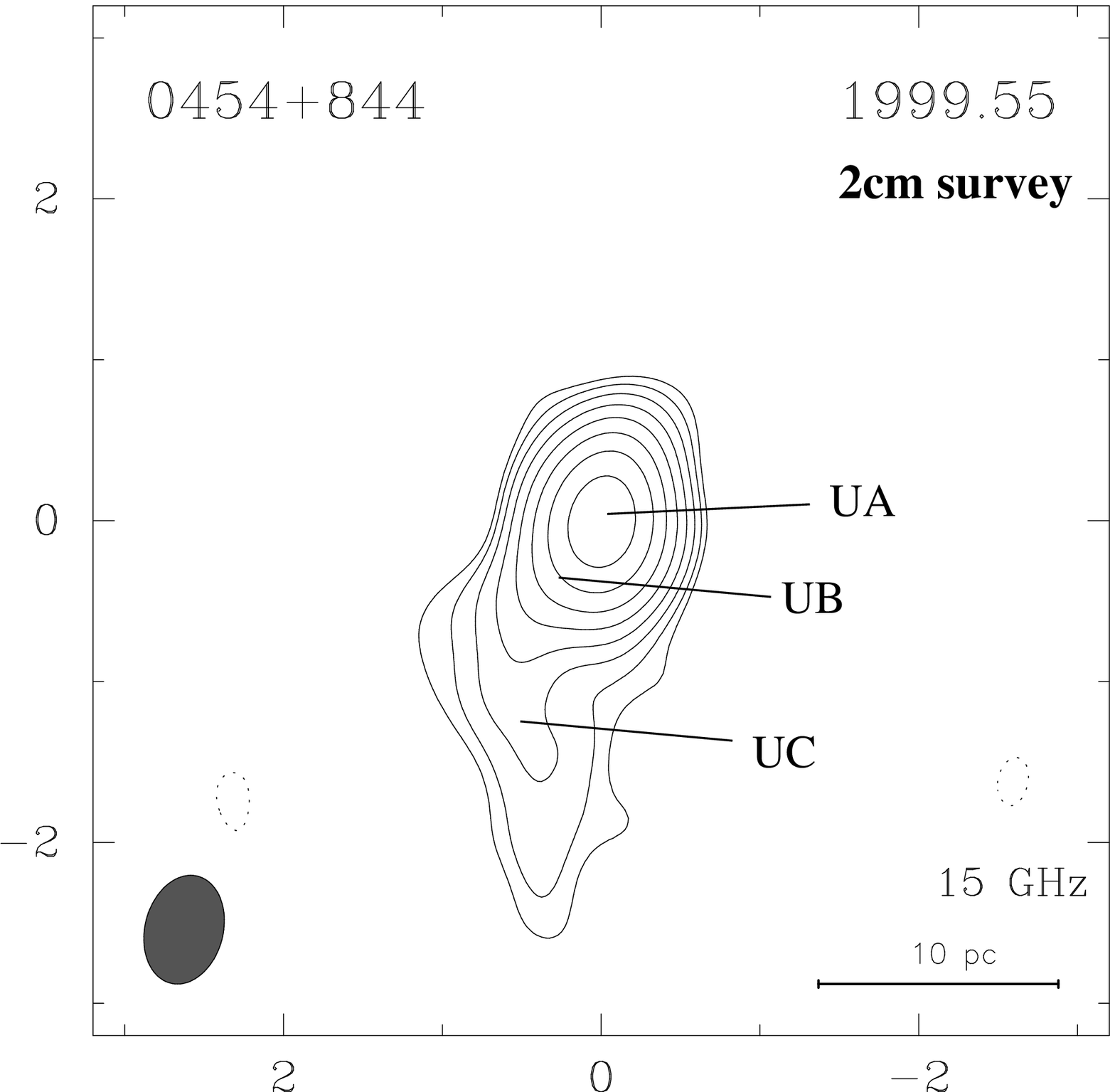} & 
\includegraphics[width=0.45\textwidth]{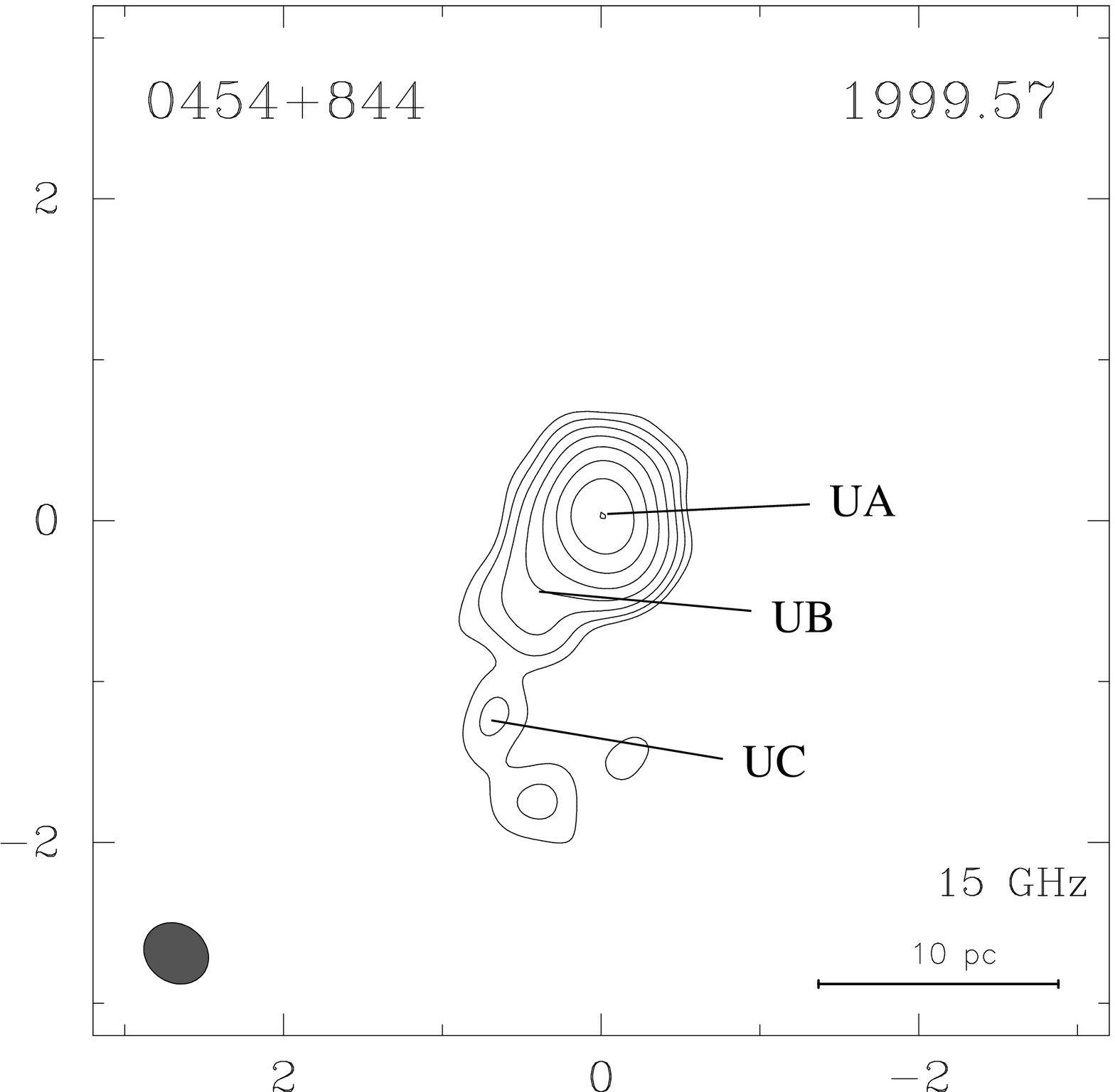} \\
\includegraphics[width=0.45\textwidth]{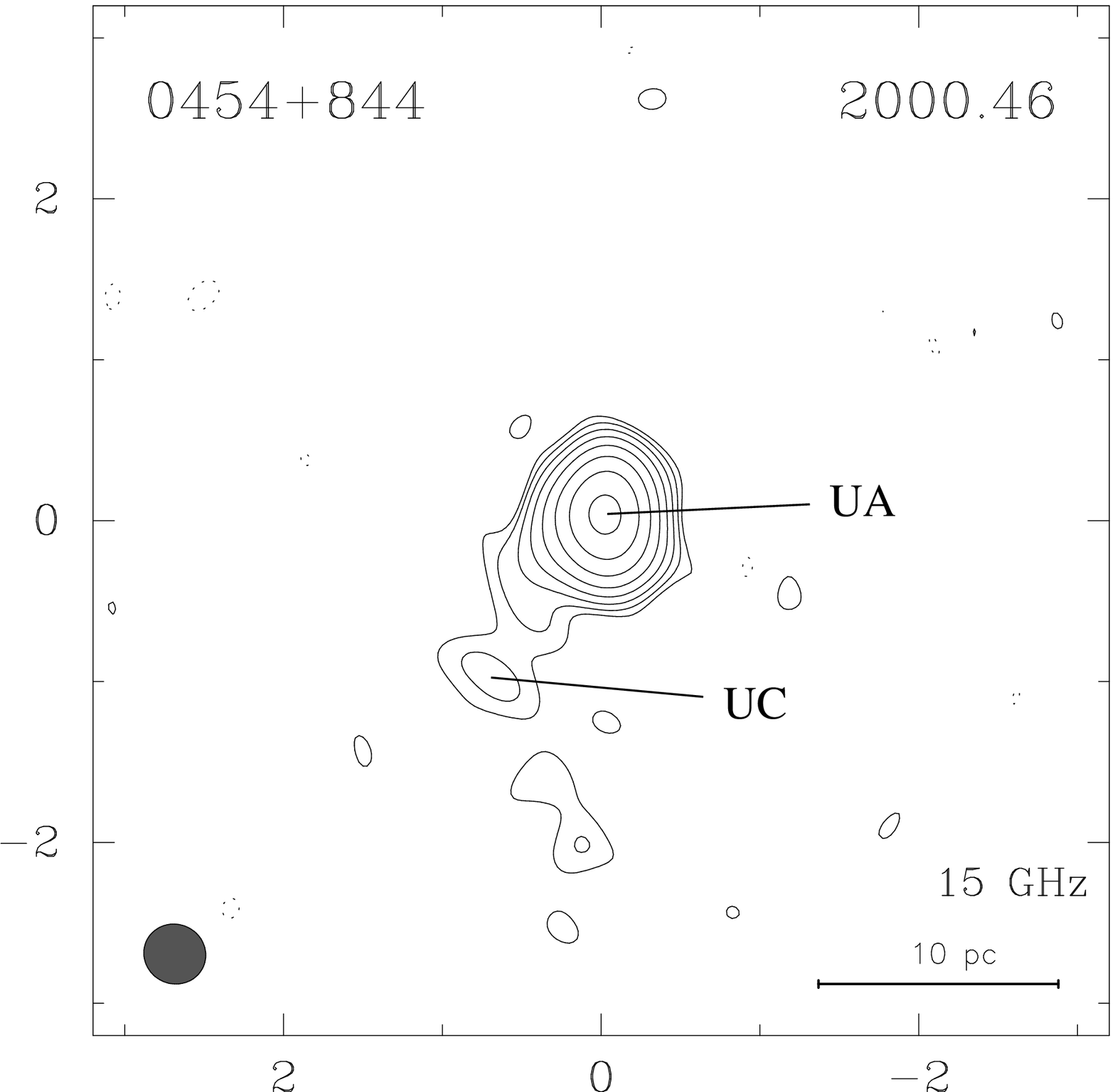} & 
\includegraphics[width=0.45\textwidth]{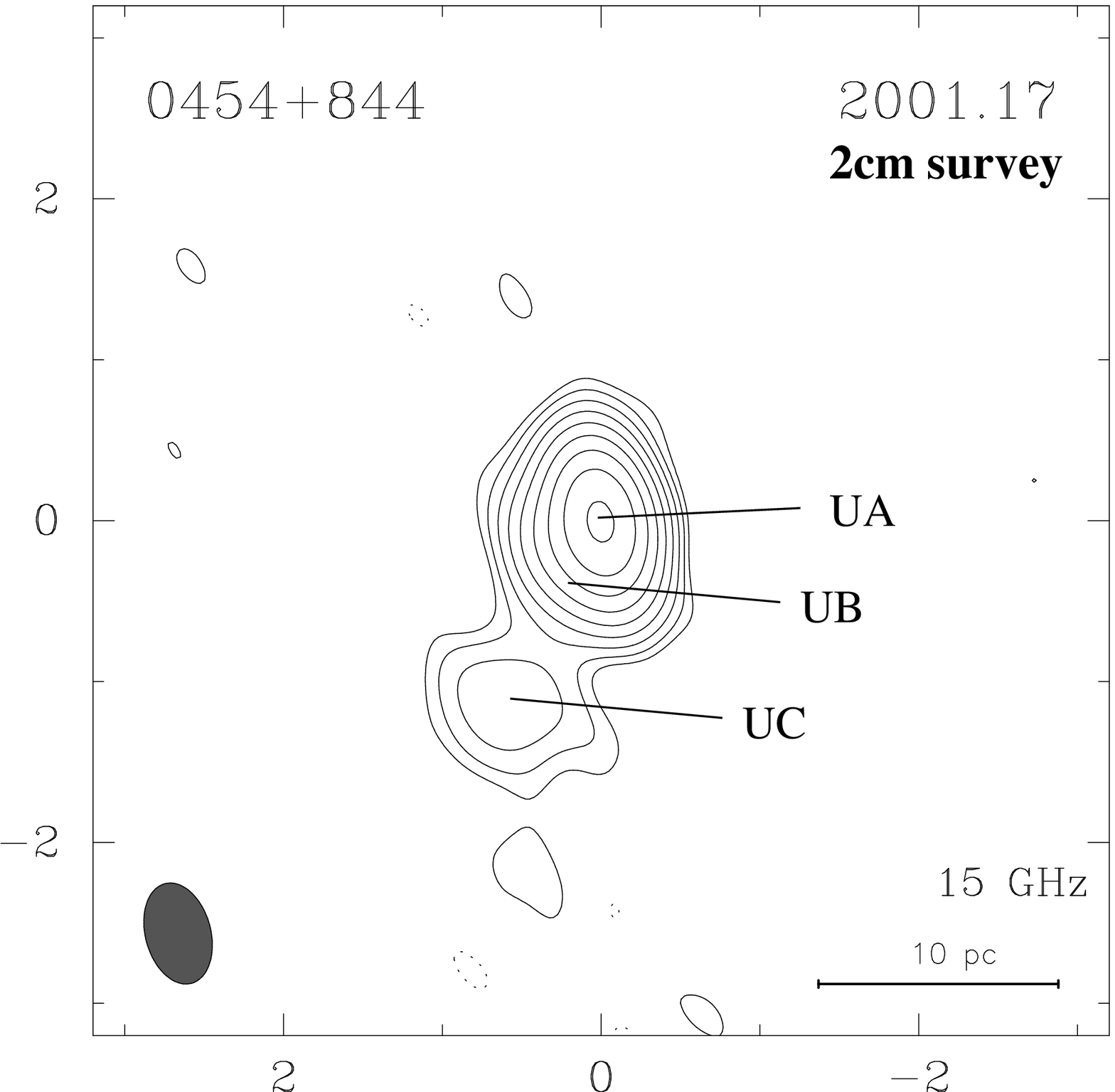} \\
\end{tabular}
\caption[]{
VLBA images of \object{BL\,0454+844}
from observations on
19 July 1999 (1999.55),  27 July 1999 (1999.57), 
15 June 2000 (2000.46),  and 4 March 2001 (2001.17).
See Tables 1--3 for contour levels, beam sizes (bottom left in the maps),
peak flux densities, and component parametrization.
Axes are relative $\alpha$ and $\delta$ in mas.
\label{fig:map0454}
}
\end{figure*}

\object{BL\,0454+844} is, at $z \gsim$1.340
(Stocke \& Rector \cite{stocke97}), 
the most distant BL Lacertae source of our sample 
with known redshift.
Its monochromatic luminosity is 
$L_{\rm 15\,GHz} \approx$(1.2 to 2.6)$\times 10^{38}$\,W (note that the previously accepted redshift 
for \object{BL\,0454+844}, $z=0.112$, implying a lower intrinsic 
luminosity by a factor of 200, which makes \object{BL\,0454+844}
the least powerful radio source of the sample).
\object{BL\,0454+844} shows, together with \object{BL\,0716+714},  
the 
 least
 structure of all the sources of the complete sample, as 
seen at 15.4\,GHz (Fig.\ \ref{fig:map0454}).
Indeed, there is no emission above 3\,$\sigma$ 
outside the inner 2\,mas (13.2\hminus\,pc) 
away from
component UA.  
Our VLBA observations showed a significant variation in 
the flux density of \object{BL\,0454+844} between 1999.55 and 1999.57, 
decreasing from 186\,mJy down to 120\,mJy (35\% change). 
It then started to slowly increase up to 131\,mJy in 2000.46
(9\% change), after which the source almost doubled its total flux
density in less than nine months (257\,mJy in 2001.17).
We fitted each observing epoch, except epoch 2000.46,  with a three-component
model (see Table \ref{table:modelfit}), which seems a reasonable
model fit of the radio source.
We find the following values for the ratio $Q$ 
(value, epoch: 3.43, 1999.55; 3.44, 1999.57; 6.37, 2000.46; 3.59, 2000.99).   
Those values seem to confirm the dominance of the core emission.
Note that $Q$ is significantly larger in 2000.46 than at the other
epochs, while the map for this epoch does not show significant 
differences with, e.g., the map in 1999.57.
This enhanced value of $Q$ in 2000.46 must be taken with caution, 
as the somewhat noisier data for this epoch resulted in a poorer 
model fit of the source structure.

At all epochs, most of the emission comes from component UA, 
likely to be the core,
which contributes $\approx$78\% of the total emission 
(except in 2000.46, where it amounts to $\approx$92\%; but see the caveat
above).
The spectral index for UA/XA is, assuming the 8.4\,GHz did not change
significantly between 1999.41 and 1999.55, $\alpha_{\rm
UA/XA}$=0.06, as expected for a core component.
Component UB belongs to the innermost regions of
\object{BL\,0454+844}, lying at a distance of 
(0.3 to 0.5)\,mas from
component UA, at an angle of 140\degr.
Component UC traces the jet region of \object{BL\,0454+844}. 
It probably consists of two or more sub-components, judging by the
fact that the radial distances span a huge range (1.2\,mas to 1.7\,mas)
in less than 20\,months.
 Our model fit indicates a backwards proper motion for
both components 
UB ($\mu = (-100\pm3)\,\mu$as/yr; $\betap = (-5.1 \pm 0.2)$\hminus),
and 
UC ($\mu = (-217\pm 140)\,\mu$as/yr; ($\betap = (-11.0 \pm 7.4)$\hminus).
However, the morphological structure of this source is so complex that 
those results must be taken with caution.

%


\subsection{\object{QSO\,0615+820}\label{subsec:0615}}

\begin{figure}[htbp]
\vspace{228pt}
\includegraphics{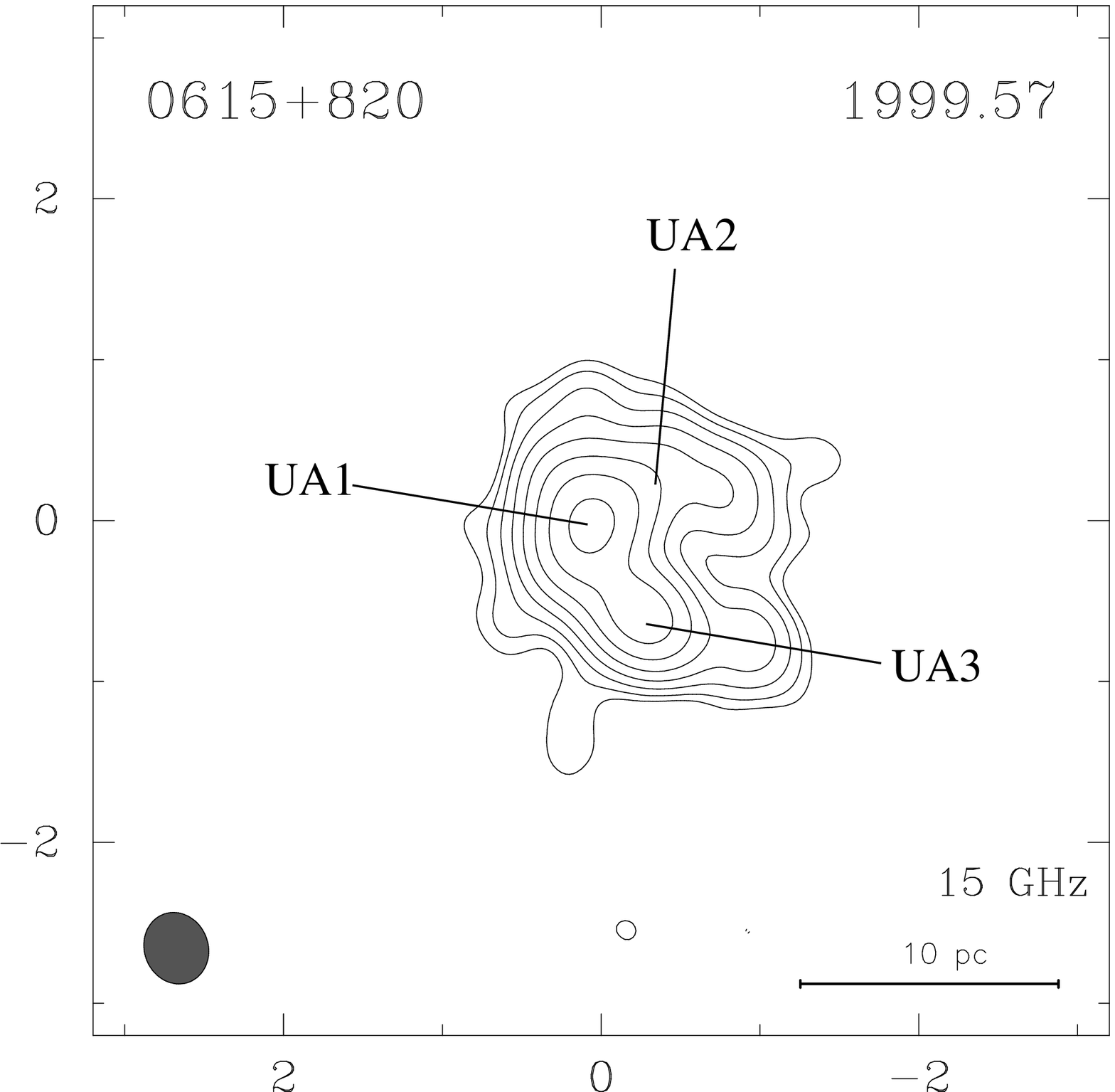}
\vspace{228pt}
\includegraphics{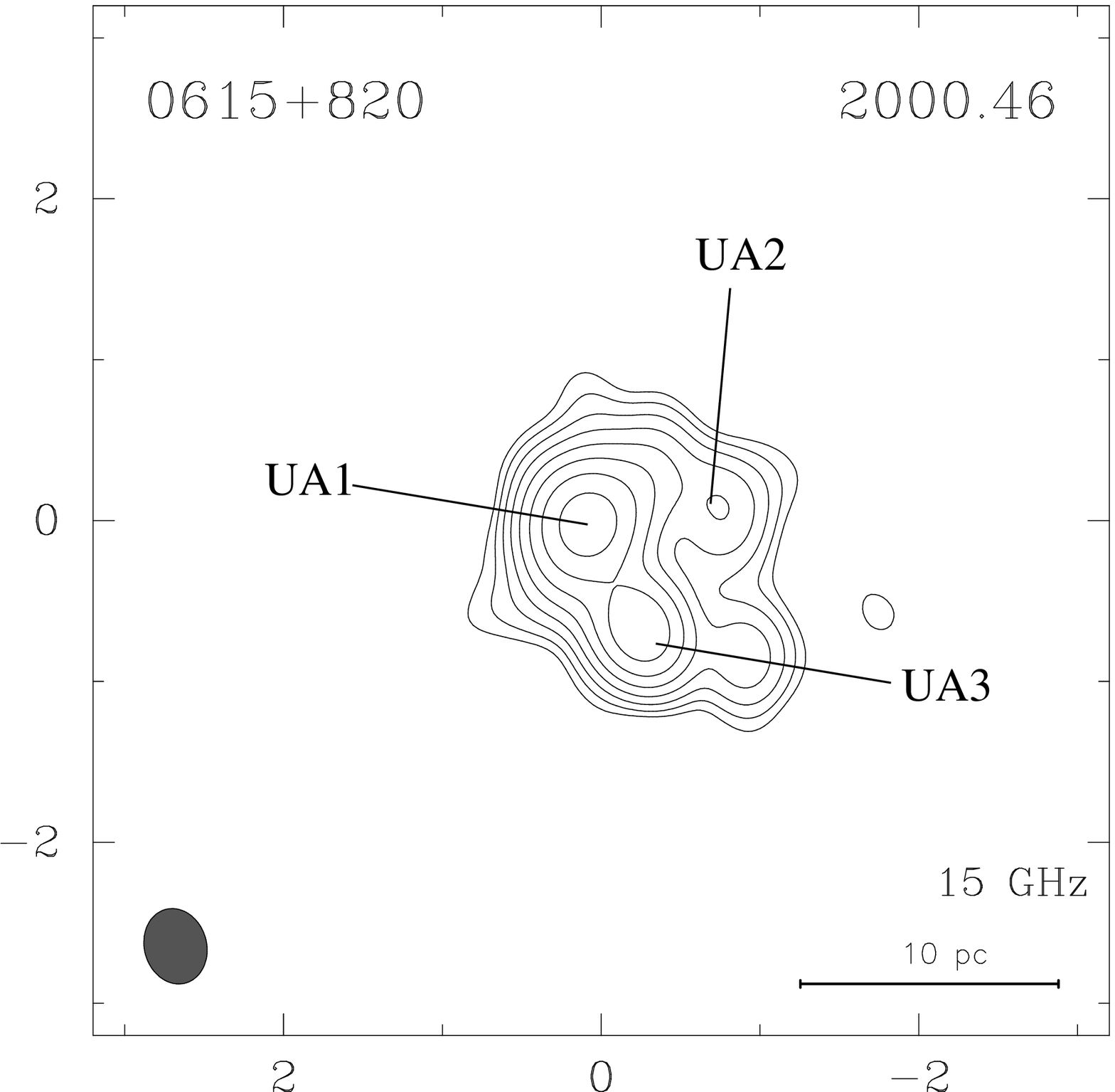}
\caption{VLBA images of \object{QSO\,0615+820}
from observations on
27 July 1999 (1999.57) and  15 June 2000 (2000.46).
Axes are relative $\alpha$ and $\delta$ in mas.
See Table 1 for contour levels,
synthesized beam sizes (bottom left in the maps), and
peak flux densities.
{See Table~\ref{table:modelfit} for component parametrization.}}
\label{fig:map0615}
\end{figure}

Our VLBA maps of \object{QSO\,0615+820} (Fig.\ \ref{fig:map0615}, $z$=0.710)
show a complex, albeit compact structure at both epochs,
with a maximum extension $\lsim 2.5$\,mas (=15.4\hminus\,pc).
The 15.4\,GHz total flux density of the source slightly decreased 
between 1999.57 and 2000.46, from 434\,mJy down to 410\,mJy
(6\% change), with a monochromatic luminosity of 
$L_{\rm 15\,GHz}$=(1.0 to 1.2)$\times 10^{38}$\,W.
Assuming that the 8.4\,GHz flux density of the source did not change
between 1999.41 and 1999.57, we obtain $\alpha=-0.26$, a moderately 
steep to flat spectrum for the milliarcsecond structure of the source.


We fitted reasonably well the milliarcsecond radio structure  of
\object{QSO\,0615+820} using three components (Table~\ref{table:modelfit}) 
within the inner 0.8\,mas (5\hminus\,pc). 
It might well be that components UA2 and UA3 consist, in turn, 
of two sub-components each, as
the maps seem to hint, but this fact is not supported by the
model fit at this stage.
The coordinates of components UA1, UA2, and UA3 in 1999.57
are so close to those obtained in Paper I at 8.4\,GHz 
(components XA1, XA2, and XA3 in 1999.41), that 
their physical association is very likely to be real.
The ratio, $Q$, of the core (UA1) to the extended emission increased steadily 
with time (value, epoch: 0.88, 1999.55; 1.04, 1999.57; 1.44, 2000.46).

This object is, given the intriguing (and underlying) substructure of
its components, worth being studied in detail. 
In particular, the core position is highly uncertain, 
and might not correspond to the position of UA1.

\subsection{\object{BL\,0716+714}\label{subsec:0716}}

\object{BL\,0716+714} 
is the only radio source of the sample
whose redshift is unknown, though the absence of any host galaxy 
seems to rule out distances $z\lsim$0.25 even for a low luminosity
object (Wagner \et\ \cite{wag96}).
The source displays intra-day variability
(Quirrenbach \et\ \cite{qui91}; Wagner \et\ \cite{wag96}).
The total flux density of \object{BL\,0716+714} at 15.4\,GHz 
halved its value between 1999.55 and 2001.17 (19 months), 
decreasing from 1254\,mJy down to 643\,mJy, in accordance with
the fact that \object{BL\,0716+714} is also a long-term variable.
(At 8.4\,GHz, the source displayed an opposite, 
increasing trend
between 1997.93 and 1999.41,  its flux growing by a factor 2.6,
from 377\,mJy up to 990\,mJy in less than 18 months
[see Paper I]).
At a redshift of 0.25, the monochromatic luminosity of 
\object{BL\,0716+714} is $L_{\rm 15\,GHz}\approx$(1.8 to 3.4)$\times 10^{37}$\,W, 
a low-to-intermediate value for a BL Lac object.

\begin{figure*}[htbp]
\begin{tabular}{@{}cc@{}}
\includegraphics[width=0.45\textwidth]{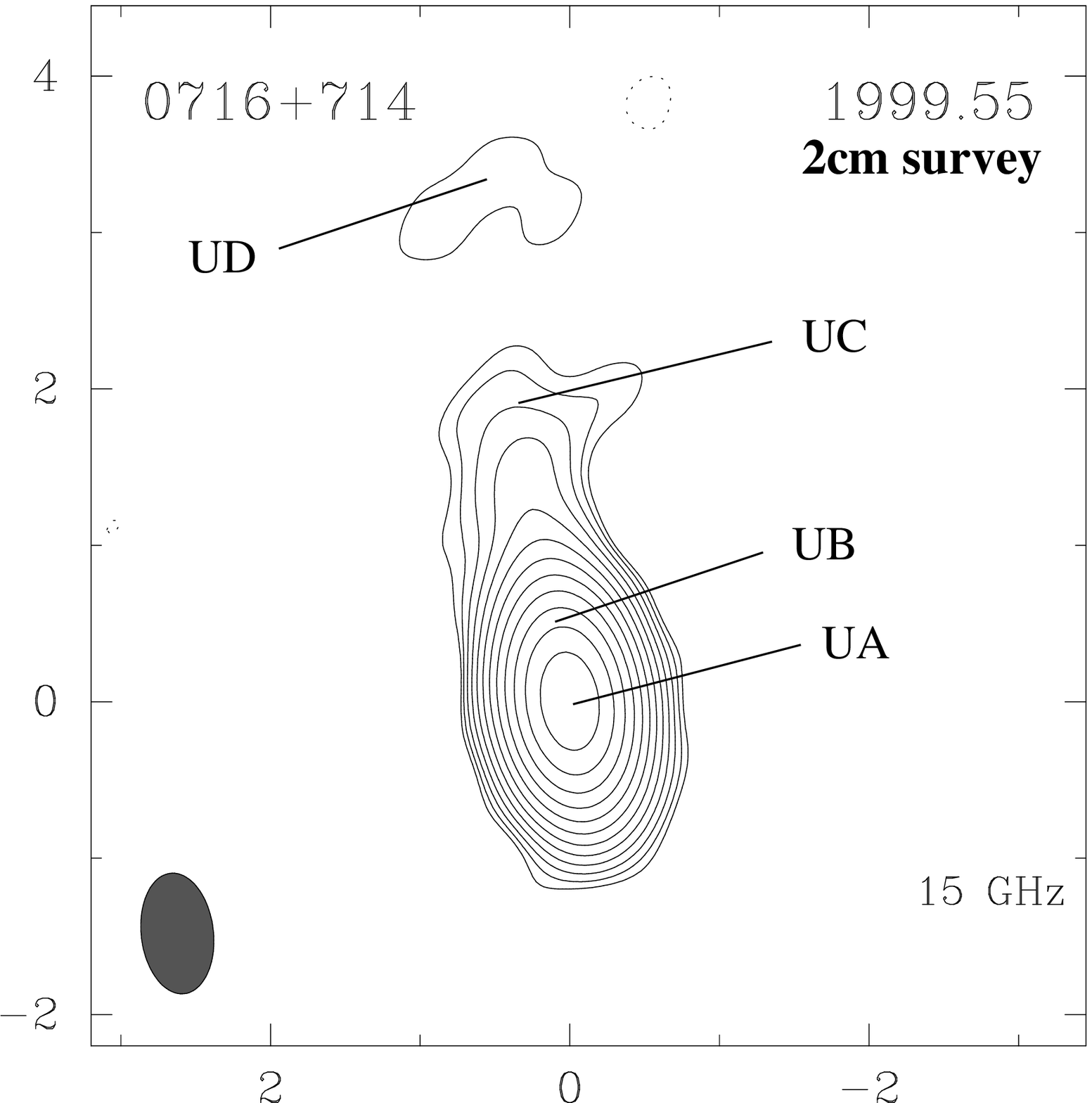} &
\includegraphics[width=0.45\textwidth]{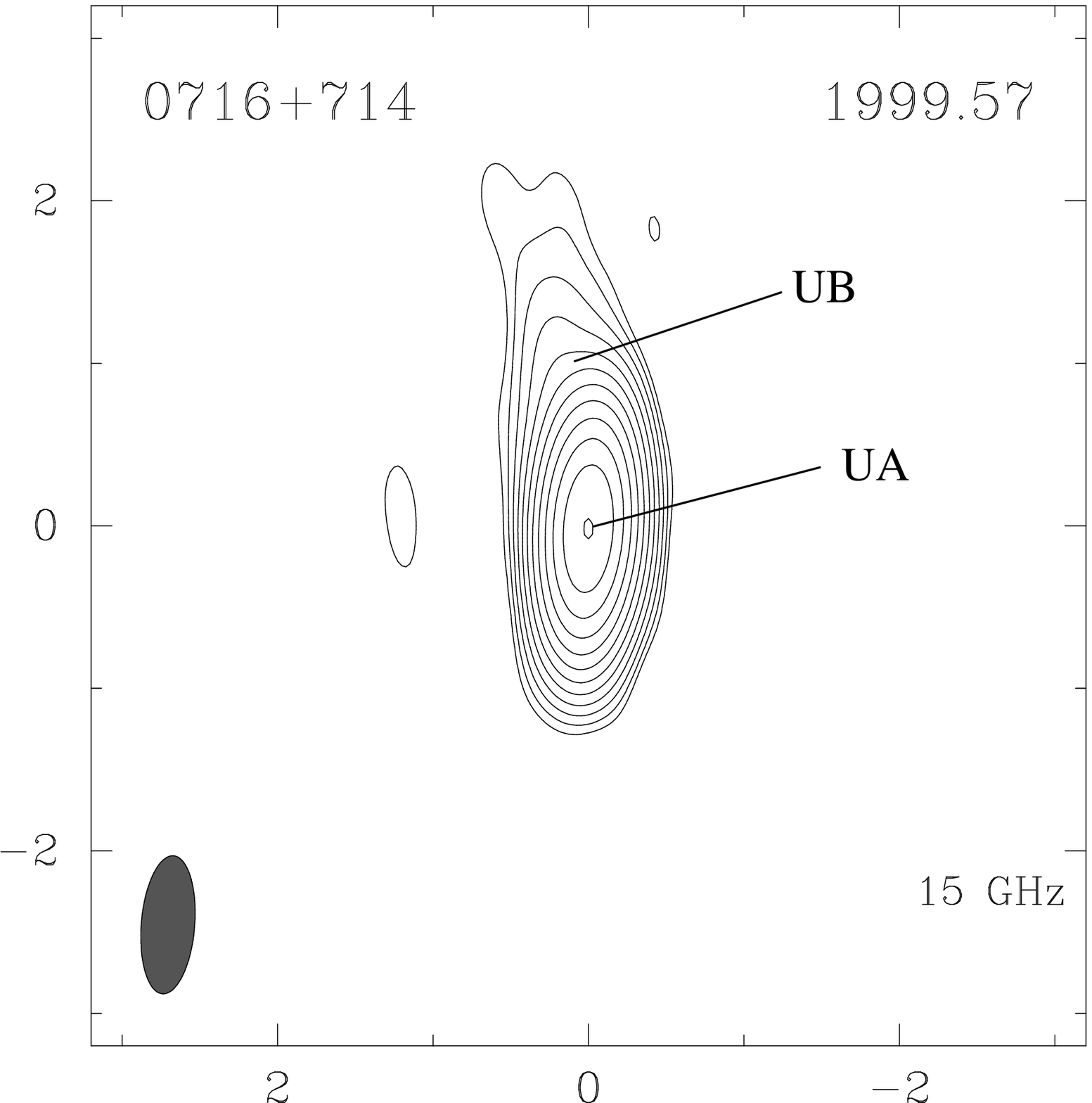} \\
\includegraphics[width=0.45\textwidth]{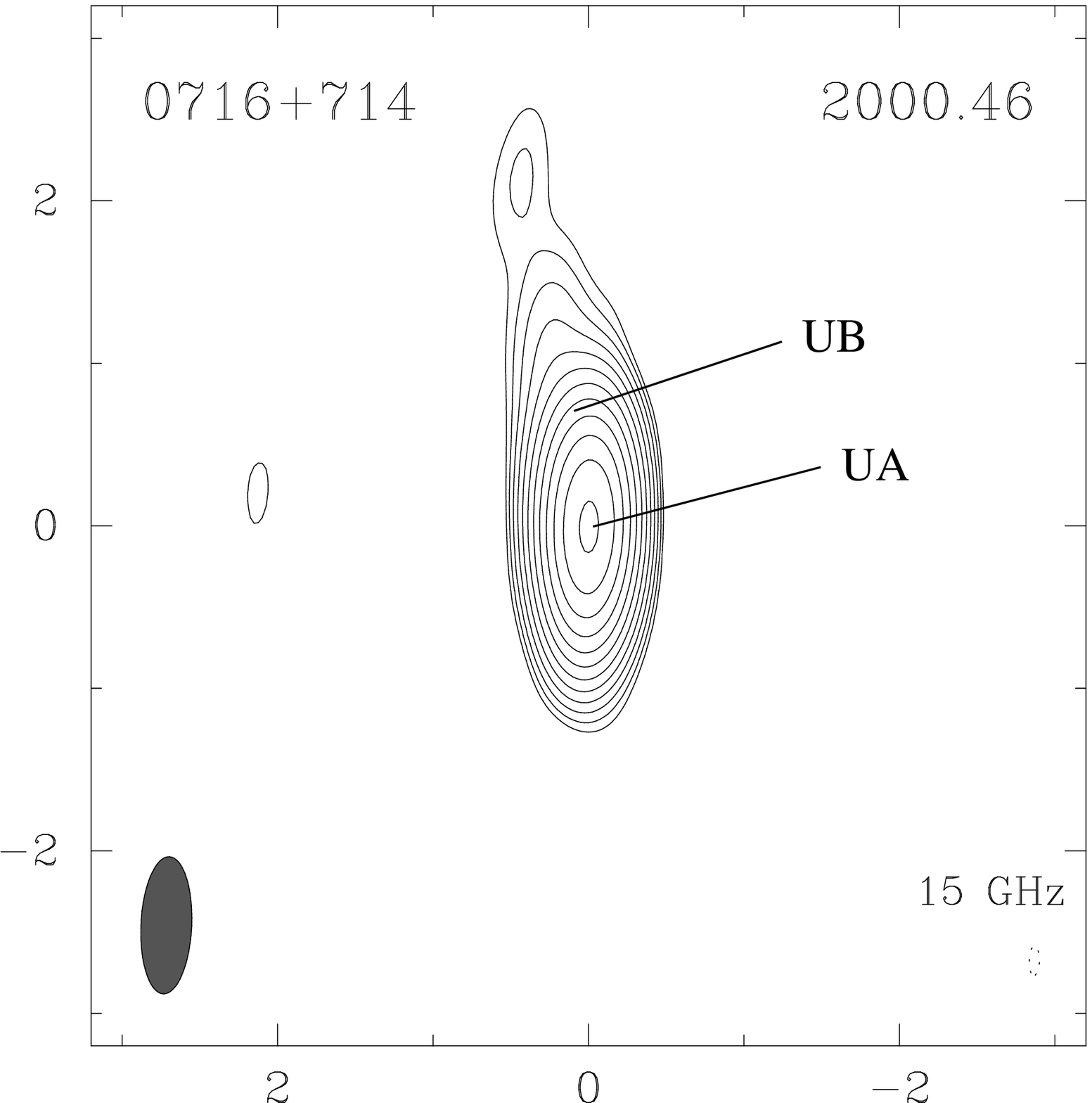} &
\includegraphics[width=0.45\textwidth]{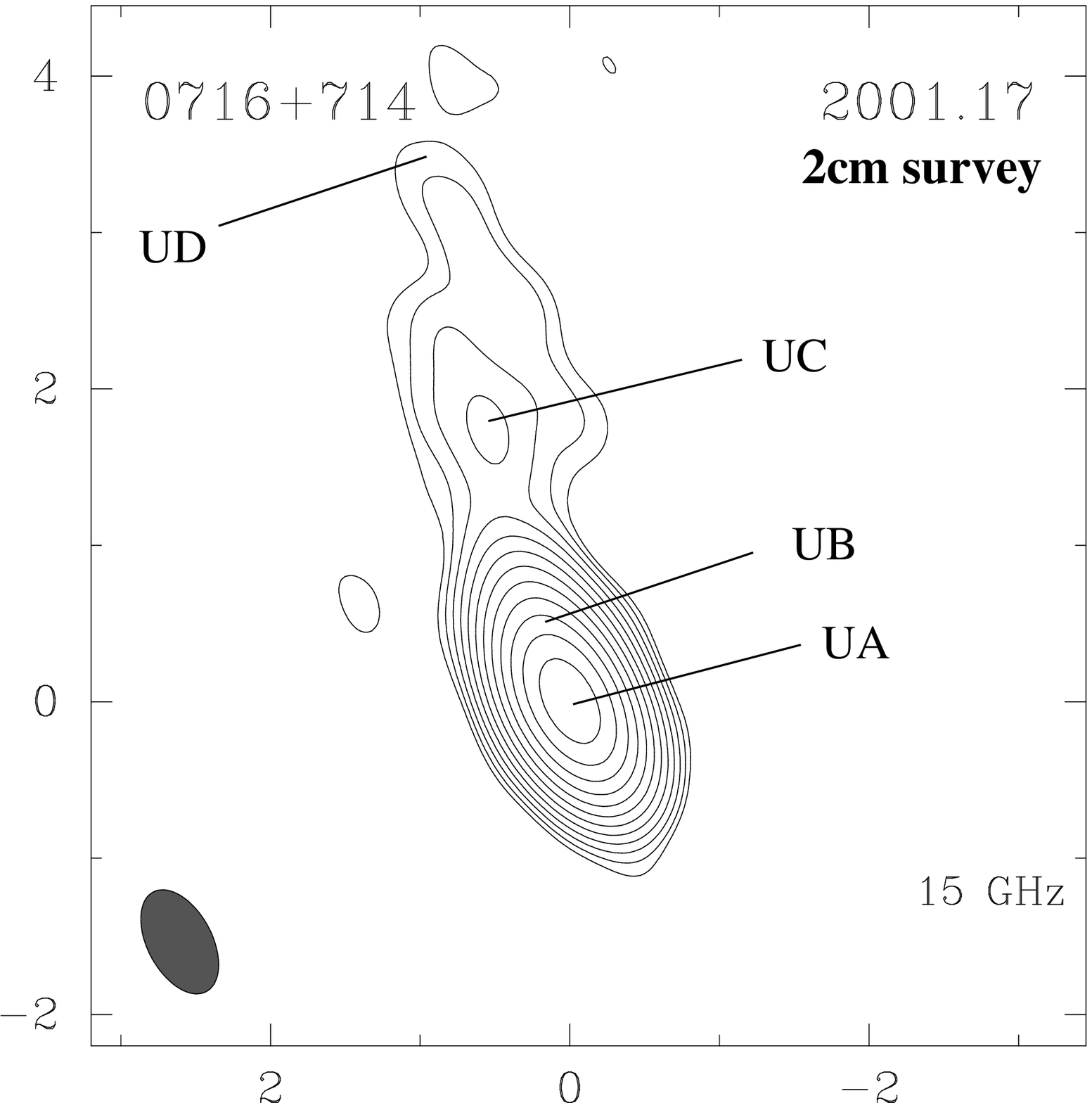} \\
\end{tabular} 
\caption[]{ 
VLBA images of \object{BL\,0716+714} from observations on
19 July 1999 (1999.55), 27 July 1999 (1999.57), 15 June 2000 (2000.46),
and  4 March 2001 (2001.17).
See Tables 1--3 for contour levels, beam sizes (bottom left in the maps),
peak flux densities, and component parametrization.
Axes are relative $\alpha$ and $\delta$ in mas.
\label{fig:map0716}
}
\end{figure*}

The 15.4\,GHz VLBA maps (Fig.\ \ref{fig:map0716}) show a core-jet structure that
extends northwards up to 4\,mas. The 
 appearance of the source at all
epochs is very similar, and rather featureless but for the last epoch. 
Our observations at 8.4\,GHz made in 1999.41 
resemble a scaled-up version of the 15.4\,GHz ones in 1999.57.
We modeled the 15.4\,GHz radio structure of \object{BL\,0716+714} in
1999.57 and 2000.46 with two components (see Table
\ref{table:modelfit}), instead of three,
as we did for our 8.4\,GHz observations (Paper I).
Nonetheless, we had to use up to four components to model the source
radio emission seen in 1999.55 and 2001.17, the two extra components
being needed to account for the extended radio emission of the jet.

Component UA, the brightest one, is at the origin and contributes 
85\% to 96\% of the total VLBA emission at all epochs. 
Its flux density evolves 
in almost exactly the same way as the total flux of
the source (the core activity drives the whole behaviour of the
source).
The ratios of the core (UA) to the extended emission were
the following: 
(value, epoch: 
14.7, 1999.55; 21.6, 1999.57; 14.0, 2000.46;
5.8, 20001.17).
Those values make of \object{BL\,0716+714} one of the 
most compact and core dominated sources of the sample.
The relatively low value of $Q$ in 2001.17 is chiefly due to the
decreasing flux density of the core
(see the model fits in Table~\ref{table:2cm-survey}).
Component UB 
probably lies close to the jet base.
It contributes 5\% to 11\% of the total 15.4\,GHz VLBA flux.
The model fit algorithm gave significantly different values for 
the position of this component at each epoch (see Table
\ref{table:modelfit}). 
This could be an indication that motions could be taking place
in this engine. 
Alternatively, it could be that UB is in fact composed of several
subcomponents.
Unfortunately, neither the 8.4\,GHz nor the 15.4\,GHz VLBA observations 
have resolved the innermost structure of \object{BL\,0716+714}.
Our 43\,GHz observations might yield 
enough resolution to 
say whether changes in its flux density
are due to the emergence of components along the core-jet
structure.
The remaining 2\% to 3\% of the 15.4\,GHz flux comes from components
UC and UD, which are not well modeled in 1999.57 and 2000.46.

Our model fit for the source 
shows an inward proper motion for component
UB (see Table~\ref{tab:proper_motion}), but 
only at the 1$\sigma$ level.
Bach \et\ (\cite{bach03}) have reanalysed multi-frequency VLBI
data obtained between 1993 and 2001, and found that the source
components display moderate superluminal motions 
(0.3\masyr; $\betap$=4.6\hminus) from the inner 
3\,mas of the jet, up to 0.8\masyr ($\betap$=12.3\hminus) for
the outer regions of the jet. 

Since \object{BL\,0716+714} is an intra-day variable, the assumption
that the 15.4\,GHz flux density of the source did not change
between 1999.41 and 1999.55 
is not well based.
Yet, if we assume the flux density did not change we obtain
$\alpha_{\small UA/XA}$=0.38, an inverted spectrum for the innermost 
structure of the source, 
in agreement with expectation if component 
UA/XA is the core.


\subsection{\object{QSO\,0836+710} (\object{4C\,71.07})\label{subsec:0836}}

\begin{figure}[htbp]
\begin{tabular}{@{}cc@{}}
\includegraphics[width=0.40\textwidth]{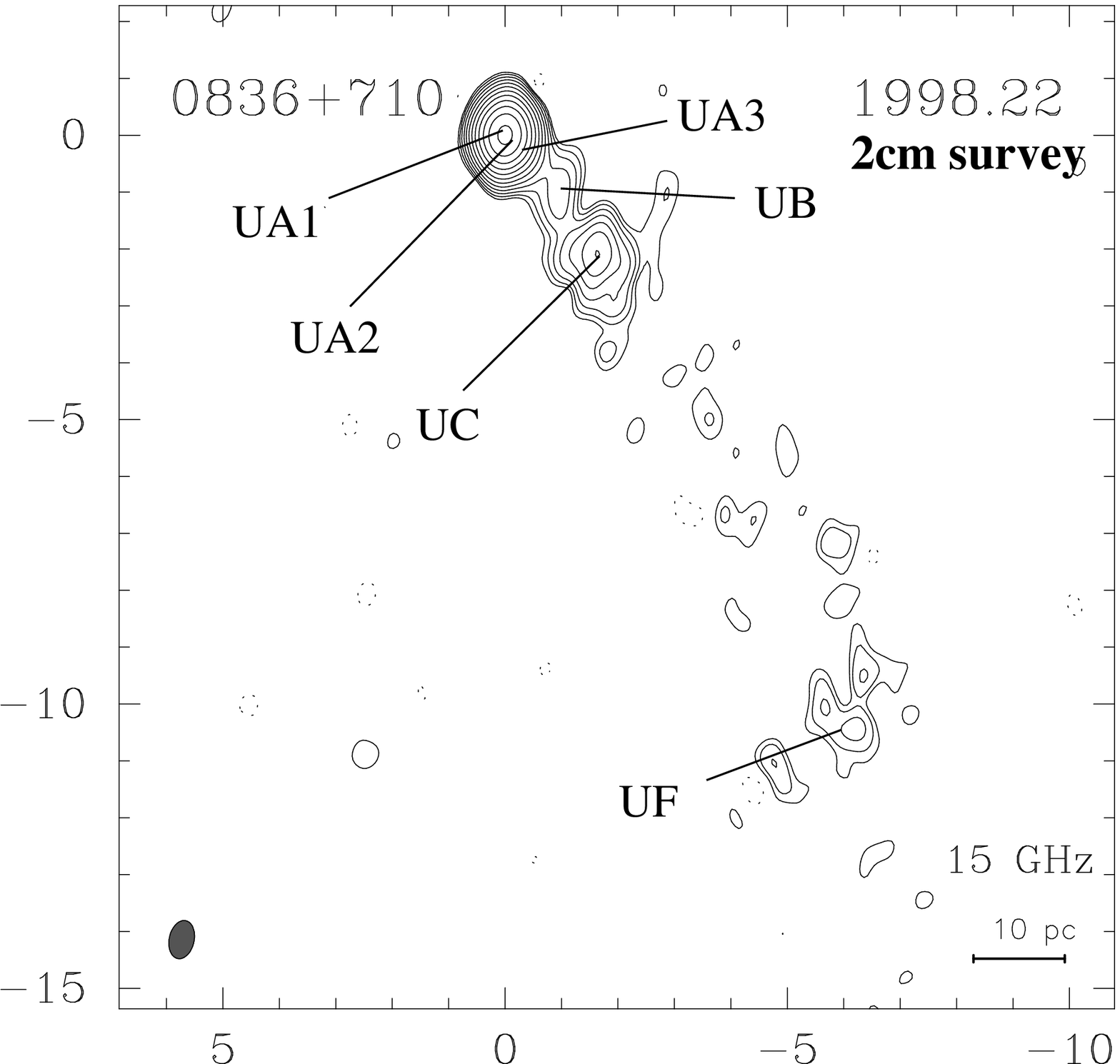} \\
\includegraphics[width=0.40\textwidth]{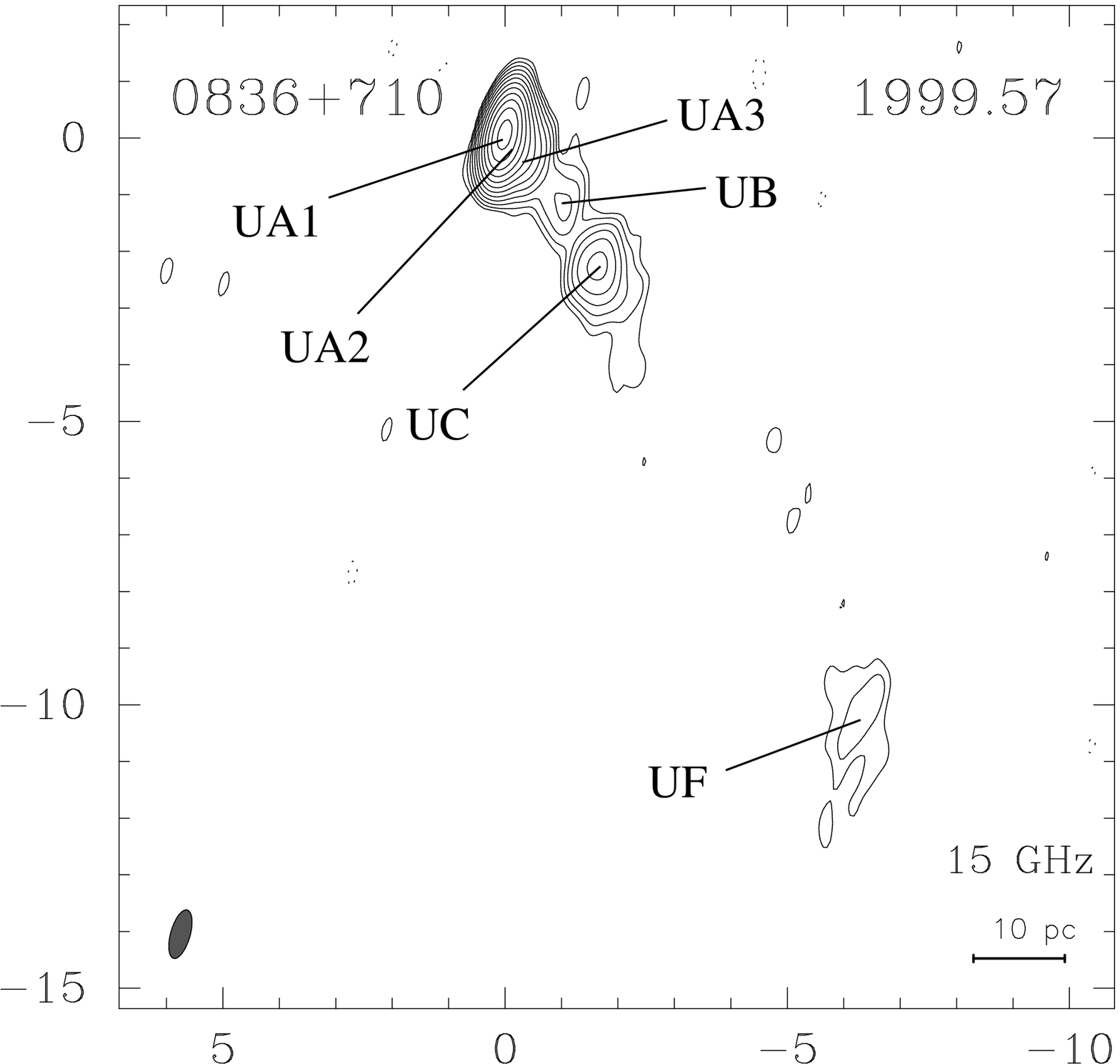} \\
\includegraphics[width=0.40\textwidth]{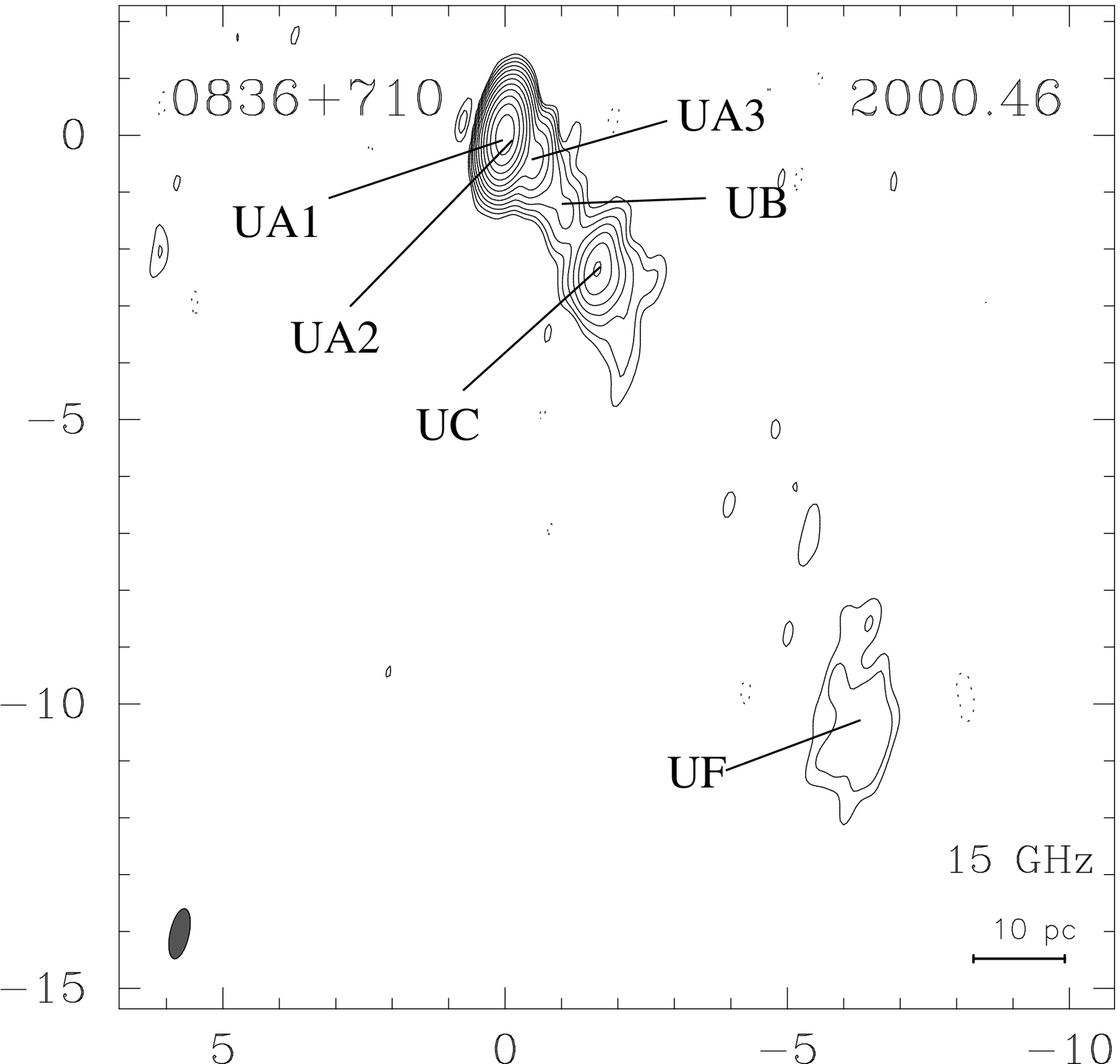} & \\
\end{tabular}
\caption[]{
VLBA images of \object{QSO\,0836+710} from observations on
20 March 1998 (1998.22), 27 July 1999 (1999.57), and  15 June 2000 (2000.46).
See Tables 1--3 for contour levels, beam sizes (bottom left in the maps),
peak flux densities, and component parametrization.
Axes are relative $\alpha$ and $\delta$ in mas.
\label{fig:map0836}
}
\end{figure}

The 15.4\,GHz VLBA maps of \object{QSO\,0836+710} 
(Fig.\ \ref{fig:map0836}; $z$=2.218, V\'eron-Cetty \& V\'eron [\cite{veron03}])
show at all three epochs a rather complex, 
one-sided core-jet structure at 
P.A.=-(140\degr to 150\degr).
We detected emission up to 12\,mas (74\hminus\,pc) away 
from the core.
The source radio emission is well 
represented by six components, five of them
within the inner 3\,mas (10\hminus\,pc) of the core, and one at a distance
of 12\,mas.

%

The 15.4\,GHz total flux density of \object{QSO\,0836+710} varied 
significantly between 1998.22 and 1999.57, 
decreasing by 23\%, from 2298\,mJy down to 1766\,mJy at a monthly rate of
1.4\%. It continued to decrease until 2000.46 (down to 1728\,mJy, a further 
2\% decrease at the much slower rate of 0.2\% per month).
\object{QSO\,0836+710} is the fourth brightest object of the complete
sample, yet it is the second most powerful emitter with
$L_{\rm 15\,GHz}\approx$(5.5 to 7.3)$\times 10^{39}$\,W.
The flux density trend of the brightest component, UA1 (likely the
core), seems to follow that of the overall source.
The values of the core-to-extended emission, $Q$, 
were 2.7 (1998.22), 2.9 (1999.57), and
2.5 (2000.46), indicative of a {relatively} compact, core-dominated source.
The flux density of component UA2 remained very stable with time, at the
220\,mJy level (10\% to 13\% of the total flux density),
but its radial distance is uncertain.
Component UA3, at a 30\,mJy level, contributes about 2\,\% of
the total flux density. 
Provided the 8.4\,GHz flux density of the (XA+XB) region did not change
significantly between 1999.41 and 1999.57, we obtain $\alpha_{\rm
UA/(XA+XB)}$=0.27.
This value indicates an inverted spectrum for the innermost region
source, and suggests that UA is likely to be the core
(namely, UA1, which dominates the emission of the region).
UB, at a distance of (1.5$\pm$0.1)\,mas ($\approx$5\hminus\,pc) 
at an angle of 140\,\degr 
sets the transition from the core region to the extended emission
of the jet. Its contribution remained constant at the 2\,\% level
at all epochs.
Similarly, the spectral indices of components UC/XC and UF/XF are
$\alpha_{\rm UC/XC}=-1.21$ and 
$\alpha_{\rm UF/XF}=-1.36$, respectively.

Compared to our 8.4 GHz VLBA maps, we now have a 
clearer view of the innermost regions of the source. 
In particular, we are able to
disentangle the fine structure of XA into
three sub-components: UA1, UA2, and UA3.
Component XB in Paper I is likely the result of a 
blending of several subcomponents at 15.4\,GHz, most probably UA3 and UB.

Our model fit suggests the existence of superluminal motions for
components UA3, UB, and UC 
($\beta_{\rm app}\approx$7\hminus; see Table~\ref{tab:proper_motion}). 
We identified components UC and UF with XC and XF, respectively 
(Paper I).
The positions of UF are consistent with no superluminal motion,
which suggests that it is a stationary feature in the jet.
UF shows an optically thin spectrum.
We detected no emission from components
XD or XE above our 3\,$\sigma$ level of 2.4\,mJy in 1999.57.
Our non-detection implies an upper limit for their synchrotron spectra
of $\alpha\lsim-1.34$ and $\alpha\lsim-2.07$.
The disappearance of the jet 
beyond component UC might well be
due to the prominent curvature displayed at 5\,GHz, as 
shown in exquisite detail with VSOP (Lobanov \et\ \cite{lobanov98}), 
and to a lesser extent in our 8.4\,GHz VLBA images (Paper I).

The  ejection of a new component at mas scales reported by Otterbein 
\et\ (\cite{ott98}) at epoch 1992.65, with an apparent superluminal motion 
of $\mu$=(0.26$\pm$0.03)\,mas/yr has been directly related to {gamma-,}
X-ray, and optical activity observed in February 1992 by these authors 
(Otterbein \et\ \cite{ott98}).
The intense radio activity shown in the inner 1.5\,mas from
our 15.4\,GHz VLBA observations seems to suggest the ejection of
at least one new component.


\subsection{\object{QSO\,1039+811}\label{subsec:1039}}

\begin{figure}[htbp]
\vspace*{227pt}
\includegraphics{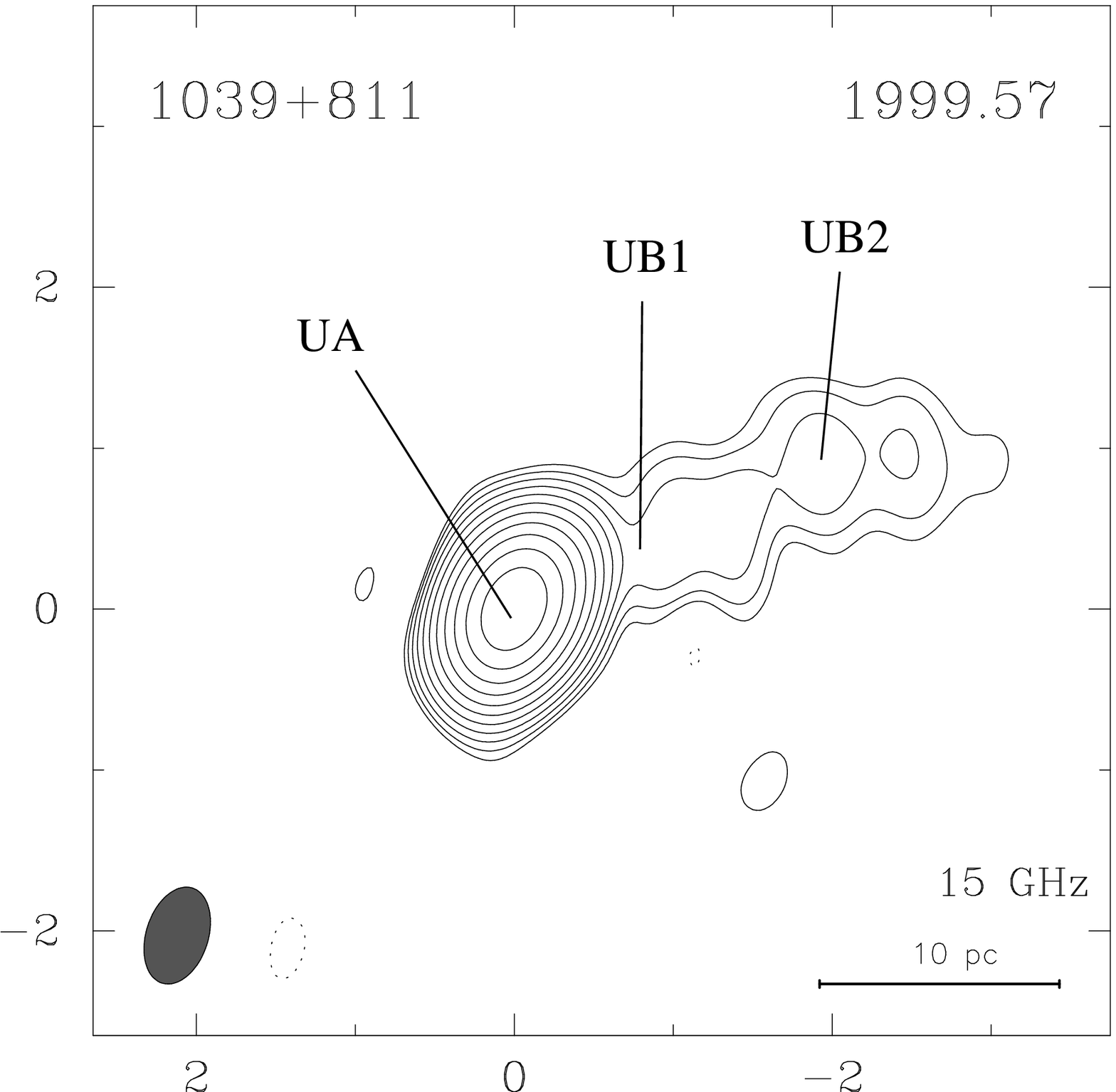}
\vspace*{227pt}
\includegraphics{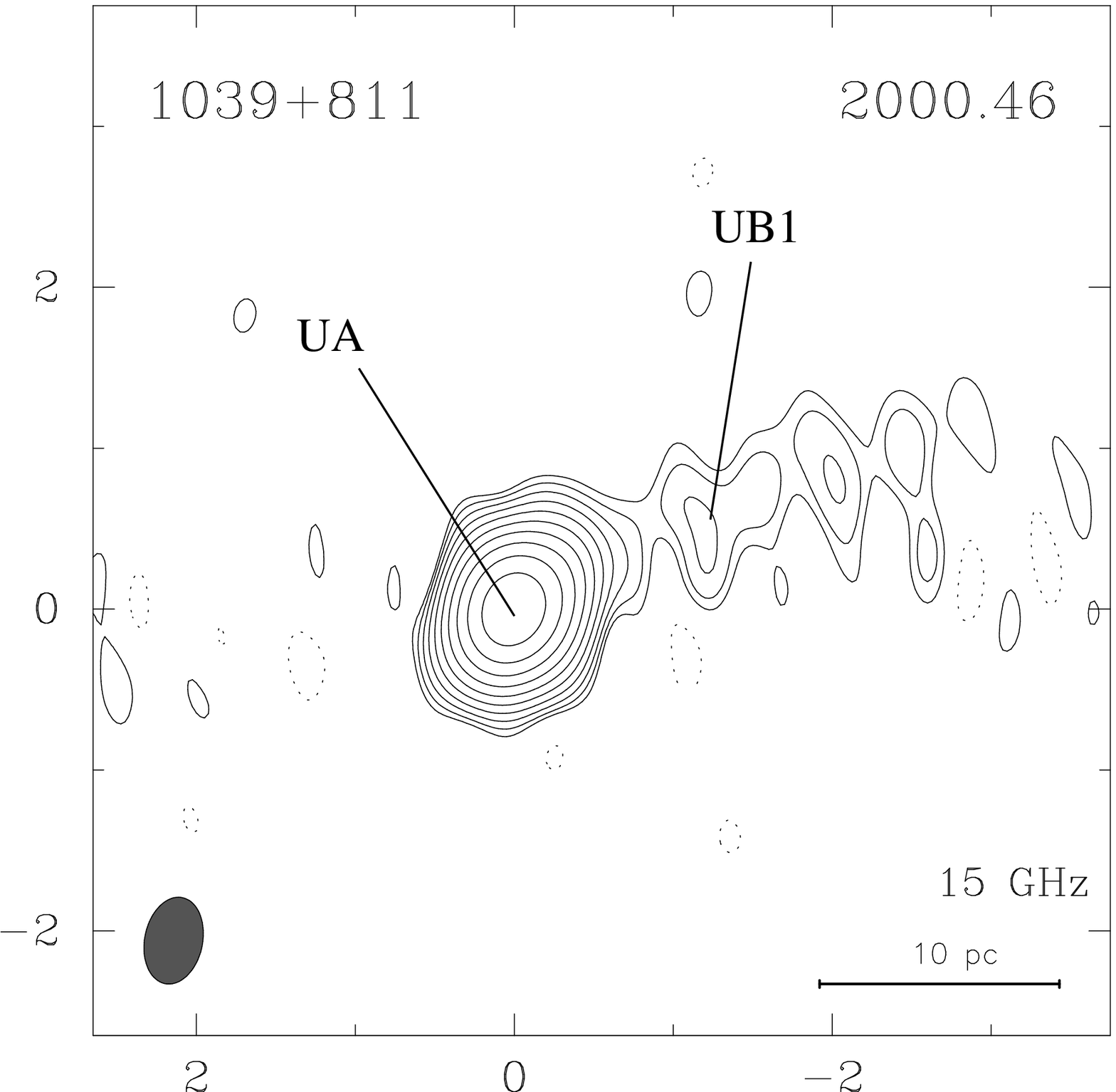}
\caption{VLBA images of \object{QSO\,1039+811}
from observations on
27 July 1999 (1999.57) and  15 June 2000 (2000.46).  
Axes are relative $\alpha$ and $\delta$ in mas.
See Table 1 for contour levels,
synthesized beam sizes (bottom left in the figure), 
and peak flux densities.
{See Table~\ref{table:modelfit} for component parametrization.}
\label{fig:map1039}
}
\end{figure}

Our VLBA maps of \object{QSO\,1039+811} (Fig.\ \ref{fig:map1039}, $z$=1.264)
show a relatively simple, one-sided core-jet structure at
an angle of -70\degr at pc-scales.
The 15.4\,GHz jet extends up to 3\,mas (20\hminus\,pc) 
from the core, a much less extended jet than seen 
at 8.4\,GHz (8\,mas long; see Paper I).
The emission is heavily concentrated within the core region 
(UA), which contributes more than 94\,\% 
of the total VLBA flux
density at 15.4\,GHz at both epochs. 
The remaining 6\,\% comes from the weak, relatively extended jet.
Correspondingly, the ratio of the core-to-jet emission, $Q$, was
as large as 21 in 1999.57, and 19 in 2000.46, making of
\object{QSO\,1039+811} the most compact, core-dominated QSO of the 
complete sample.
The jet shows some substructure, which we are able to model 
as two components in the first epoch (components UB1 and UB2, 
50\,mJy).
These components are model fitted as just one blended component
at the second epoch, having almost the same flux density
of UB1+UB2.
Although our maps show that the jet morphology
seems to have experienced dramatic changes between 1999.57 and 2000.46,
its interpretation should be
taken with caution. In particular, our second epoch is noisier
than the first one, thus resulting in a lower confidence of the position
of the features inside the jet.

%

Our model fit results do not show any evidence 
of backwards motion in the jet, in contrast to what was reported
for components XB and XC at 8.4\,GHz 
(Paper I; also for XD, XE, and XF, without counterparts at 15.4\,GHz).
As noted in Paper I, such apparently contracting motion could be related 
to changes in the core region, e.g., 
the emergence of a component can induce, in its early stage of ejection, 
an apparent backwards motion of the rest of the jet components at cm wavelengths
(e.g., Marcaide \et\ \cite{marca85}, Guirado \et\ \cite{gui98}, Ros \et\ \cite{ros99}).
Our model fit also did not show the counterpart for component 
XB seen in 1999.41, at $r \approx 0.4$\,mas of the core. 
We tried to set an upper limit to the position of such a component in our
first epoch, 1999.57, by forcing a model fit of the core with two components.
Although we managed to model the core with two components, the model fit
was no better, and therefore we found that a second component is not
needed to explain the emission from the core.
We suggest that if a second such component does exist, 
it should have stayed at roughly the same distance ($\approx$0.2\,mas) at
both epochs, and with an estimated uncertainty in its flux density
of 100\%. 

The 15.4\,GHz total flux density of \object{QSO\,1039+811} 
decreased by 12\,\% between 1999.57 and 2000.46, 
changing from (926$\pm$2) \,mJy down to (826$\pm$3)\,mJy,
which corresponds to a monochromatic luminosity of 
$L_{\rm 15\,GHz}$=(7.4 to 8.3)$\times 10^{38}$\,W.
The flux decrease is mainly due to the decrease in flux density of
UA, from (883$\pm$2)\,mJy to (785$\pm$2)\,mJy (11\% change).
Assuming that the 8.4\,GHz flux density of the source did not change
between 1999.41 and 1999.57, we obtain $\alpha$=0.08, a flat
spectrum for the milliarcsecond structure of \object{QSO\,1039+811}.
The inverted spectrum of component UA ($\alpha_{\rm UA/XA}$=0.34)
supports our suggestion that this component is the core.


\subsection{\object{QSO\,1150+812}\label{subsec:1150}}

\begin{figure}[htbp]
\vspace*{227pt}
\includegraphics{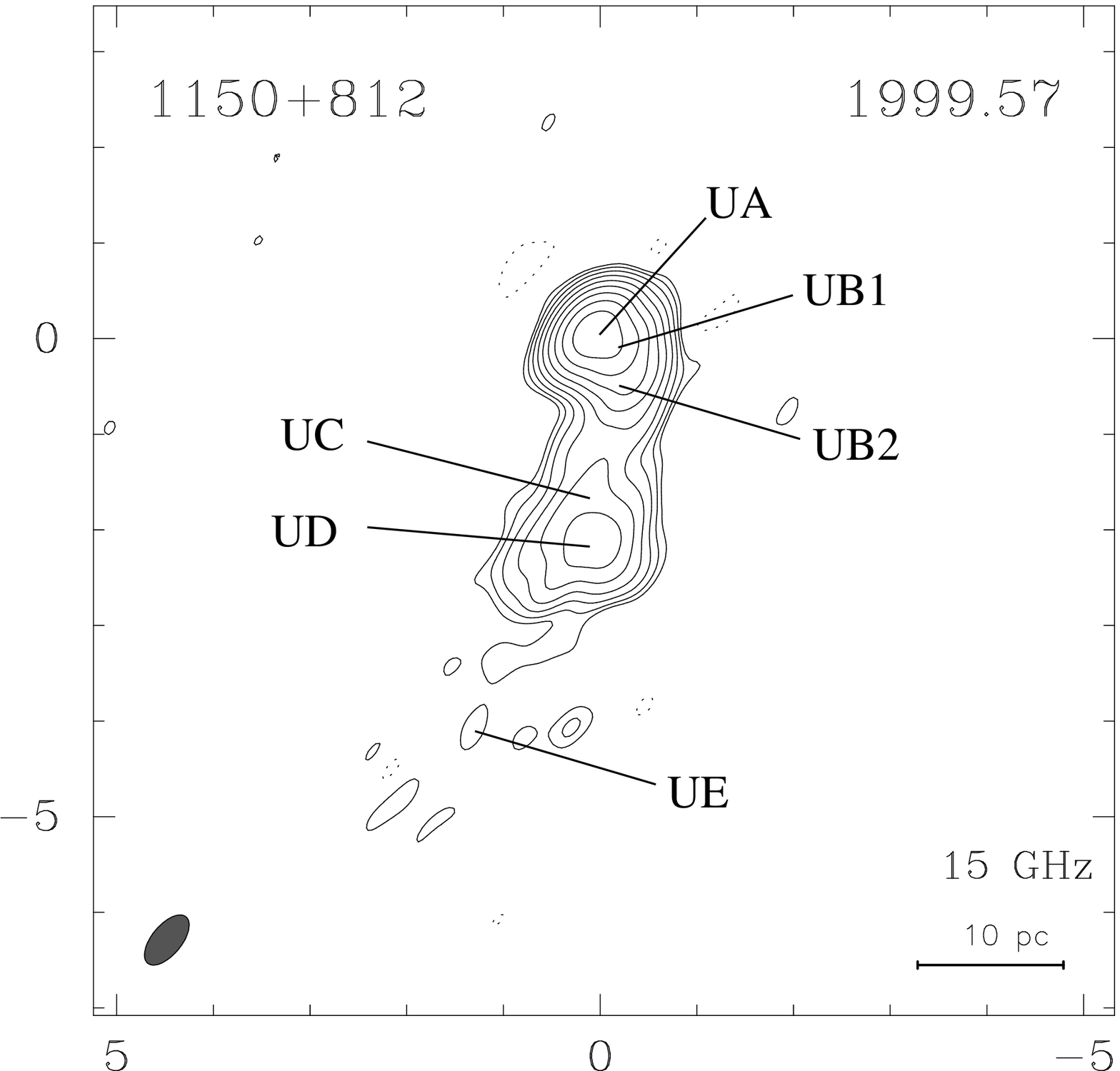}
\vspace*{227pt}
\includegraphics{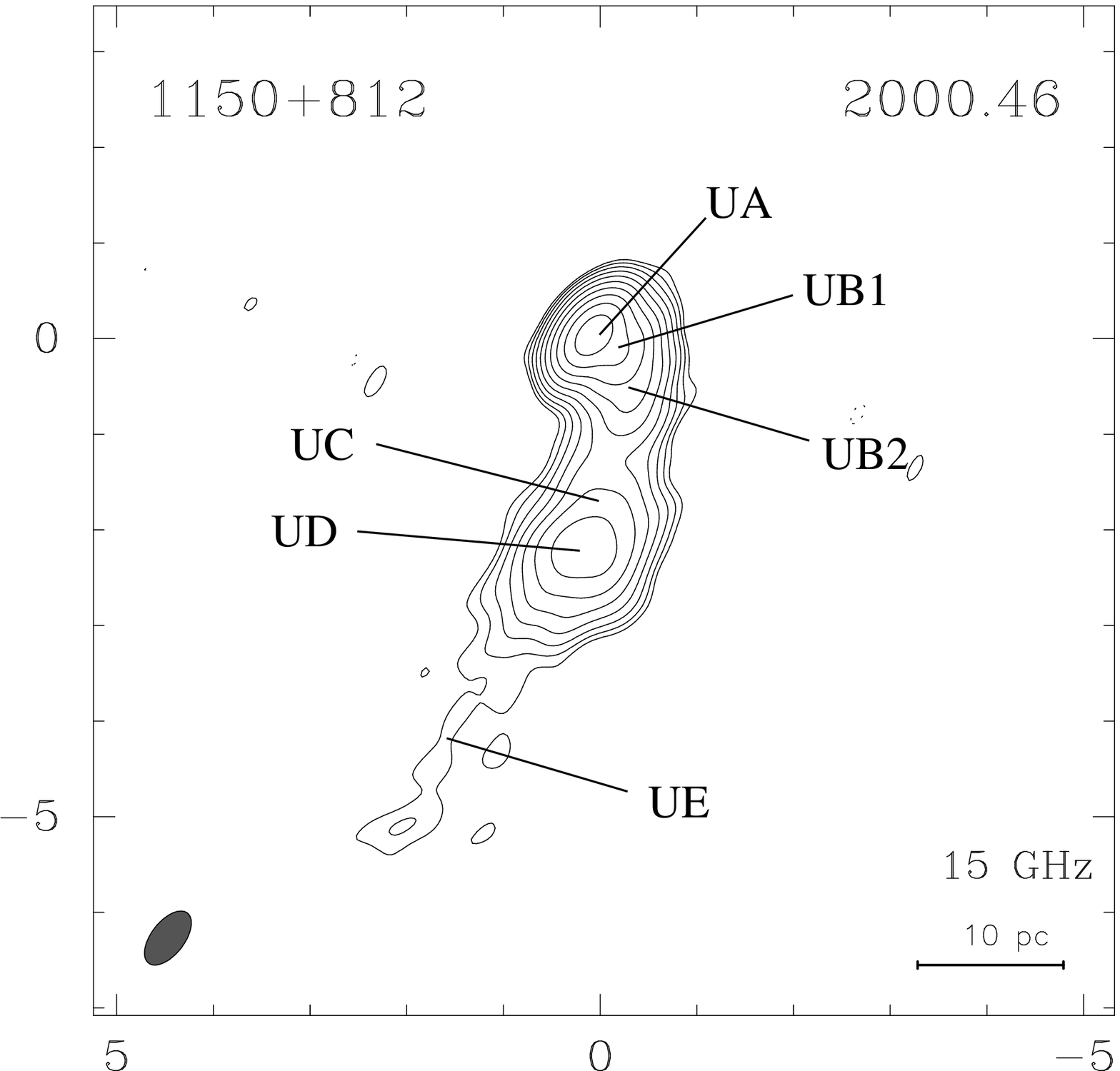}
\caption{VLBA images of \object{QSO\,1150+812}
from observations on
27 July 1999 (1999.57) and  15 June 2000 (2000.46).  
Axes are relative $\alpha$ and $\delta$ in mas.
See Table 1 for contour levels,
synthesized beam sizes (bottom left in the maps), and
peak flux densities.
{See Table~\ref{table:modelfit} for component parametrization.}
\label{fig:map1150}
}
\end{figure}

Our 15.4\,GHz VLBA maps of \object{QSO\,1150+812}
(Fig.\ \ref{fig:map1150}, $z$=1.250) 
show a one-sided core-jet structure directed southwards
up to 4\,mas (27\hminus\,pc) from the core.
There is also a hint for very faint emission extending up to about 5\,mas, 
as our previous 8.4\,GHz VLBA observations clearly confirm (Paper I).
The 15.4\,GHz total flux density of \object{QSO\,1150+812}
slightly decreased from 946\,mJy in 1999.57 down to 907\,mJy in 2000.46 
(4\% change). 
Its monochromatic luminosity is
$L_{\rm 15\,GHz}\approx$(8.0 to 8.3)$\times 10^{38}$\,W, 
almost the same as the luminosity of \object{QSO\,1039+811}, 
due to their very close values in redshift and flux density.
Assuming that the 8.4\,GHz flux density of the source did not change
between 1999.41 and 1999.57, we obtain $\alpha=-0.40$, a moderately 
steep spectrum for the milliarcsecond structure of \object{QSO\,1150+812}.

%

From the maps, there is no clear evidence of significant
morphological changes in either the core or the jet structure.
Nevertheless, our model fit does show such evidence.  
We modeled the source (Table \ref{table:modelfit}) 
with six components at each observing epoch.
The flux density of the brightest component, UA, 
decreased from 405\,mJy in 1999.57
down to 355\,mJy ($\sim$12\,\%) in 2000.46, yielding 
\deltamin$\approx$2.1 to 3.8.
The ratio $Q$ decreased from 1.46 in 1999.57 to 1.39 in
2000.46.
We detected component UB1 at a distance of
$r\approx$0.33 at an angle of 
$\theta$=(235\degr$\pm$5)\degr
at both epochs.
Component UB2 is at a distance 
$r\approx$0.65\,mas to 0.74\,mas
at an angle of 220\degr at both epochs.
We suggest that component XB in Paper I,  at distance $r\approx$0.5\,mas,
is a blend of components UB1 and UB2.
UB2 seems to trace the curving of the jet, changing from an angle of
235\degr (component UB1) to a value of 
170\degr, as displayed by the southern components of the jet.
Component UC seems to correspond to XC in Paper I.
We obtain  $\alpha_{\rm UC/XC}=-1.53$, if UC and XC correspond
to the same physical feature of the jet. 
The increase in flux density for this component is 28\% 
between 1999.57 and 2000.46.
Component UD is at a distance $r = (2.25\pm0.05)$\,mas 
at an angle $\theta$= 175\degr$\pm$5\degr. 
Very similar values of $r$ and $\theta$ are found for XC (Paper I).
Its flux density remained very stable. 


\subsection{\object{BL\,1749+701}\label{subsec:1749}} 

\begin{figure*}[htbp]
\begin{tabular}{@{}cc@{}}
\includegraphics[width=0.40\textwidth]{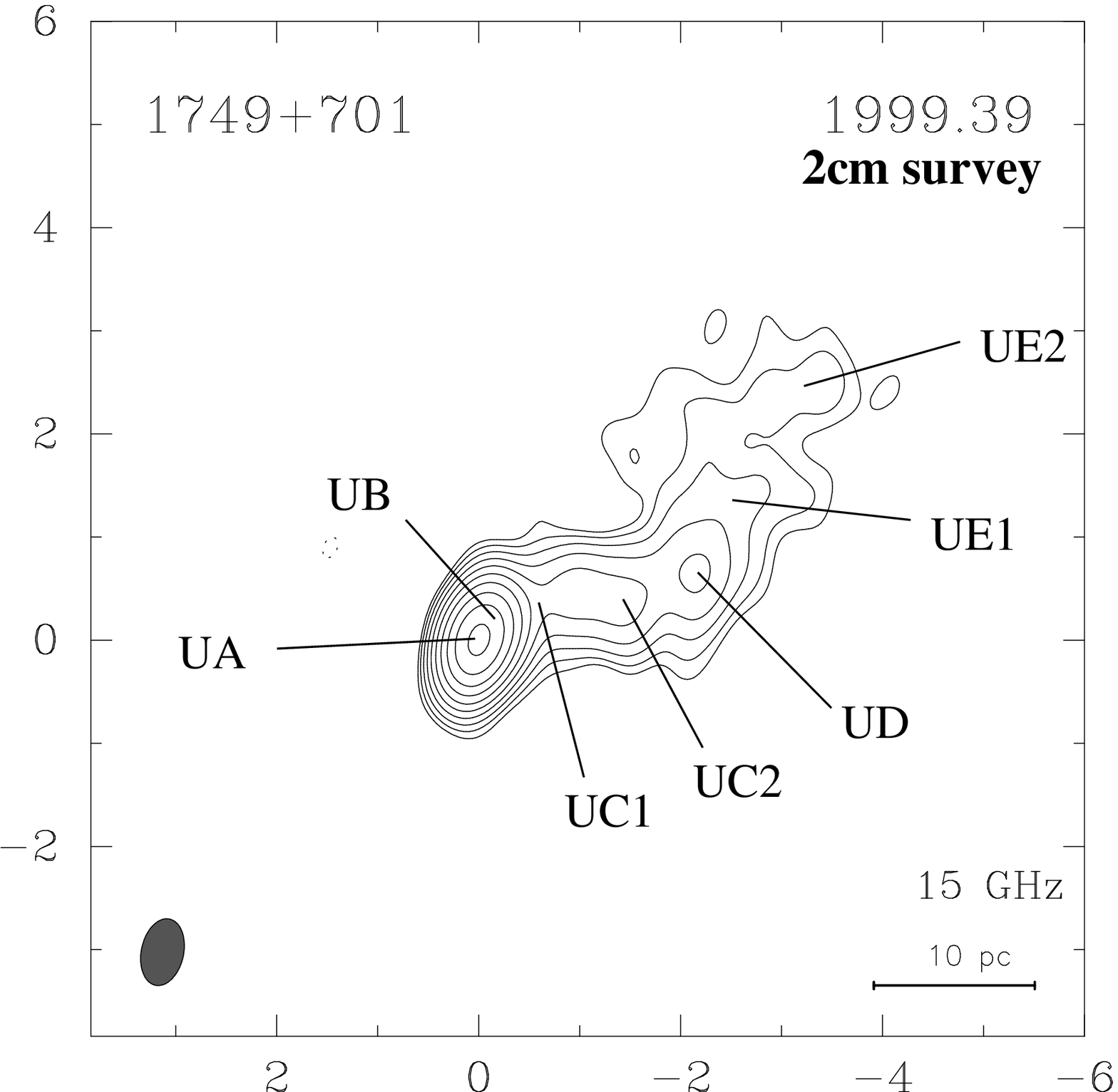} &
\includegraphics[width=0.40\textwidth]{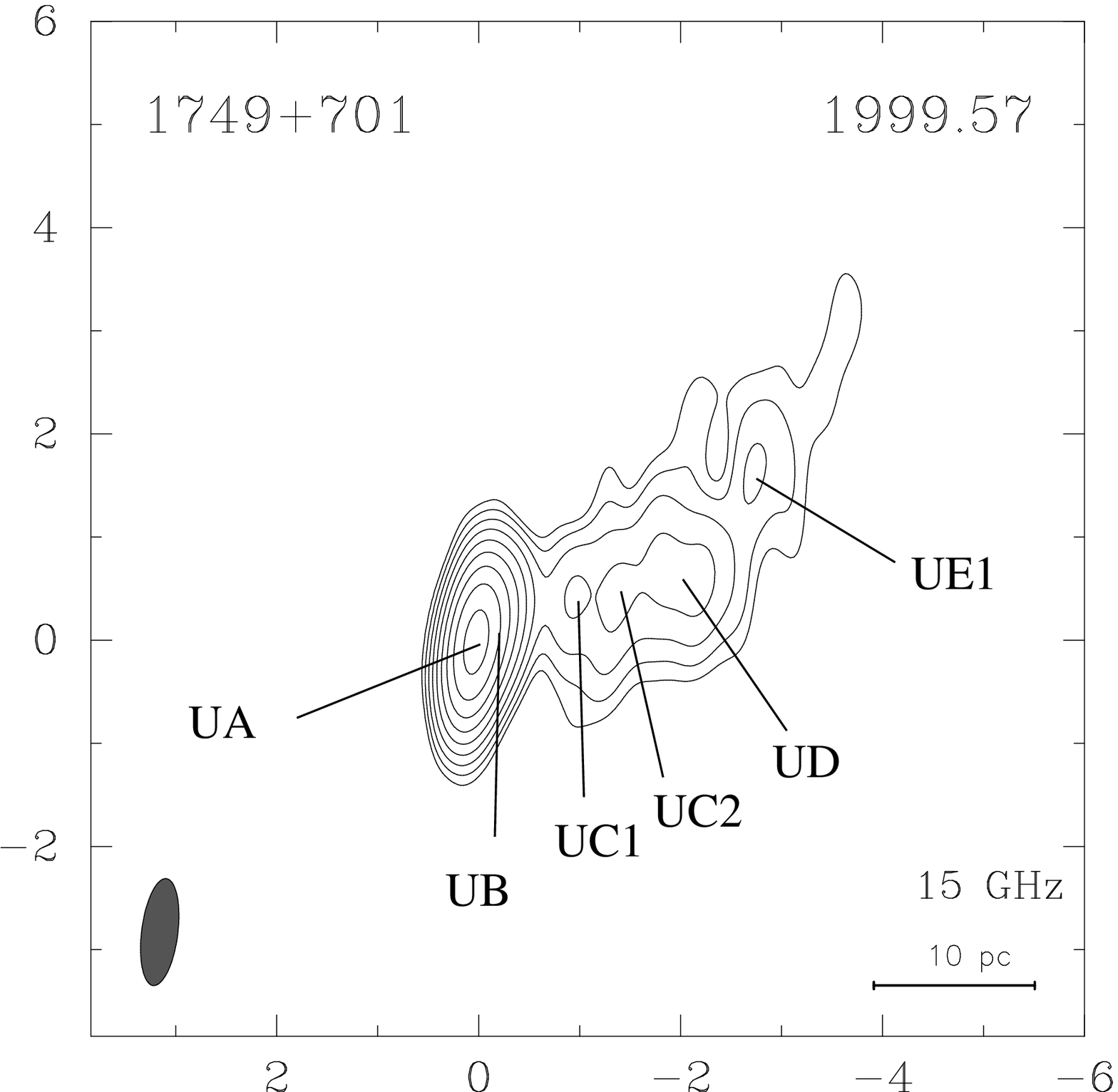} \\
\includegraphics[width=0.40\textwidth]{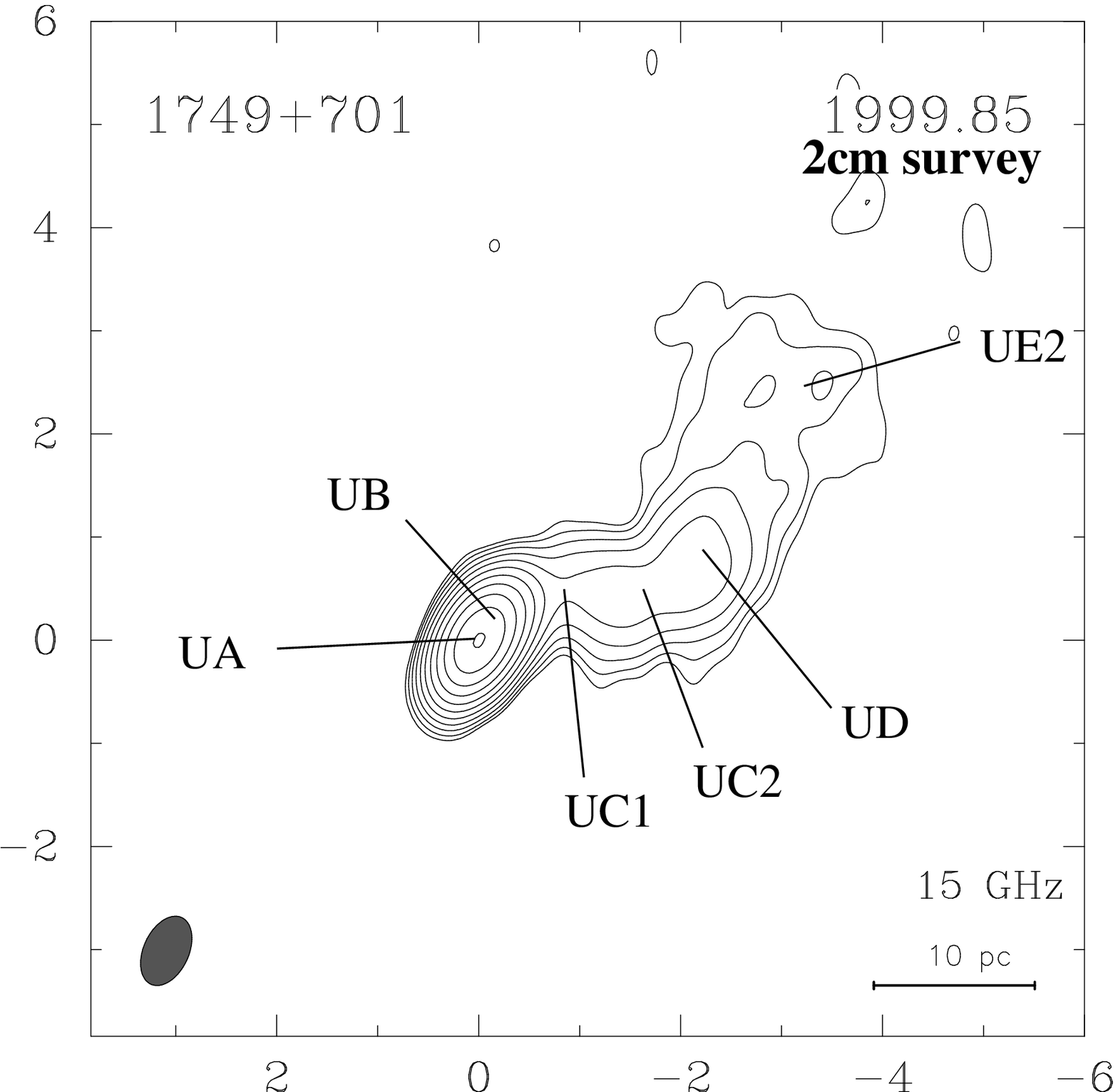} &
\includegraphics[width=0.40\textwidth]{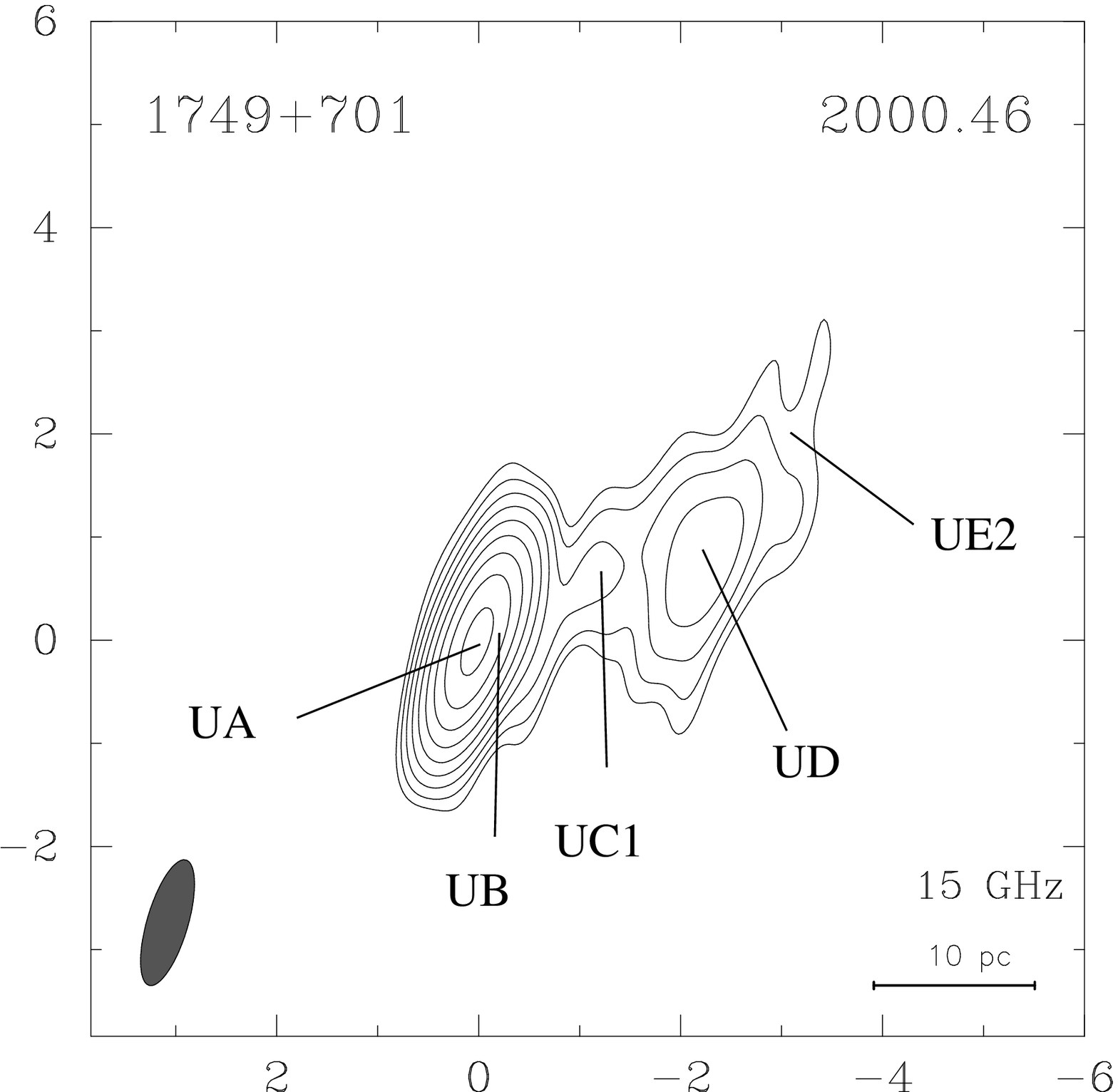} \\
\includegraphics[width=0.40\textwidth]{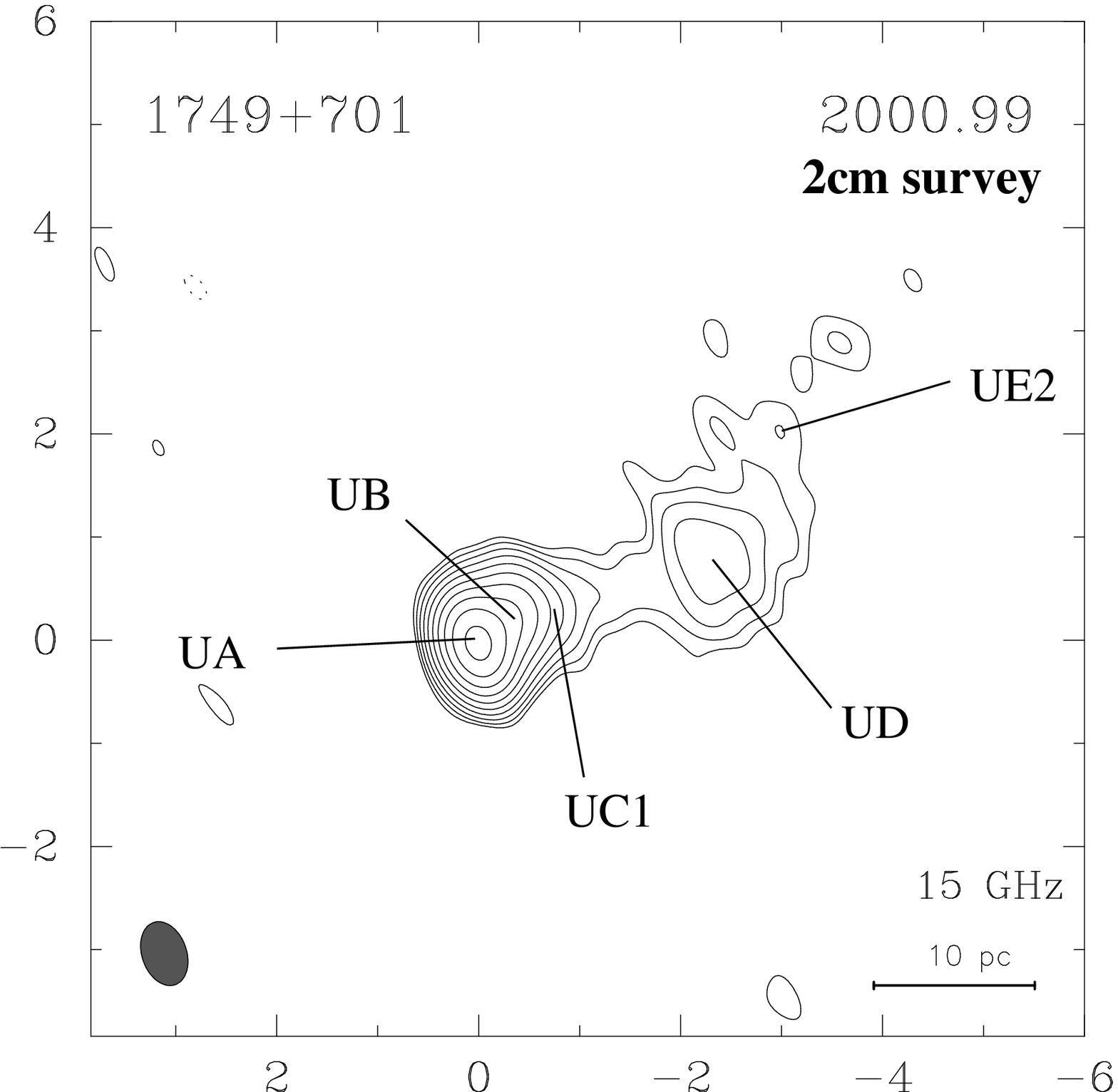} & \\
\end{tabular}
\caption[]{
VLBA images of \object{BL\,1749+701} from observations on
21 May 1999 (1999.39), 27 July 1999 (1999.57), 6 November 1999 (1999.85),
15 June 2000 (2000.46),  and 28 December 2000 (2000.99).
See Tables 1--3 for contour levels, beam sizes (bottom left in the maps),
peak flux densities, and component parametrization.
Axes are relative $\alpha$ and $\delta$ in mas.
\label{fig:map1749}
}
\end{figure*}

The 15.4\,GHz VLBA observations of \object{BL\,1749+701} 
(Fig.\ \ref{fig:map1749}; $z$=0.770) show a 
core-jet structure directed to the west/northwest that extends up to 
4\,mas.  We identify in the images up to seven Gaussian components
(Table~\ref{table:modelfit}).  In all images, the central region (UA+UB) 
contributes 70\% to 80\,\% of the total radio flux.  
This region extends up to a distance of 0.4\,mas (2.5\hminus\,pc), 
and we tentatively identify the brightest component, UA, with the core.
The rest of the components, UC through UE, correspond to the jet, 
which extends up to $r\approx$4\,mas (25\hminus\,pc).
UC, at a distance of ($1.15\pm0.15$)\,mas, 
is likely the base of the jet.
The jet is oriented to the west/northwest at an angle
-69\degr$\pm$9\degr for the inner 2.5\,mas
(16\hminus\,pc; components UC and UD), and then changes its direction northwards
($\theta$=-55\degr$\pm$3\degr) around a distance of 
(3.0$\pm$0.2)\,mas, keeping this orientation until the radio jet fades away.

%

The 15.4\,GHz total flux density of \object{BL\,1749+701}
oscillated around a mean value of 416\,mJy between 1999.39 and
2000.99 (19 months), with variations usually not exceeding 7\%.
The corresponding monochromatic luminosity is 
$L_{\rm 15\,GHz}\approx$(1.2 to 1.3)$\times 10^{38}$\,W.
The ratio of the core (taken as UA+UB) to the extended emission, $Q$,
increased with time as follows: 
2.4 (1999.39), 2.6 (1999.57), 2.9 (1999.85), 3.9 (2000.46), and
4.6 (2000.99), in agreement with the increasing activity of the core region.
Assuming that the 8.4\,GHz total flux density of the source did not change
between 1999.39 and 1999.41 (one week), we obtain $\alpha$=0.0, 
showing a perfect flat spectrum for the source.
The spectral index of component UA/XA is $\alpha_{\rm
UA/XA}$=0.19, an inverted spectrum, which
supports the suggestion that this is the core of the source.

Our previous 8.4\,GHz VLBA observations of \object{BL\,1749+701}
(Paper I) showed the existence of two components, XB and XC, at distances
of (0.45$\pm$0.05)\,mas and (1.2$\pm$0.1)\,mas of the core (at P.A.$\approx$-60\degr), 
respectively.
Component UB, at a distance of (0.3$\pm$0.1)\,mas, likely corresponds
to XB.  Similarly, component UC1, at (1.15$\pm$0.15)\,mas, 
is likely the counterpart of XC. 
In addition, we find another component between XC and XD, which we
labeled UC2, at a distance of (1.6$\pm$0.2)\,mas from
the core.
This component is seen in the first three epochs (1999.39 through
1999.85), but was not detected later.
This behaviour points towards real morphological changes in the inner 
10\hminus\,pc of \object{BL\,1749+701}, most likely
the ejection and passage of components from the core.
Components UD and UE belong to the weak, extended jet.
We identified  UD with component XD (Paper I), based on their similar
coordinates.
Our model fit shows evidence for significant proper motions of
components UB and UD, and of marginal significance for UC2 (see 
Table~\ref{tab:proper_motion}).
The apparent proper motion of component UD
coincides with the reported separation rate 
by Gabuzda \et\ (\cite{gab92}) for their component K2. 
We hope to unambiguously confirm or rule out the existence of this
proper motion from our astrometric analysis. 
Component XE (Paper I) could be a blend of two components, namely
UE1 and UE2.
At the position of UE1 the jet suddenly changes its direction of
propagation, 
towards the north ($r$=(3.0$\pm$0.2)\,mas; $\theta\approx-58\degr$). 
In our 15.4\,GHz observations of 1999.57, this component seems to correspond to XE 
(epoch 1997.93; Paper I).
We fail, however, to fit this component at epochs later than 1999.57. 
Our model fit shows a backwards superluminal motion 
for component UE2 (Table~\ref{tab:proper_motion}).
We are skeptical about this motion, given the strong flux and morphological
variability of the source beyond component UD.
We do not detect any emission above 3$\sigma$ at angular distances 
$\gsim$4.0\,mas from the core, implying a spectral index 
$\alpha_{UF/XF}$=-(2.1 to 3.5).

\subsection{\object{QSO\,1803+784}\label{subsec:1803}}

\begin{figure*}[htbp]
\begin{tabular}{@{}cc@{}}
\includegraphics[width=0.45\textwidth]{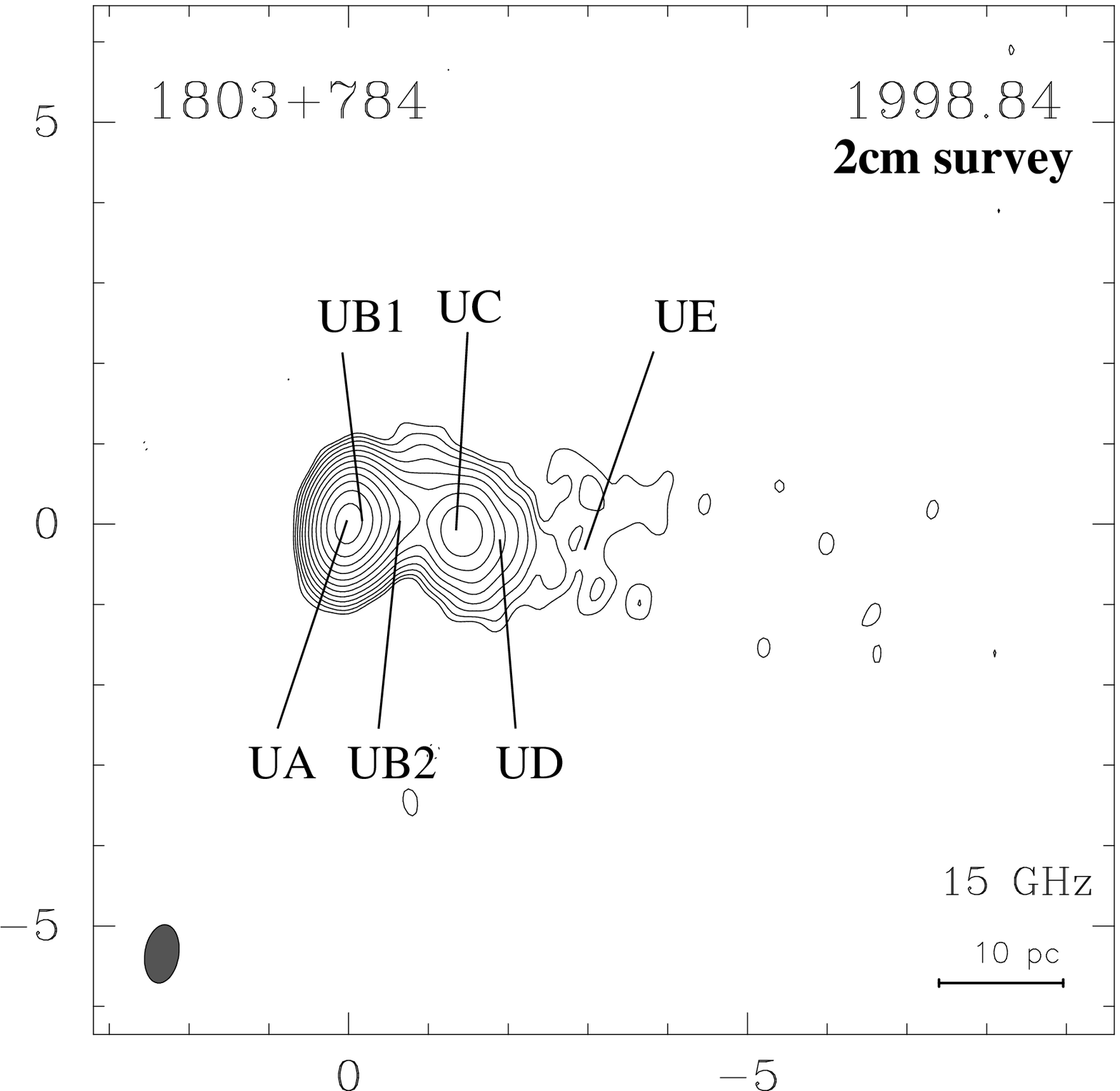} &
\includegraphics[width=0.45\textwidth]{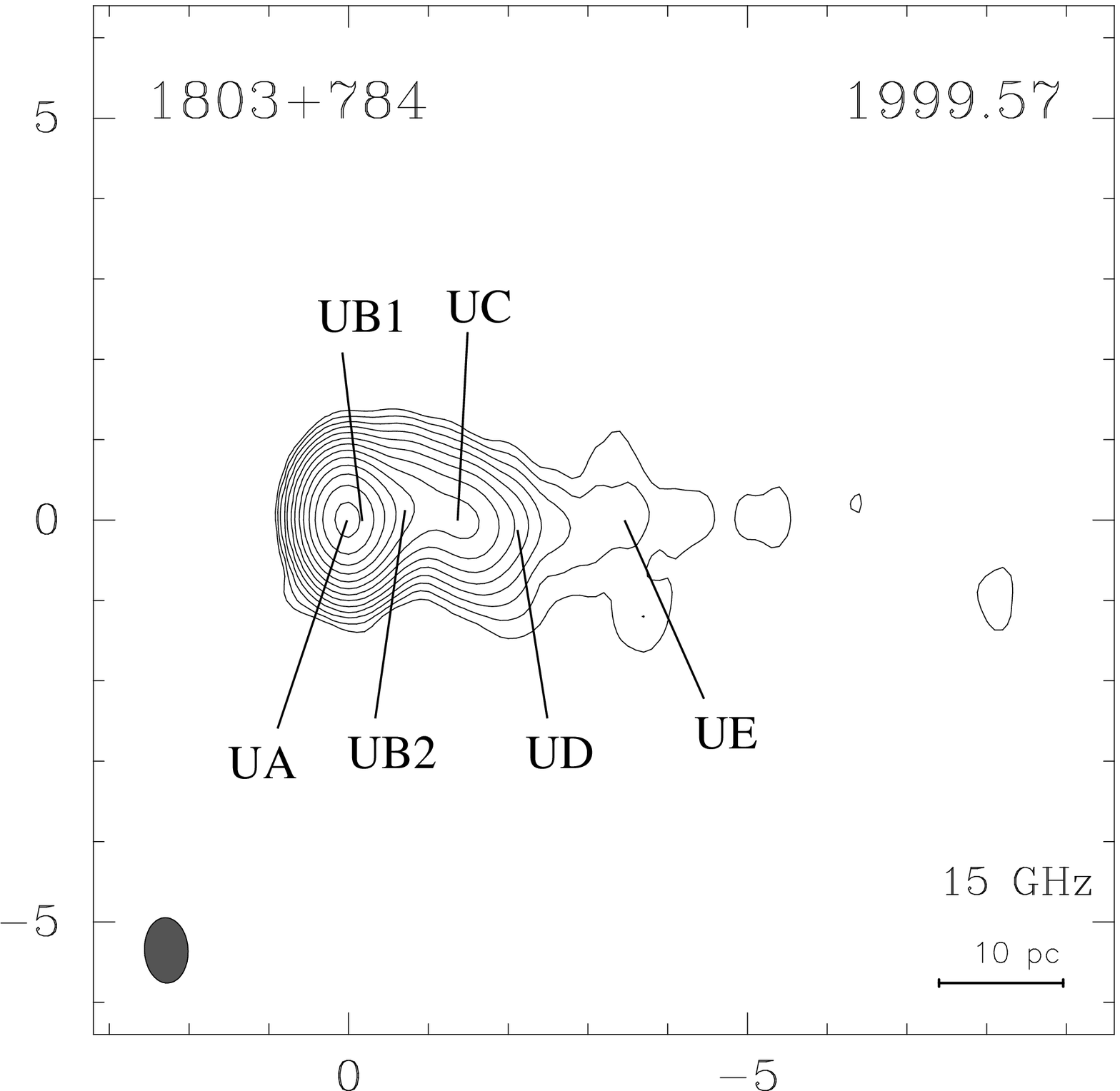} \\
\includegraphics[width=0.45\textwidth]{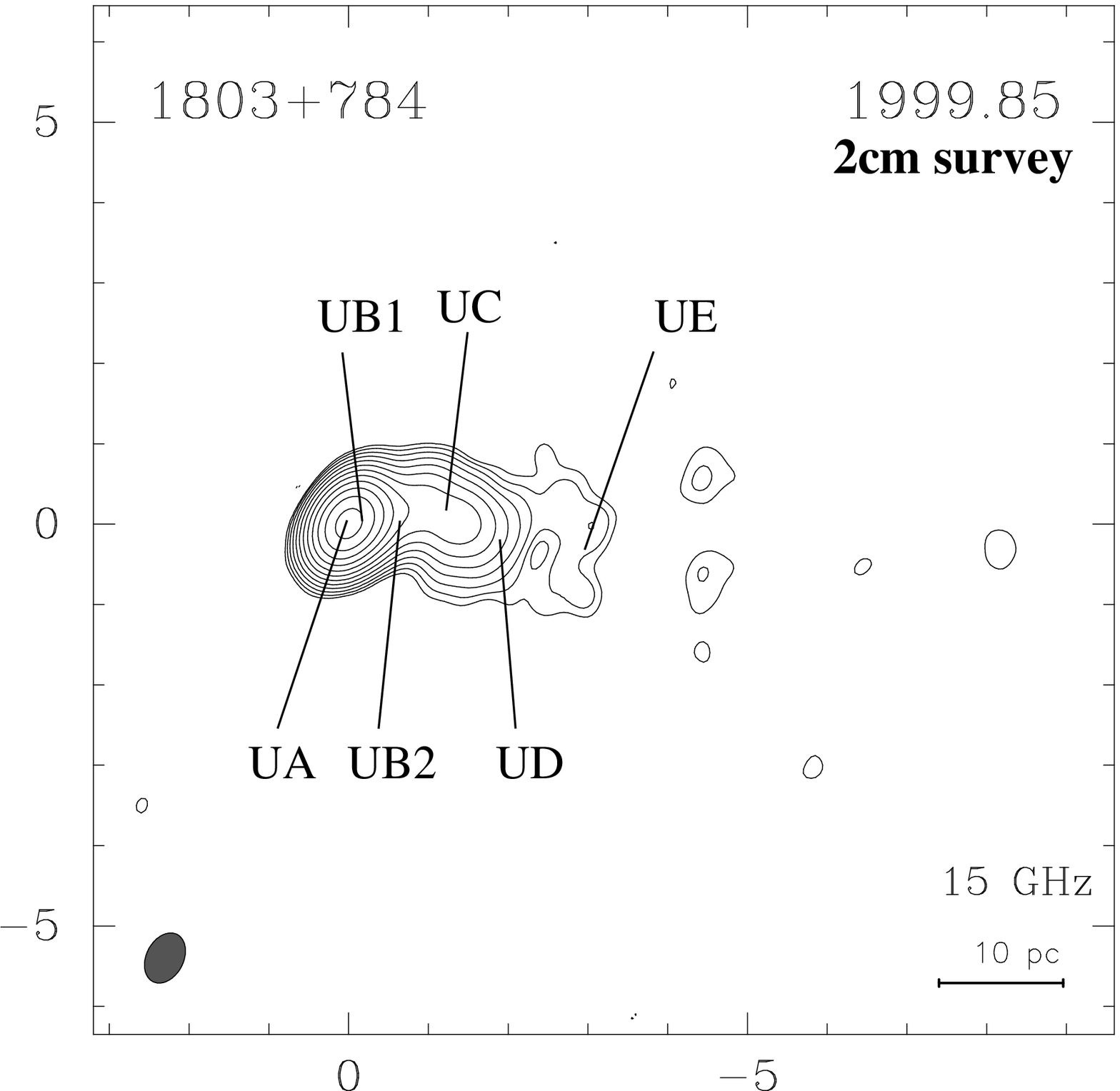} &
\includegraphics[width=0.45\textwidth]{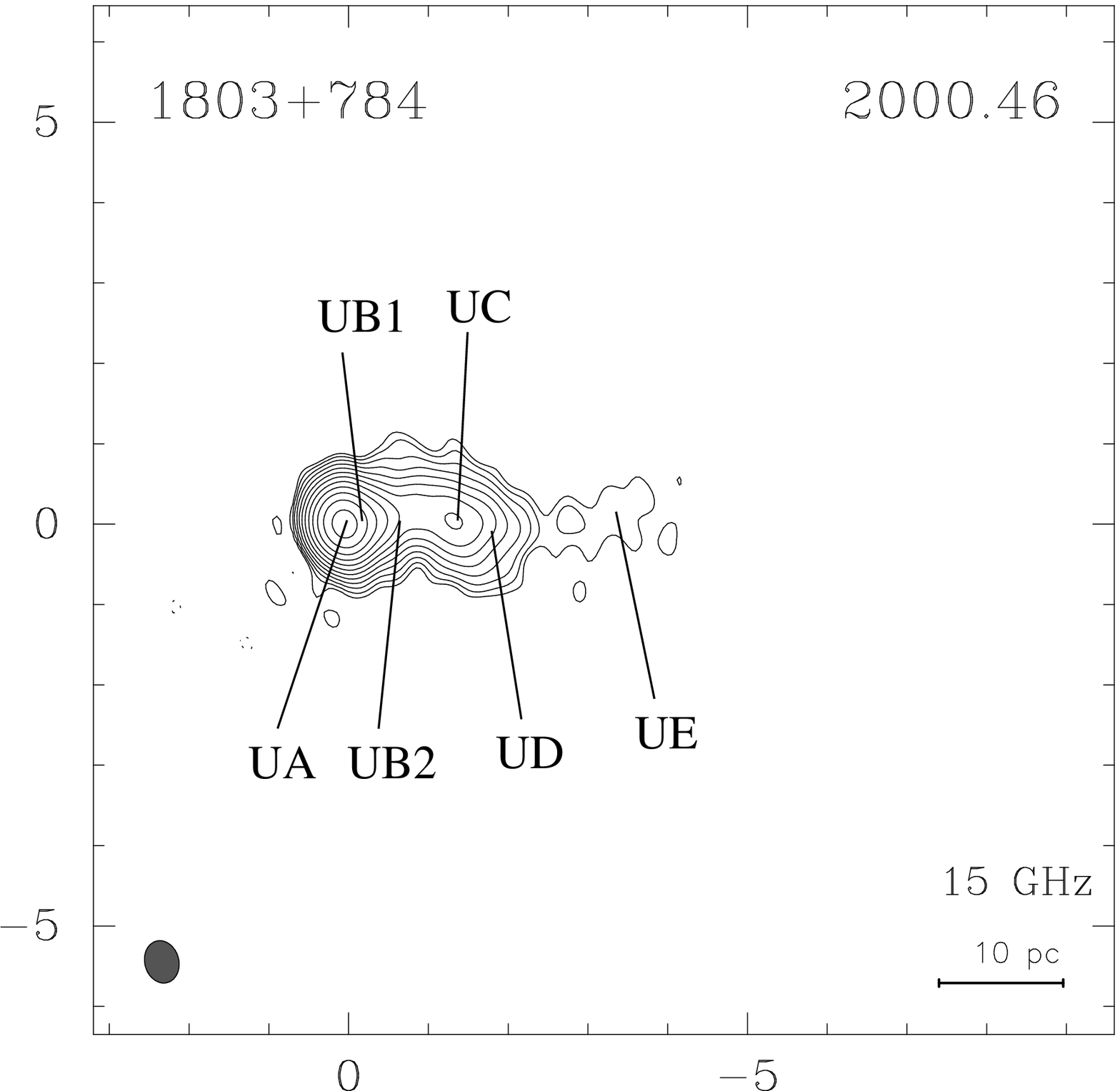} \\
\end{tabular}
\caption[]{
VLBA Images of \object{QSO\,1803+784} from observations on
1 November 1998 (1998.84), 27 July 1999 (1999.57), 
6 November 1999 (1999.85),  and  15 June 2000 (2000.46).
See Tables 1--3 for contour levels, beam sizes (bottom left in the maps),
peak flux densities, and component parametrization.
Axes are relative $\alpha$ and $\delta$ in mas.
\label{fig:map1803}
}
\end{figure*}

The 15.4\,GHz VLBA maps of \object{QSO\,1803+784} 
(Fig.\ \ref{fig:map1803}; $z$=0.680, V\'eron-Cetty \& V\'eron 
[\cite{veron03}]) 
reveal the same basic emitting structure seen
at 8.4\,GHz (Paper I); a one-sided core-jet 
oriented westward (P.A.$\approx-90\degr$). 
(Note that 1803+784 has been previously classified as a BL Lac object, but
is now considered to be a QSO [V\'eron-Cetty \& V\'eron \cite{veron03}]). 
There are, however, significantly different features that appear
in our model fit (Table \ref{table:modelfit}), compared
to our 8.4\,GHz observations.
We characterize reasonably well the source with six
components, instead of the eight needed at 8.4\,GHz, 
because of significant absorption at 15.4\,GHz of the jet emission 
at $r\gsim$4\,mas (we barely detect emission above 
3$\sigma$, which cannot be adequately model fitted).
From our model fits, it follows that the inner 0.8\,mas
(4.9\hminus\,pc) structure needs at least three components (UA, UB1, and UB2)
to describe this region adequately describe this region,  
in contrast with the two components found at 8.4\,GHz
(XA and XB). 
We identify the brightest component, UA (likely the core), with
component XA. 
Components UB1 
and UB2 
indicate the transition of the core to the jet region, and we  
suggest that component XB could be a blend of the two components
seen at 15.4\,GHz. 
The source structure at $r\gsim$1\,mas is well characterized at 15.4\,GHz
by three components,
UC, UD, and UE. 
Whilst the positions of UC and UD agree well with those of XC and XD, respectively
(Paper I), we did not find clear counterparts for component XE,
--at least for the first three epochs-- 
nor for XF, XG, and XH.

%

The 15.4\,GHz total flux density of \object{QSO\,1803+784}
changed significantly from epoch to epoch
(up to 26\,\% between 1988.84 and 1999.57). 
Its monochromatic luminosity is
$L_{\rm 15\,GHz}\approx$(5.3 to 6.6)$\times 10^{38}$\,W.
The innermost regions of \object{QSO\,1803+784} ($r\lsim$1\,mas,
equivalent to a linear distance of 6.4\hminus\,pc)
encompass components UA, UB1, and UB2.
The contribution of this innermost region to the total flux density
amounts to 80\% to 94\%, depending on the epoch.
(If we include component UC/XC, then the above figures become
93\% to 98\%, in agreement with the percentages reported in Paper I.)
The total flux density variability of  \object{QSO\,1803+784}  closely follows that 
of the brightest component, UA, and is also partially modulated by fluctuations
in the flux density of UB1 and UB2.
The ratio of the core (taken as UA+UB1) to the extended emission, $Q$, 
confirms that the activity of \object{QSO\,1803+784} is core-dominated:
3.3 (1998.84), 5.2 (1999.57), 4.1 (1999.85), and 3.3 (2000.46). 

Assuming that the 15.4\,GHz total VLBI flux density of the source did not
vary noticeably between 1999.41 and 1999.57, we obtain
$\alpha$=0.41, a rather
inverted spectrum for the milliarcsecond structure of \object{QSO\,1803+784}.
Similarly, we obtain the following indices for each component:
$\alpha_{\rm UA/XA}$=0.39,
$\alpha_{\rm (UB1+UB2)/XB}$=1.0, 
$\alpha_{\rm UC/XC}=-0.04$,
$\alpha_{\rm UD)/XD}\approx-0.70$, and
$\alpha_{\rm UE/XE}\approx$0.52.
The spectral indices of UA and UB correspond with those expected for
the innermost regions of the core, and those of UC and UD with what is
expected in the inner-to-outer jet region. 
The value of $\alpha_{\rm UE/XE}$ is not expected for an optically
thin component. 
While the non-simultaneity of the 8.4\,GHz and 15.4\,GHz could have some
effect, it could also be that our component UE is indeed the result of
the blending of XE and XF, in which case we obtain
$\alpha_{\rm UE/(XE+XF)}\approx-0.63$, a more typical value for an
extended jet component.
Our non-detection of significant emission around the positions of
components XG and XH allows us to set upper limits for the
spectral indices of these components (5$\sigma$):
$\alpha_{\rm UG)/XG}\lsim-2.6$, and
$\alpha_{\rm UH/XH}\lsim-2.4$.

The study of this source by VLBI is likely to give us
relevant information on the kinematics and morphological changes
of their innermost regions. 
The dramatic flux density variations 
displayed by the innermost regions of  \object{QSO\,1803+784} 
along with the identification 
of a complex structure close to the core point to a possible
ejection of components from the central engine.
Another remarkable aspect of \object{QSO\,1803+784} is the
stationarity of a component at (1.3$\pm$0.1)\,mas.
Schalinski (\cite{sch90}) suggested the existence 
of such a (stationary) component at 1.2\,mas 
from the core.
Krichbaum \et\ (\cite{kri93}, \cite{kri94}) used 
43\,GHz VLBI observations to show the existence of 
traveling components between the core and this component,
setting it at a distance of 1.4\,mas from the core.
Both our 15.4\,GHz and 8.4\,GHz observations show this component
(UC/XC) to be at (1.45$\pm$0.05)\,mas, and no sign of any component
at 1.2\,mas existed between 1998.84 and 2000.46. 
We therefore suggest that there does indeed exist a
stationary component at $\approx$1.4\,mas (9\hminus\,pc) from the core.

\subsection{\object{QSO\,1928+738} (\object{4C\,73.18})\label{subsec:1928}}

\begin{figure}[htbp]
\vspace*{222pt}
\includegraphics{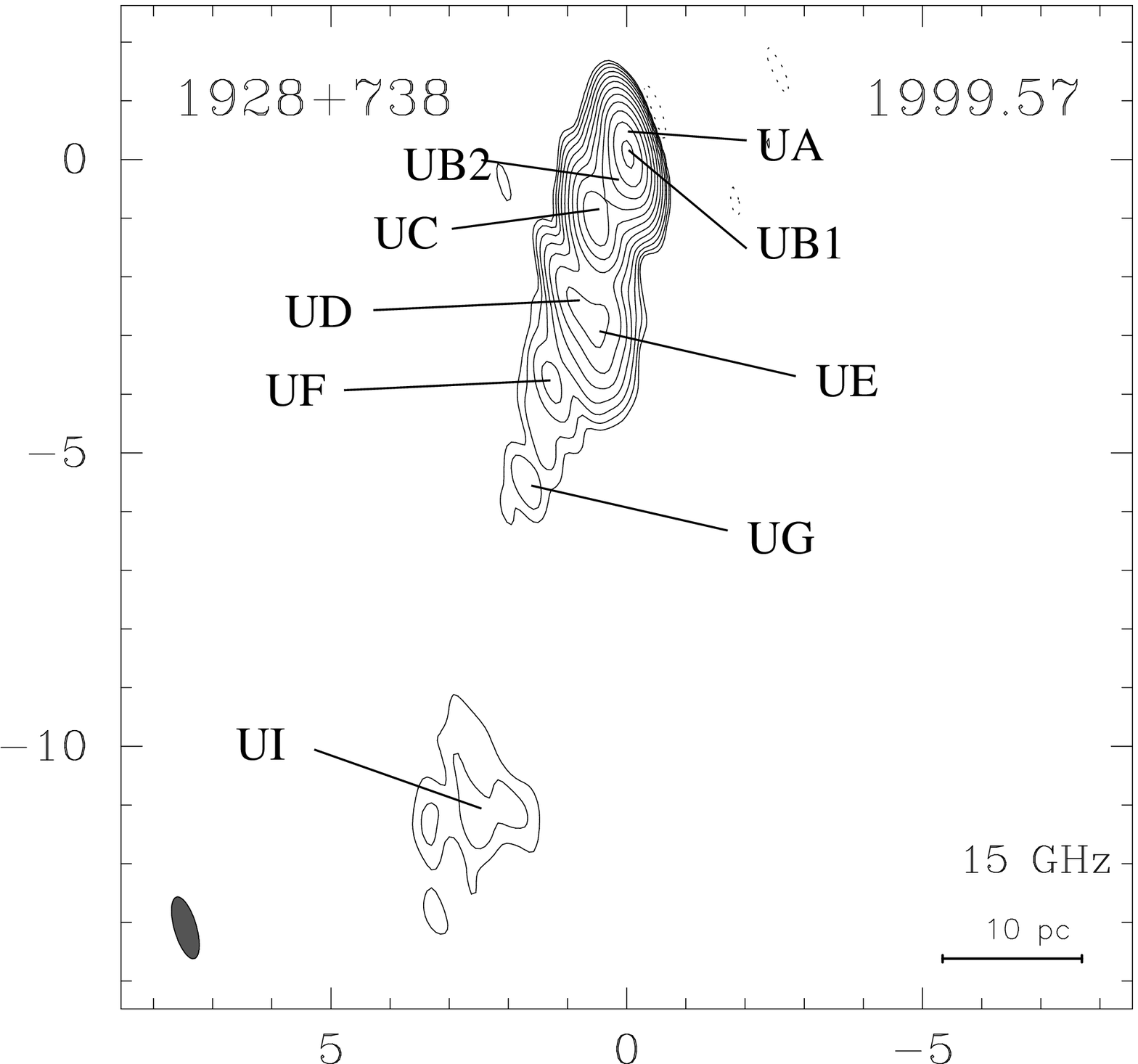}
\vspace*{222pt}
\includegraphics{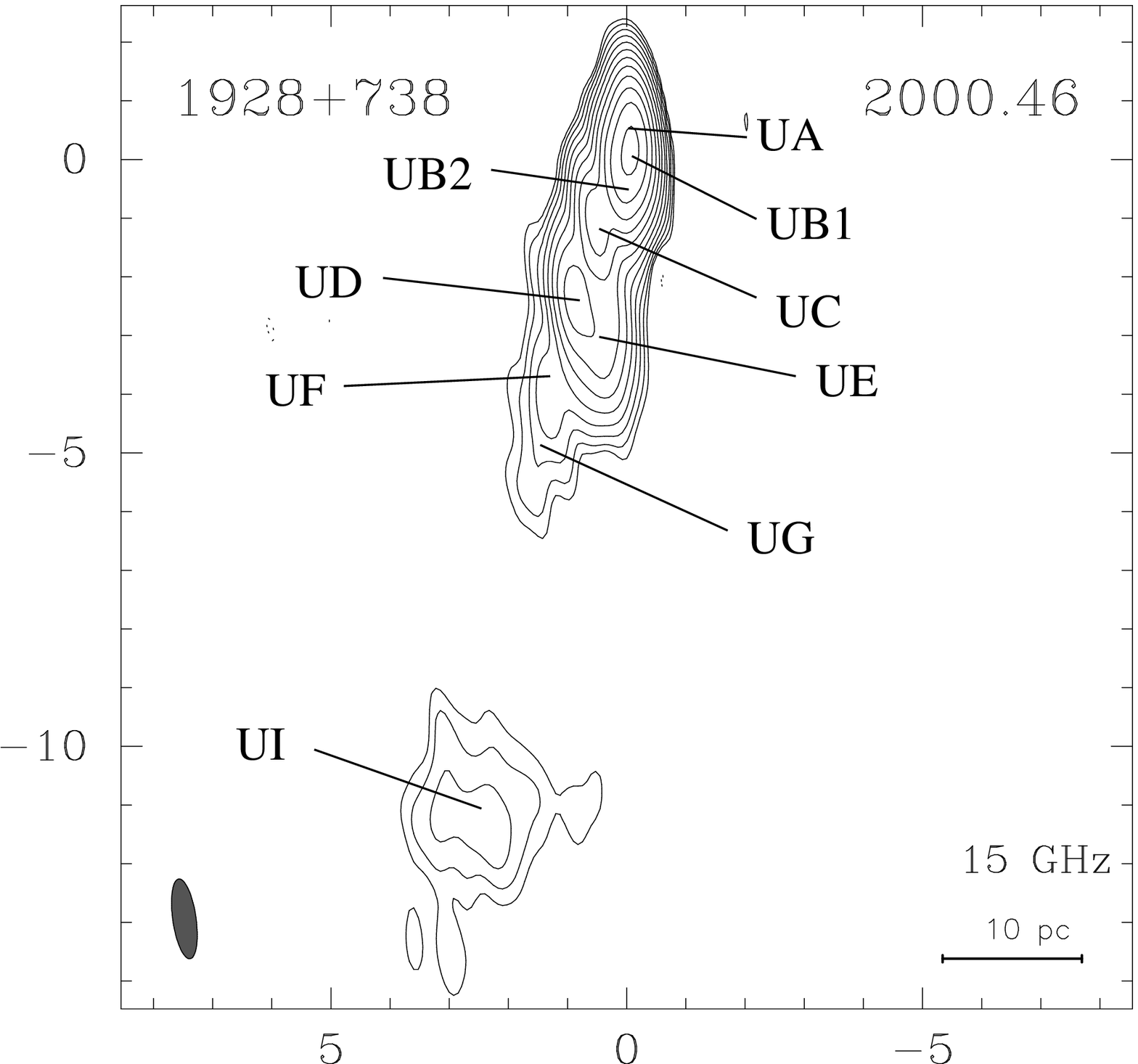}
\caption{VLBA images of \object{QSO\,1928+738} (\object{4C\,73.18})
from observations on
27 July 1999 (1999.57) and  15 June 2000 (2000.46).  
Axes are relative $\alpha$ and $\delta$ in mas.
See Table 1 for contour levels,
synthesized beam sizes (bottom left in the maps), 
and peak flux densities.
{See Table~\ref{table:modelfit} for component parametrization.}
\label{fig:map1928}
}
\end{figure}

\object{QSO\,1928+738} (Fig.~\ref{fig:map1928}, $z$=0.3021)
is the most extensively studied radio source of the complete S5 polar cap
sample, and the one that shows the richest structure.
Our 15.4\,GHz VLBA maps display a core-jet structure, 
with the jet southward-oriented (at an angle
of 160\,\degr), and extending up to 11.5\,mas
(49\hminus\,pc).
The 15.4\,GHz total flux density of \object{QSO\,1928+738}
decreased from 3006\,mJy down to 2889\,mJy (4\% change). 
Assuming that the 8.4\,GHz flux density of the source did not 
vary significantly 
between 1999.41 and 1999.57, we obtain $\alpha$=0.02, a remarkably
flat spectrum for the milliarcsecond structure of
\object{QSO\,1928+738}.
We fitted reasonably well the brightness radio structure 
of \object{QSO\,1928+738} with nine components 
(Table \ref{table:modelfit}). 
We model the innermost region ($r\lsim$1\,mas, 
corresponding to a linear distance of $4.3\,h^{-1}$\,pc),
with three components: UA, UB1, and UB2.
Component UC ($r$=1.2\,mas at P.A.=152\degr)
indicates the transition from the innermost to the outermost 
structure of the jet.

%

UA, the brightest component of \object{QSO\,1928+738} 
at both epochs, is not at the origin
of coordinates, but to the north at 
$r$=(0.35$\pm$0.05)\,mas  
from the origin, at an
angle of -18\degr$\pm$2\degr.
Component UB1 is at the origin, and component UB2 is at 
$r$=(0.35$\pm$0.05)\,mas at an angle of 145\degr$\pm$5\degr.
The total intensity emission is dominated by this innermost region.
Indeed, the combined emission from components UA, UB1, and UB2 
amounts to 71\% (70\%)
of the total at the first (second) epoch.
The inclusion of the closest jet components up to a distance
of 3\,mas increases those figures to
96\% and 94\% on 1999.57 and 2000.46, respectively.
The ratio of the emission from the core region (taken here as UA+UB1)
to the extended emission was  $Q$=1.1 in 1999.57, and $Q$=0.9
in 2000.46. Although the compactness of the source is thus moderate, 
the overall source activity seems to be driven by the core, as 
the total flux density closely follows the core behaviour.

The spectral index of the innermost component, UA, is
$\alpha_{\rm UA/XA}$=0.04, as one would expect if UA were the core.
Note that though 
the total intensity of \object{QSO\,1928+738} only varied 
by 4\% between 1999.57 and 2000.46, 
the flux density of each component varied much more strongly.
The decrease in flux density of UA, together with its 
change in position (a shift of 0.13\,mas northward),
could indicate that a component was ejected around or before
1999.57. 
Assuming this is the case, our model fit for UA at the second
epoch suggests that the core is at least 0.4\,mas
to the north of component UB1, in agreement
with previous estimates (e.g., Guirado \et\ \cite{gui98},
Ros \et\ \cite{ros99}).

Component XC (Paper I) is likely a blend of components
UB2 and UC. 
As mentioned earlier, component UC is a signpost in the
inner-to-intermediate structure of the jet. From 
this position onwards, the jet displays
 a noticeable change in its direction, 
from 150\degr at $r$=1.25$\pm$0.05\,mas (UC) 
to 160\degr at $r=$2.55$\pm$0.05\,mas (UD). 
Component UE at $r$=3.05$\pm$0.05\,mas at $\theta$=173\degr, shows
a sudden change in the main orientation of the jet, 
which at the distance of component UF 
($r$=3.95$\pm$0.05\,mas) recovers its initial orientation 
($\theta$=162\degr$\pm$2\degr).
Component UG ($r=$5.45$\pm$0.15\,mas; $\theta$=166\degr$\pm$2\degr) 
signals
a first termination of the jet at a linear distance of
25\hminus\,pc.
The 15.4\,GHz jet reappears at almost twice that distance
($r$=11.55$\pm$0.05\,mas, corresponding to $50\,h^{-1}$\,pc, at
an angle of 167\degr). 
We identify this component, UI, with component XI (Paper I),
which is found at essentially the same distance and angle, 
and suggests it is a stationary component of the jet.
The extended jet structure of \object{QSO\,1928+738} 
at 8.4\,GHz (Paper I) is richer in components than 
the one seen at 15.4\,GHz. In particular, we fail to detect any 
significant emission beyond component UI ($r$=11.6\,mas), 
while the 8.4\,GHz VLBA jet extends at least out to 
$(25\pm1)$\,mas.

The activity of \object{QSO\,1928+738} is so intense
that the task of identifying components from
VLBI observations taken at different epochs and frequencies
becomes challenging.
For example, the astrometric studies carried out
by Guirado \et\ (\cite{gui98}) and 
Ros \et\ (\cite{ros99}) indicated that 
the core position of \object{QSO\,1928+738} is 
to the north of the reference point chosen in earlier
works.
Ros \et\ (\cite{ros99}) estimated the offset between 
the true core position and their 8.4\,GHz observations 
to be about 1.5\,mas to 2\,mas. 
Our previous 8.4\,GHz observations (Paper I) showed component
XA to be at $r = 0.65\pm0.15$\,mas northward. 
The shift we find for its counterpart at 15.4\,GHz, UA, 
is, however, significantly  smaller ($r=0.35\pm0.05$\,mas).
If, as Guirado \et\ and Ros \et\ suggested, the ``true'' core
is very much self-absorbed, we should expect to see
a 15.4\,GHz VLBI component northward of the 8.4\,GHz 
component detected in Paper I (i.e., UA northward of XA). 
We fail, however, to detect such a component. 
A likely explanation is the potential confusion in the identification of 
components at 8.4\,GHz and 15.4\,GHz,
despite our relatively close campaigns in 1999.

Another remarkable point is the proper motion of components 
in the jet of \object{QSO\,1928+738}. 
Ros \et\ (\cite{ros99}) reported a mean proper motion of 
(0.32$\pm$0.10)\,mas/yr for components in the inner 2.5\,mas
of the jet.  Our model fit for the innermost regions of 
\object{QSO\,1928+738} does not, at this stage, show evidence for 
such large motions.
Murphy \et\ (\cite{murphy03}) have recently reported a wide range 
of proper motions in \object{QSO\,1928+738}, from nearly stationary
(0.02\,mas/yr or 0.5\,$c$) to relatively fast (0.82\,mas/yr or 19\,$c$),
based on VSOP observations.  
Unfortunately, there is no reference in their work
to make it possible to identify which regions show superluminal motions.
The authors mention that their model fitting shows that a relativistic, 
variable speed-ejecta ballistic model is preferred over a relativistic
helical-jet model. 
If this is the case, then it is no surprise that component identification
is so difficult for the source, especially if observations are 
widely spaced in time.
Our quasi-simultaneous VLBA observations at 15.4\,GHz and 43\,GHz 
will shed light on the above two controversial issues 
(component identification and proper motion), 
and might be very useful in setting stringent limits on the putative
core position. 

\subsection{\object{BL\,2007+777}\label{subsec:2007}}

%
\begin{figure*}[htbp]
\begin{tabular}{@{}cc@{}}
\includegraphics[width=0.45\textwidth]{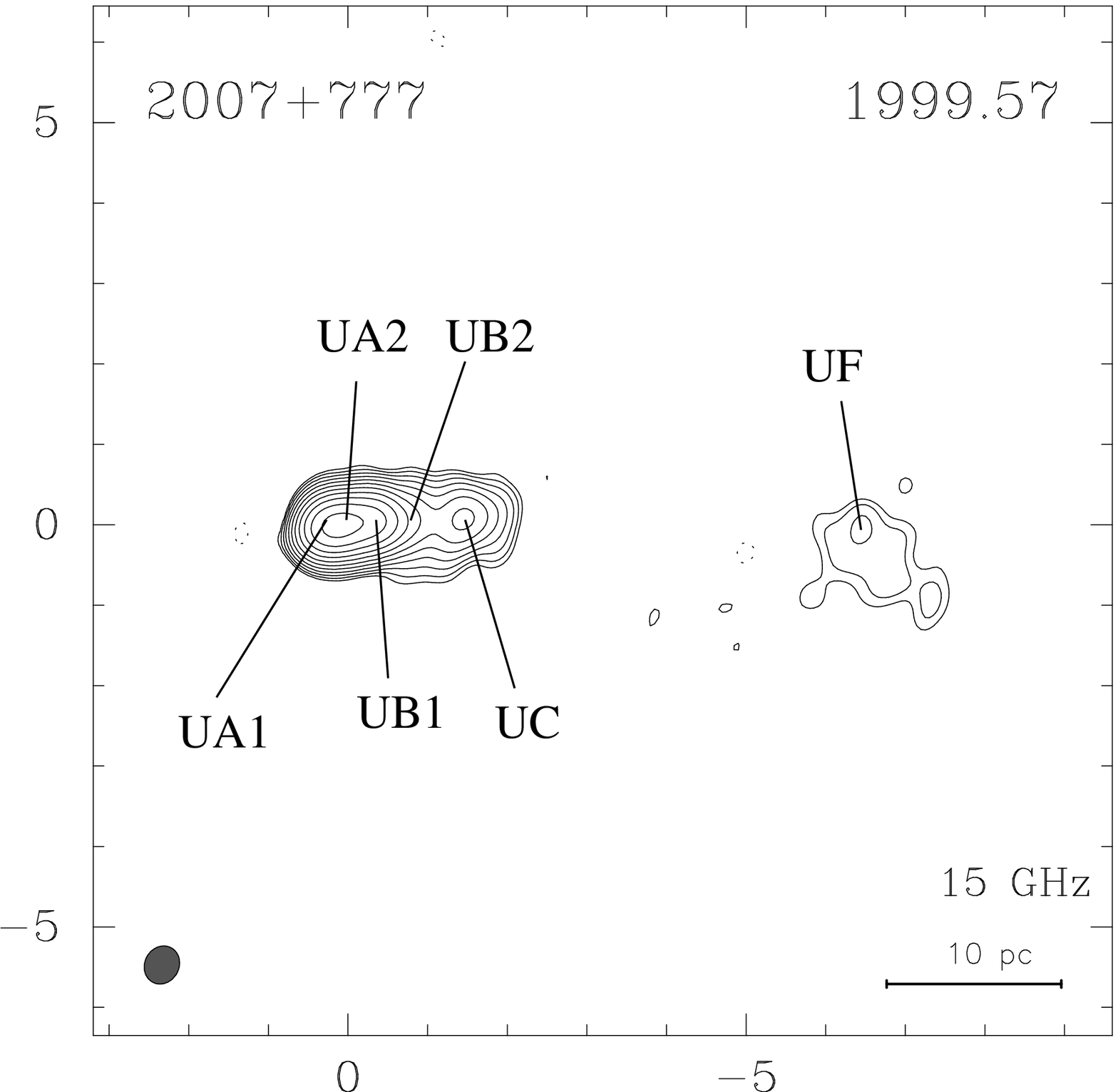} &
\includegraphics[width=0.45\textwidth]{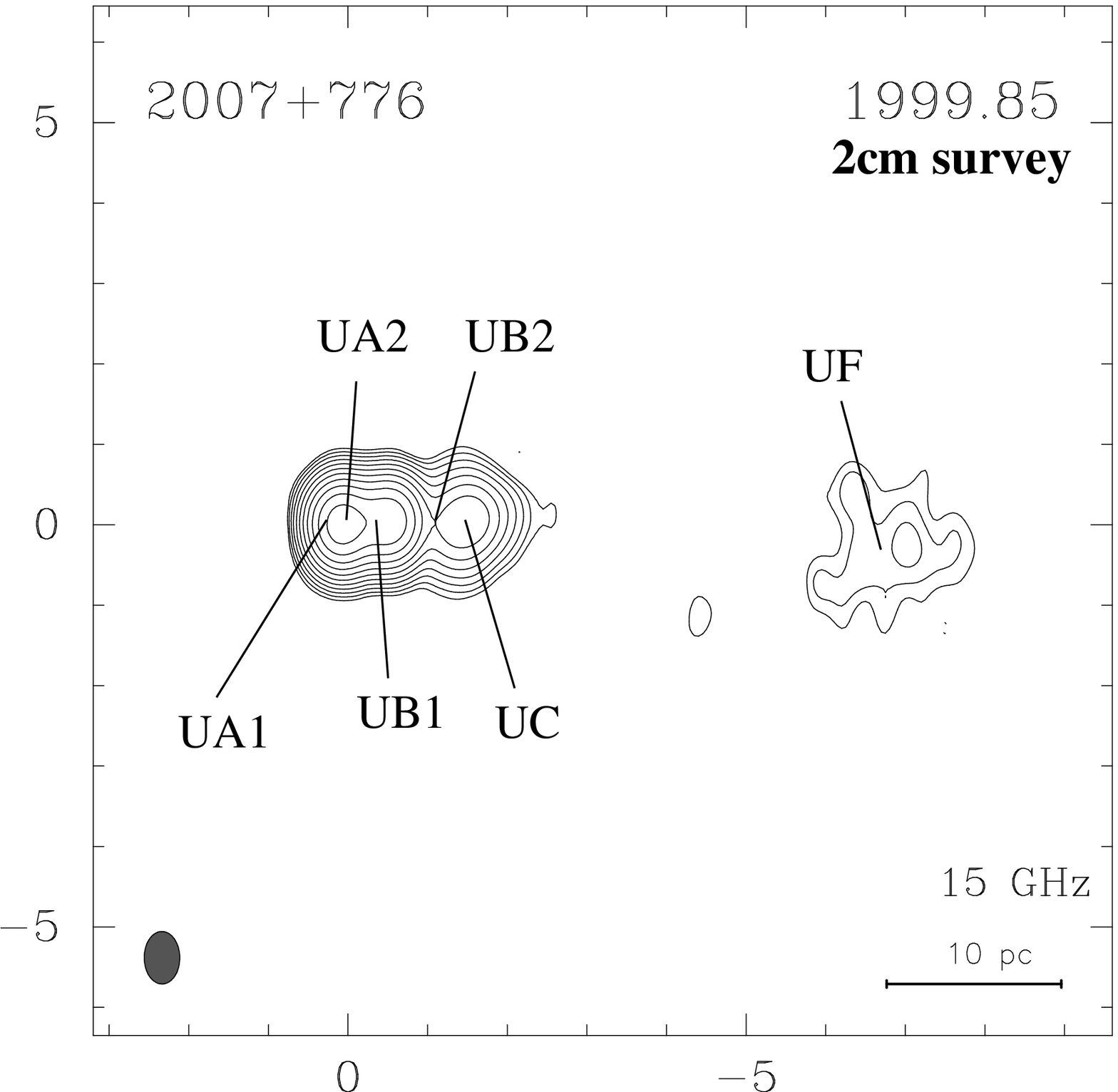} \\
\includegraphics[width=0.45\textwidth]{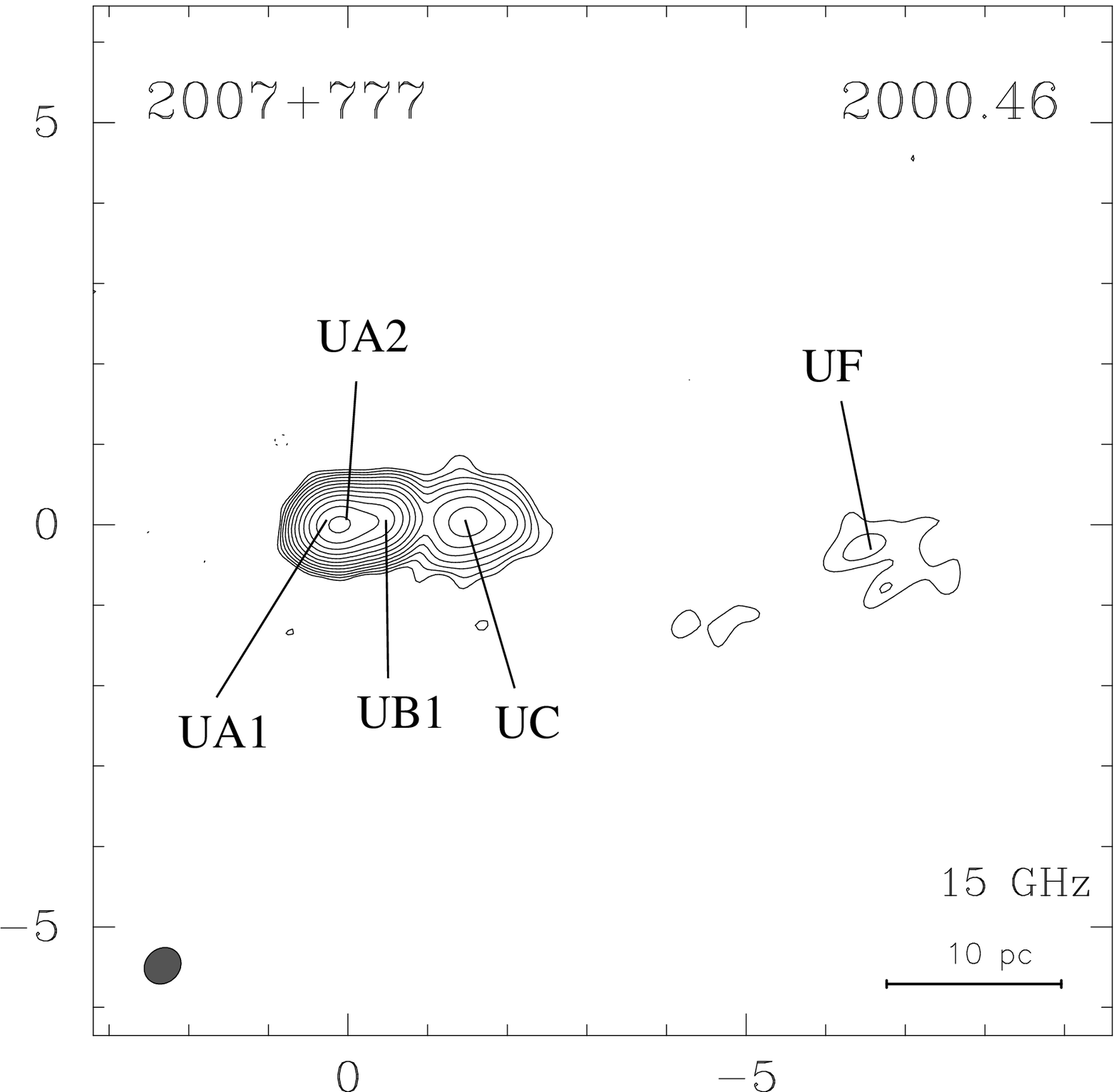} &
\includegraphics[width=0.45\textwidth]{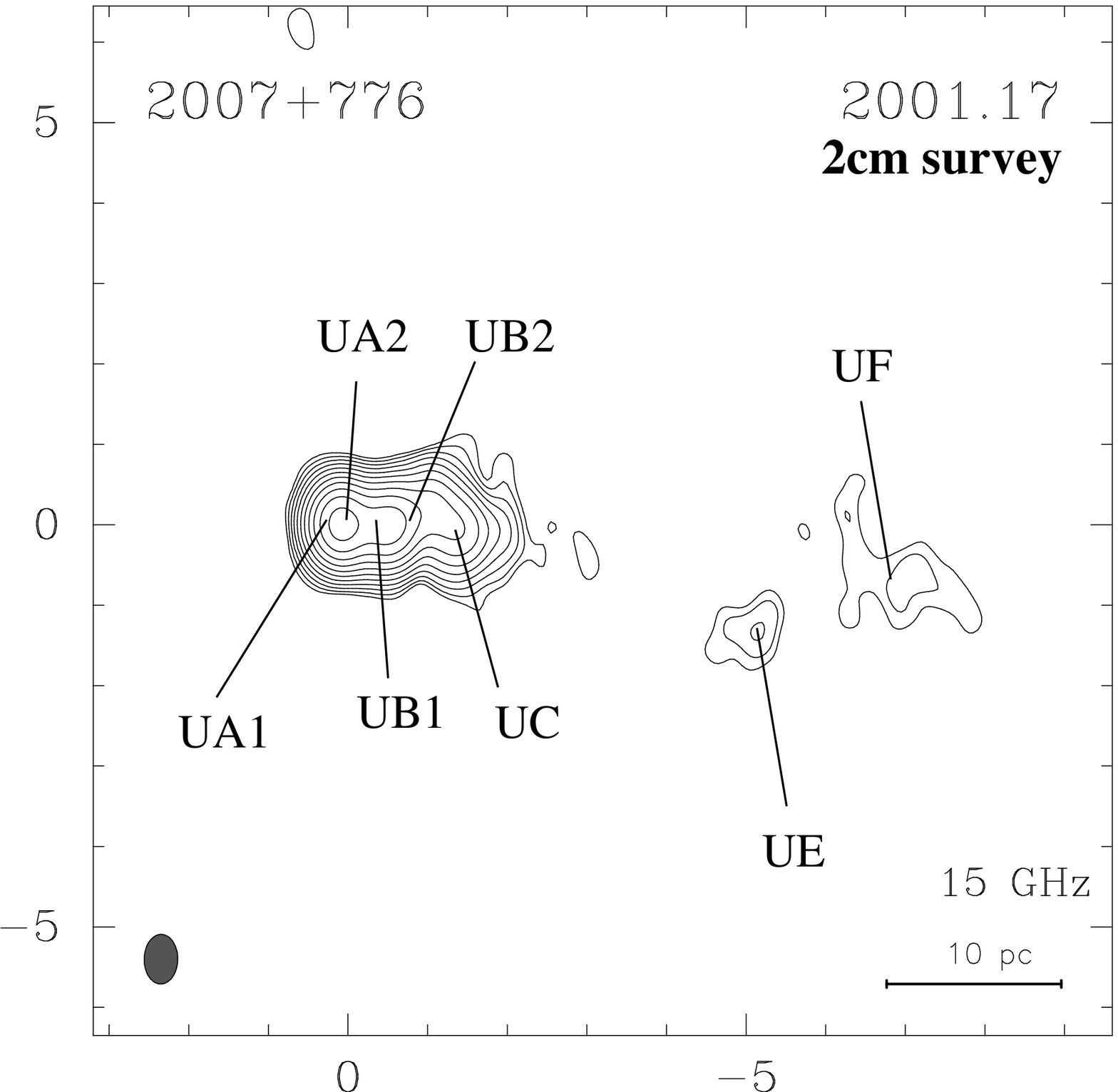} \\
\end{tabular}
\caption[]{
VLBA images of \object{BL\,2007+777} from observations on
27 July 1999 (1999.57), 6 November 1999 (1999.85),
15 June 2000 (2000.46), and 4 March 2001 (2001.17).
See Tables 1--3 for contour levels, beam sizes (bottom left in the maps),
peak flux densities, and component parametrization.
Axes are relative $\alpha$ and $\delta$ in mas.
\label{fig:map2007}
}
\end{figure*}

Our VLBA maps of \object{BL\,2007+777}
(Fig.\ \ref{fig:map2007}, $z$=0.342)
display a one-sided core-jet structure oriented westward.
The jet extends up to 6.9\,mas (31\,\hminus\,pc).
The 15.4\,GHz total flux density of \object{BL\,2007+777}
seems to have steadily increased between 1999.57 and 2000.46,
from 825\,mJy up to 1258\,mJy (52\%). 
From then on, the source started to fade down to 948\,mJy in 2001.17,
corresponding to a decrease of 25\% in less than 
nine months.  Its monochromatic luminosity is quite modest
($L_{\rm 15.4\,GHz}\approx$(4.3 to 6.6)$\times 10^{37}$\,W); 
it is the least powerful source after \object{BL\,0454+844}.

We fitted its brightness radio structure with six components
at all epochs except 2000.46, where only five components were 
model fitted
(Table \ref{table:modelfit}). 
We need up to four components (UA1, UA2, UB1, and UB2) to adequately describe
the inner 1-mas region ($4.6\,h^{-1}$\,pc).
In contrast, the modeling of this region 
at 8.4\,GHz required only two components, 
one at the origin (XA), and a second one at $r=0.55\pm0.05$\,mas (XB). 
The brightest 15.4\,GHz component at all epochs is UA1, which is not
at the origin, but shifted 
$0.3\pm0.1$\,mas eastward.
This component went undetected at 8.4\,GHz (Paper I).
If this finding is confirmed by our 43\,GHz observations, we
have another case, along with those of
\object{QSO\,0212+735} and \object{QSO\,1928+738}, 
in which the core position is shifted
from the reference point normally used, the peak-of-brightness.
Component UA2 is at the origin, component UB1 is 
at 
$r$=0.35$\pm$0.05\,mas at angle -85\degr, 
and component UB2 is at $r$=0.85$\pm$0.25\,mas at angle 
-83\degr$\pm$7\degr. 
Our model fit shows a significant proper motion for component UB1, 
and marginally significant ones for UA1, UC, and UF (Table~\ref{tab:proper_motion}).
The model fitting procedure merged UB1 and UB2 into just one component
in 2000.46. 
Note that the flux density of UB1 in 2000.46 is more than 
50\% greater than at the other three epochs, and close to 
the common flux of UB1+UB2 in the closest epochs. 
Hence we suggest that the motion of an undetected component, 
probably traveling from UA2 toward UC (and passing through UB1), 
could be responsible for the blending.
This could, in turn, explain the apparent paradox of a forward superluminal 
motion of UB1, and a backward one for UC.
The ratio $Q$ of the emission from the core region (taken as UA1+UA2) 
to the extended emission shows a moderate dominance of the core  
which modulates the overall emission:
1.59 (1999.57), 1.63 (1999.85), 1.71 (2000.46), and 1.21 (2001.17).
The total intensity emission is strongly dominated by that of the innermost region
(the inner 5\hminus\,pc).
Indeed, the combined emission from components UA1, UA2, UB1, and UB2
amounts to 90\% of the total at both epochs.
Component UC contributes 3\% to 9\%, 
and component UF 2\% to 5\%.  
Assuming that the 8.4\,GHz flux density of the milliarcsecond
structure of \object{BL\,2007+777}  did not change
appreciably between 1999.41 and 1999.57, we obtain $\alpha$=-0.16, a 
moderately steep spectrum.
Despite the non-varying global contribution to the flux density 
from the inner 1 mas structure
of \object{BL\,2007+777}, significant flux density variations do exist
for each component.

Component UC, at a distance of 
$1.28\pm0.10$\,mas 
at P.A. -92\degr$\pm$5\degr
is likely to correspond to component XC in Paper I.
We fail, however, to see any clear counterpart to component XD 
($r = 1.65\pm0.05$\,mas at angle -92\degr$\pm$6\degr; Paper I).
We encounter  a similar situation with respect to component XE
($r = 4.75\pm0.05$\,mas at -100\degr), for which we only find a
clear counterpart in our last epoch (UE, at $r\approx$5.1\,mas at P.A.
$\approx-106\degr$).
This component indicates a curvature in the jet of 
\object{BL\,2007+777}, as confirmed by our previous 8.4\,GHz 
VLBA observations (paper I), and VSOP observations
at 5\,GHz (Peng \et\ \cite{peng00}, Jin \et\ \cite{jin01}).
From the images, there are hints for the existence of component UE
at the other epochs, with marginal detections just above 
3$\sigma$,
but which our model fit failed to reproduce. 
If real, the inferred spectral index 
is $\alpha_{\rm UE/XE}\lsim-1.9$.
The 15.4\,GHz jet reappears at an angular distance of
$r = 6.5\pm0.4$\,mas, 
at an angle of $\approx-95\degr$ (component UF). 
The similar component at 8.4\,GHz (XF in Paper I), is at 
$r\approx$6.7\,mas, which agrees relatively well. 
In the maps, there seems to be more than just one, albeit
weak, component around 
6\,mas to 7\,mas
 westward. 
Although the model fit could not confirm our suspicion, 
we suggest UF is the blending of at least two sub-components, which
would explain the large variations we find in its radial coordinate.

Our non-detection of a component around 
1.7\,mas to 2.0\,mas
might be somewhat uncomfortable, since this component 
seems to have been detected also by 5\,GHz VSOP
observations close to those reported in Paper I (Krichbaum \et\ 2000).
Although the use of these data (taken at different wavelengths and
with different arrays) casts some doubts on their identification, 
we propose the following explanation:
Component UA1 is the stationary core of \object{BL\,2007+777}, 
which lies at $r\gsim$0.2\,mas to 0.4\,mas eastward from the point normally
used as reference.
At lower frequencies, UA1 is not seen alone, but as a blended
component (UA), at the origin. 
If we simply subtract the shift, the position of the ``unseen'' component would 
correspond with that of UC.

\section{Summary\label{sec:summary}}

We observed the thirteen extragalactic radio sources  
of the S5 polar cap sample at 15.4\,GHz with the Very Long Baseline Array, 
on 27 July 1999 (1999.57) and 15 June 2000 (2000.46).
We present the maps from those two epochs, along
with maps obtained from observations of the 2\,cm VLBA survey
for some of the sources of the sample, making a total of 40 maps.
Our observations correspond to the first two epochs at 15.4\,GHz of a program
directed to study the absolute kinematics of the radio source 
components of the members of the sample, by means of phase delay
astrometry at 8.4\,GHz, 15.4\,GHz, and 43\,GHz.
We summarize our main findings from our 
VLBA imaging at 15.4\,GHz of the sources of the complete 
sample as follows:

\begin{itemize}

\item
All of the 13 radio sources display at 15.4\,GHz prominent 
one-sided jet structures, 
except for \object{QSO\,0615+820}. This source shows the most
intriguing radio structure of all the sources of the sample, 
displaying a complex structure despite being 
the most compact QSO of the sample ($l\lsim$15.4\hminus\,pc).

\item
The imaging at 15.4\,GHz yielded a better resolved view
of the sources, compared to our previous imaging
at 8.4\,GHz (Paper I). 
In this way, we were able to disentangle the 
inner milliarcsecond structure of some of the sources, thus resolving
out components that appeared blended at 8.4\,GHz.
For most of the sources we identified the brightest 
feature in each radio source with the core. 
These identifications are supported by the spectral index
estimates for those brightest features
($\alpha$ ranges from 0.04 up to 1.99) except for 
\object{QSO\,0615+820}, for which we find 
$\alpha$=-0.69 for the apparent core of the source.
Higher frequency observations might be
of great help to disentangle the radio brightness structure of this
challenging object.
The putative core of \object{QSO\,1150+812} also shows a moderately steep spectrum
($\alpha\approx -0.13$).

\item
The sources cover a wide range of luminosity values that spans
almost four orders of magnitude: 
from the least luminous object, \object{BL\,0454+844} 
($L_{\rm 15\,GHz}\approx$(0.6 to 1.3)$\times 10^{36}$\,W), 
up to the most luminous one, 
\object{QSO\,0212+735}, with a monochromatic luminosity 
of $L_{\rm 15\,GHz} = (9.25\pm0.05)\times 10^{39}\,$W.
The luminosity of \object{QSO\,0212+735} is so 
impressive that its weakest
component (UG, at the not less impressive distance of 85\hminus\,pc 
from the core) has $L_{\rm 15\,GHz}\approx$1.2$\times 10^{38}\,$W, 
about 100 times more powerful than \object{BL\,0454+844}, and 
competes in power with several other sources of the complete sample.

\item
Most of the sources show core-dominance in the overall
emission, as given by the core-to-extended ratio, $Q$.
Indeed, $Q$ is significantly greater than unity 
for most sources and epochs. 
Nevertheless, there is a large spread in the values of $Q$, 
indicating a corresponding spread in the core dominance.
The four sources with the largest core dominance are
\object{BL\,0716+714} ($Q$ = 6 to 22),
\object{QSO\,1039+811} ($Q$ = 19 to 21), 
for \object{BL\,0454+844} ($Q$ = 9 to 16), and
\object{QSO\,0016+731} ($Q$ = 4 to 11).
The rest of the sources show moderate-to-large core dominance, 
with values in the range 1.3 to 5.2. 
The two sources with the lowest $Q$ values are
\object{QSO\,0212+735} ($Q$ = 0.8 to 1.0)
and 
\object{QSO\,0615+820} ($Q$ = 0.9 to 1.4).
While \object{QSO\,0212+735} has a low value due 
to the emission strength of its innermost jet structure, 
the case of 
\object{QSO\,0615+820} is slightly different. Here, the
halo-like, albeit relatively compact, emission competes in strength 
with the component we tentatively
identify with the core (UA1).


\item 
Three of the sources have the most inverted spectrum component
shifted with respect to the origin in the map, which approximately 
coincides with the peak-of-brightness at both 15.4\,GHz and 8.4\,GHz. 
At 15.4\,GHz, the offsets of these inverted components are:
\object{QSO\,0212+735}, $\Delta r$=0.6\,mas; 
\object{QSO\,1928+738}, $\Delta r$=0.35$\pm$0.05\,mas; 
\object{BL\,2007+777}, $\Delta r$=0.25$\pm$0.05\,mas.
Our 8.4\,GHz observations in 1999.41 (Paper I), close
to our first 15.4 GHz observations (1999.57), 
also show offsets for  
\object{QSO\,0212+735} ($\Delta r\,\approx$\,0.6\,mas) 
and \object{QSO\,1928+738} ($\Delta r\,\approx$\,0.5\,mas), but
not for \object{BL\,2007+777}, whose peak-of-brightness 
coincides with the origin in the map.
In the standard synchrotron self-absorption theory, 
the high-frequency components are closer to the central machine
than the low-frequency ones. 
Within their positional uncertainties, we find that the source morphologies 
of those objects, as seen at 8.4 and 15.4\,GHz, are satisfactorily explained 
by the standard scenario.

\end{itemize}

\begin{acknowledgements}
MAPT is supported by the Spanish National program Ram\'on y Cajal.
We are very grateful to Alan Roy for a careful reading of our manuscript, and 
for his idiomatic and grammar corrections.
We are also grateful to the 2cm VLBA survey team for kindly allowing us
to use part of their data.
This work has been partially financed by Grants AYA2001-2147-C02-01 
and AYA2001-2147-C02-02 of the Spanish DGICYT, 
and by the European Grant IHP-MCFI-99-1.
The NRAO is a facility of the National Science Foundation 
operated under cooperative agreement by Associated Universities, Inc.
This research has made use of the NASA Astrophysics Data System.
\end{acknowledgements}

\end{document}